{}
{}

\documentclass[a4paper,12pt]{article}

\newif\ifpublic\publicfalse

\publictrue

\usepackage{amsmath}            
\usepackage{amssymb}            
\usepackage{graphicx}           
\usepackage[nosort]{cite}
\usepackage{mathrsfs}
\ifpublic\else\usepackage{showkeys}\fi
\usepackage[bookmarks=true,hyperfigures=true]{hyperref}
\usepackage[bulletsep]{collref}

\usepackage{fixmath}

\expandafter\def\expandafter\bfseries\expandafter{\bfseries\ifmmode\else\boldmath\fi}
\expandafter\def\expandafter\mdseries\expandafter{\mdseries\ifmmode\else\unboldmath\fi}
\expandafter\def\expandafter\normalfont\expandafter{\normalfont\ifmmode\else\unboldmath\fi}

\allowdisplaybreaks[3]

\numberwithin{equation}{section}

\usepackage[font=small,labelfont=bf,width=0.85\textwidth]{caption}

\RequirePackage{amsmath}
\makeatletter
\newcommand{\brk@ord}{\bBigg@{0}}
\newcommand{\brk@ordl}{\mathopen\brk@ord}
\newcommand{\brk@ordr}{\mathclose\brk@ord}
\newcommand{\brk@ordm}{\mathrel\brk@ord}
\newcommand{\brk@var}{\brk@ord}
\newcommand{\brk@varl}{\left}
\newcommand{\brk@varr}{\right}
\newcommand{\brk@varm}{\mathrel\brk@var}
\newcommand{\brk@altname}[3]{\expandafter\def\csname#2\expandafter\@gobble\string#1\endcsname{#1[#3]}}
\newcommand{\brk@usearg}[3]{%
  \def\brk@star{*}\def\brk@blank{}\def\brk@arg{#1}%
  \ifx\brk@arg\brk@blank\def\brk@arg{brk@ord}\fi%
  \ifx\brk@arg\brk@star\def\brk@arg{brk@var}\fi%
  \csname\brk@arg #2\endcsname#3}

\newcommand{\DeclareMathBrackets}[3]{
  \newcommand{#1}[2][]{\brk@usearg{##1}{l}{#2}##2\brk@usearg{##1}{r}{#3}}
  \brk@altname{#1}{big}{big}\brk@altname{#1}{lr}{*}}
\newcommand{\DeclareMathBiBrackets}[4]{
  \newcommand{#1}[3][]{\brk@usearg{##1}{l}{#2}##2#3##3\brk@usearg{##1}{r}{#4}}
  \brk@altname{#1}{big}{big}\brk@altname{#1}{lr}{*}}
\newcommand{\DeclareMathBiMBracketsStar}[4]{
  \newcommand{#1}[3][]{\brk@usearg{##1}{l}{#2}##2\brk@usearg{##1}{m}{#3}##3\brk@usearg{##1}{r}{#4}}
  \brk@altname{#1}{bi}{big}}
\newcommand{\DeclareMathBiBracketsStar}[4]{
  \newcommand{#1}[3][]{\brk@usearg{##1}{l}{#2}##2\brk@usearg{##1}{}{#3}##3\brk@usearg{##1}{r}{#4}}
  \brk@altname{#1}{big}{big}}

\makeatother

\DeclareMathBrackets{\brk}{(}{)}
\DeclareMathBrackets{\vev}{\langle}{\rangle}
\DeclareMathBrackets{\bra}{\langle}{|}
\DeclareMathBrackets{\ket}{|}{\rangle}
\DeclareMathBrackets{\sbra}{\langle}{|}
\DeclareMathBrackets{\sket}{|}{\rangle}
\DeclareMathBrackets{\cbra}{[}{|}
\DeclareMathBrackets{\cket}{|}{]}
\DeclareMathBrackets{\sbrk}{[}{]}
\DeclareMathBrackets{\set}{\{}{\}}
\DeclareMathBrackets{\abs}{|}{|}
\DeclareMathBrackets{\eval}{.}{|}
\DeclareMathBiBrackets{\comm}{[}{,}{]}
\DeclareMathBiBrackets{\acomm}{\{}{,}{\}}
\DeclareMathBiBrackets{\gcomm}{[}{,}{\}}
\DeclareMathBiMBracketsStar{\braket}{[}{|}{\}}

\def\showkeysrefformat#1{{\normalfont\tiny\ttfamily#1}}
\makeatletter
\def\SK@@ref#1>#2\SK@{%
 {\@inlabelfalse\leavevmode\vbox to\z@{%
 \vss\SK@refcolor\rlap{\vrule\raise .75em%
  \hbox{\showkeysrefformat{#2}}}}}}
\makeatother

\usepackage[a4paper,text={160mm,247mm},centering]{geometry}

\newcommand{\order}[1]{\mathcal{O}(#1)}          

\newcommand{\grp}[1]{\mathrm{#1}}

\def\eqn{\eqref}

\DeclareMathOperator{\tr}{tr}

\newcommand{\superN}{\mathcal{N}}
\newcommand{\trans}{{\scriptscriptstyle\mathsf{T}}}

\newcommand{\sfrac}[2]{{\textstyle\frac{#1}{#2}}}
\newcommand{\half}{\sfrac{1}{2}}

\DeclareMathOperator{\pathord}{P}

\newcommand{\gen}[1]{\mathrm{#1}}

\newcommand{\genfield}[1]{\mathbb{#1}}

\newcommand{\der}{\mathrm{d}}

\newcommand{\cdel}{\mathscr{D}}
\newcommand{\del}{\partial}
\newcommand{\sdel}{D}
\newcommand{\scdel}{\mathcal{D}}
\newcommand{\scdelF}{\scdel}
\newcommand{\scdelB}{\scdel}
\newcommand{\scdelC}{\bar{\scdel}}

\newcommand{\sgauge}{\mathcal{A}}
\newcommand{\spre}{\mathcal{B}}

\newcommand{\sscalar}{\Phi}

\newcommand{\sfstr}{\mathcal{F}}

\newcommand{\swilson}{\mathcal{W}}

\newcommand{\sviel}{e}
\newcommand{\svielB}{e}
\newcommand{\svielF}{\der\theta}
\newcommand{\svielC}{\der\bar\theta}
\newcommand{\dual}{\mathord{\ast}}

\newcounter{numidx}
\newcommand{\idxtspin}[1]{{\setcounter{numidx}{\number`#1}\addtocounter{numidx}{-65}\hat{%
  \ifcase\arabic{numidx}\alpha\or\beta\or\gamma\or\epsilon\else\zeta\fi}}}
\newcommand{\idxtvec}[1]{{\setcounter{numidx}{\number`#1}\addtocounter{numidx}{-109}\hat{%
  \ifcase\arabic{numidx}\mu\or\nu\or\rho\or\sigma\or\kappa\or\lambda\else\xi\fi}}}
\newcommand{\idxall}[1]{{\mathcal{#1}}}
\newcommand{\idxconf}[1]{{\setcounter{numidx}{\number`#1}\addtocounter{numidx}{-97}%
  \ifcase\arabic{numidx}\delta\or\rho\or\kappa\or\sigma\else\omega\fi}}
\newcommand{\Real}{\mathbb{R}}
\newcommand{\Comp}{\mathbb{C}}
\newcommand{\Quat}{\mathbb{H}}

\newcommand{\nln}{\nonumber\\}

\def\[{\begin{equation}}
\def\]{\end{equation}}

\providecommand{\href}[2]{#2}
\newcommand{\arxivlink}[1]{\href{http://arxiv.org/abs/#1}{arxiv:#1}}

\makeatletter
\def\mr@ignsp#1 {\ifx\:#1\@empty\else #1\expandafter\mr@ignsp\fi}%
\newcommand{\multiref}[1]{\begingroup
\xdef\mr@no@sparg{\expandafter\mr@ignsp#1 \: }%
\def\mr@comma{}%
\@for\mr@refs:=\mr@no@sparg\do{\mr@comma\def\mr@comma{,}\ref{\mr@refs}}%
\endgroup}
\renewcommand{\eqref}[1]{(\multiref{#1})}
\makeatother

\makeatletter
\newcommand{\namedref}[2]{\hyperref[#2]{#1~\ref*{#2}}}
\newcommand{\secref}{\@ifstar{\namedref{Section}}{\namedref{Sec.}}}
\newcommand{\appref}{\@ifstar{\namedref{Appendix}}{\namedref{App.}}}
\newcommand{\tabref}{\@ifstar{\namedref{Table}}{\namedref{Tab.}}}
\newcommand{\figref}{\@ifstar{\namedref{Figure}}{\namedref{Fig.}}}
\makeatother

\let\oldbib=\thebibliography
\def\thebibliography{\phantomsection\addcontentsline{toc}{section}{\refname}\oldbib}

\let\oldtoc=\tableofcontents
\def\tableofcontents{\phantomsection\addcontentsline{toc}{section}{\contentsname}\oldtoc}

\RequirePackage{verbatim}

\makeatletter
\newwrite\mpi@out
\def\mpi@write#1{\immediate\write\mpi@out{#1}}
\immediate\openout\mpi@out\jobname.mp
\mpi@write{\@percentchar generated from `\jobname'}%
\AtEndDocument{\mpostdone}

\def\mpostdone{
  \immediate\closeout\mpi@out%
  \ifpublic\else%
    \immediate\write18{mpost -tex=latex \jobname.mp}
  \fi%
  \gdef\mpostdone{}
}

\newcommand{\mpi@putlineno}{%
  \mpi@write{\@percentchar---------------------------------------}%
  \mpi@write{\@percentchar l.\the\inputlineno}%
}

\newcommand{\mpi@verbatim}{
  \@bsphack
  \let\do\@makeother\dospecials
  \catcode`\^^M\active
  \def\verbatim@processline{\mpi@write{\the\verbatim@line}}%
  \verbatim@start
}

\newenvironment{mpostcmd}{%
  \mpi@putlineno%
  \mpi@verbatim%
}%
{\mpi@write{}\@esphack}

\newenvironment{mpostfile}[1]{%
  \mpi@putlineno%
  \mpi@write{filenametemplate "#1";}%
  \mpi@write{beginfig(0)}%
  \mpi@verbatim%
}%
{\mpi@write{endfig;}\@esphack}

\newcommand{\includegraphicsex}[2][]{%
  \xdef\mpi@tmp{#2}%
  \IfFileExists{\mpi@tmp}%
    {\includegraphics[#1]{\mpi@tmp}}%
    {\textbf{??}\typeout{file \mpi@tmp{} missing}}%
}

\makeatother

\makeatletter
\newsavebox{\apb@box}\newlength{\apb@width}
\newcommand{\autoparbox}[2][c]{\sbox{\apb@box}{#2}%
 \settowidth{\apb@width}{\usebox{\apb@box}}%
 \parbox[#1]{\apb@width}{\usebox{\apb@box}}}
\newcommand{\includegraphicsbox}[2][]{\autoparbox{\includegraphicsex[#1]{#2}}}
\makeatother

\providecommand{\hypersetup}[1]{}
\providecommand{\texorpdfstring}[2]{#1}

\hypersetup{plainpages=false}
\hypersetup{pdfpagemode=UseNone}
\hypersetup{bookmarksnumbered=true}
\hypersetup{pdfstartview=FitH}
\hypersetup{colorlinks=false}
\hypersetup{citebordercolor={.5 1 .5}}
\hypersetup{urlbordercolor={.5 1 1}}
\hypersetup{linkbordercolor={1 .7 .7}}

\makeatletter
\let\@keywords\@empty
\let\@subject\@empty
\providecommand{\keywords}[1]{\gdef\@keywords{#1}}
\providecommand{\subject}[1]{\gdef\@subject{#1}}
\def\thetitle{\@title}
\def\theauthor{\@author}
\def\thesubject{\@subject}
\def\thedate{\@date}
\def\thekeywords{\@keywords}
\makeatother
\AtBeginDocument{
\hypersetup{pdftitle={\thetitle}}%
\hypersetup{pdfauthor={\theauthor}}%
\hypersetup{pdfsubject={\thesubject}}%
\hypersetup{pdfkeywords={\thekeywords}}%
}

\begin{mpostcmd}
prologues:=2;

verbatimtex
\documentclass{article}
\usepackage{amsfonts,amssymb,latexsym}
\begin{document}
etex
\end{mpostcmd}

\begin{mpostcmd}
picture copyrightline,copyleftline;
copyrightline := btex \copyright\ \textsf{2015 Niklas Beisert} etex;
copyleftline := btex $\circledast$ etex;
def putcopyspace =
label.bot(btex \vphantom{gA} etex scaled 0.1, lrcorner(currentpicture));
enddef;
def putcopy =
label.ulft(copyrightline scaled 0.1, lrcorner(currentpicture)) withcolor 0.9white;
label.urt(copyleftline scaled 0.1, llcorner(currentpicture)) withcolor 0.9white;
currentpicture:=currentpicture shifted (10.5cm,14cm);
enddef;
\end{mpostcmd}

\begin{mpostcmd}
def pensize(expr s)=withpen pencircle scaled s enddef;
xu:=1cm;
pair pos[];
path paths[];
def midarrow (expr p, t) =
fill arrowhead subpath(0,arctime(arclength(subpath (0,t) of p)+0.5ahlength) of p) of p;
enddef;
\end{mpostcmd}

\begin{mpostfile}{FigKappa.mps}
randomseed:=992;
paths[1]:=((-1.5xu,-0.5xu){dir -10}..{dir -50}(+2xu,+0.0xu)) rotated -5;
paths[2]:=
for i=0 upto 10:
( (point (i/10) of paths[1]) + ((unitvector direction (i/10) of paths[1] rotated 90)*4pt*((uniformdeviate(2))-1)))
if i<10: .. fi
endfor;
linecap:=butt; draw paths[1] pensize(10pt) withcolor 0.85white; linecap:=rounded;
draw paths[1] pensize(1.5pt);
draw paths[2] pensize(0.5pt);
ahlength:=5pt;
midarrow(paths[1],0.6);
label.lft(btex $\tilde Z$ etex, point 0 of paths[1]);
label.top(btex $\delta_\kappa\tilde Z$ etex, point 2 of paths[2]);
putcopyspace;putcopy;
\end{mpostfile}


\begin{mpostfile}{FigPoly.mps}
def curve(expr a,b,w)=
a{(b-a) rotated w}..{(b-a) rotated -w}b
enddef;
pos[1]:=( 1.86778xu, 1.78537xu);
pos[2]:=( 2.28603xu, 0.87497xu);
pos[3]:=( 1.29332xu, 0.03263xu);
pos[4]:=( 0.68781xu, -0.10141xu);
pos[5]:=(-0.08606xu, 0.58860xu);
pos[6]:=( 0.22305xu, 2.05081xu);
pos[7]:=( 0.94058xu, 1.78942xu);
paths[1]:=curve(pos[1],pos[2],uniformdeviate(60)-30);
paths[2]:=curve(pos[2],pos[3],uniformdeviate(60)-30);
paths[3]:=curve(pos[3],pos[4],uniformdeviate(60)-30);
paths[4]:=curve(pos[4],pos[5],uniformdeviate(60)-30);
paths[5]:=curve(pos[5],pos[6],uniformdeviate(60)-30);
paths[6]:=curve(pos[6],pos[7],uniformdeviate(60)-30);
paths[7]:=curve(pos[7],pos[1],uniformdeviate(60)-30);
linecap:=butt;
for i=1 upto 7:
draw paths[i] pensize(10pt) withcolor (0.80white+(uniformdeviate(1),uniformdeviate(1),uniformdeviate(1))*0.1);
endfor
linecap:=rounded;
for i=1 upto 7:
draw point 0 of paths[i]..
for k=1 upto 9:
( (point (k/10) of paths[i]) + ((unitvector direction (k/10) of paths[i] rotated 90)*2pt*((uniformdeviate(2))-1)))..
endfor point 1 of paths[i];
endfor
putcopyspace;putcopy;
\end{mpostfile}

\begin{mpostfile}{FigSpline.mps}
def curve(expr a,b,w)=
a{(b-a) rotated w}..{(b-a) rotated -w}b
enddef;
pos[1]:=( 1.86778xu, 1.78537xu);
pos[2]:=( 2.28603xu, 0.87497xu);
pos[3]:=( 1.29332xu, 0.03263xu);
pos[4]:=( 0.68781xu, -0.10141xu);
pos[5]:=(-0.08606xu, 0.58860xu);
pos[6]:=( 0.22305xu, 2.05081xu);
pos[7]:=( 0.94058xu, 1.78942xu);

paths[10]:=pos[1]--pos[2]--pos[3]--pos[4]--pos[5]--pos[6]--pos[7]--cycle;
ang:=111.5;
for i=1 upto 7:
paths[i]:=curve(point (i-1) of paths[10], point i of paths[10], ang);
ang:=angle(direction (i-0.5) of paths[10])-angle(direction (i+0.5) of paths[10])-ang;
endfor
linecap:=butt;
for i=1 upto 7:
draw paths[i] pensize(10pt) withcolor (0.80white+(uniformdeviate(1),uniformdeviate(1),uniformdeviate(1))*0.1);
endfor
linecap:=rounded;
for i=1 upto 7:
draw point 0 of paths[i]..
for k=1 upto 9:
( (point (k/10) of paths[i]) + ((unitvector direction (k/10) of paths[i] rotated 90)*2pt*((uniformdeviate(2))-1)))..
endfor point 1 of paths[i];
endfor
putcopyspace;putcopy;
\end{mpostfile}

\begin{mpostfile}{FigCorner.mps}
def curve(expr a,b,w)=
a{(b-a) rotated w}..{(b-a) rotated -w}b
enddef;
pos[1]:=( 1.86778xu, 1.78537xu) rotated -60 scaled 1.5;
pos[2]:=( 2.28603xu, 0.87497xu) rotated -60 scaled 1.5;
pos[3]:=( 1.29332xu, 0.03263xu) rotated -60 scaled 1.5;
paths[1]:=curve(pos[1],pos[2],30);
paths[2]:=curve(pos[2],pos[3],-30);
linecap:=butt;
draw paths[1] pensize(10pt) withcolor (0.8,0.9,0.8);
draw paths[2] pensize(10pt) withcolor (0.8,0.8,0.9);
linecap:=rounded;
paths[3]:=((-unitvector direction 1 of paths[1])--(0,0)) scaled 0.5xu shifted (point 1 of paths[1]);
paths[4]:=((0,0)--(unitvector direction 0 of paths[2])) scaled 0.5xu shifted (point 0 of paths[2]);
drawarrow paths[3];
drawarrow paths[4];
label.rt(btex $(p_-,\dot\theta_-,q_-)$ etex, point 0 of paths[3]);
label.top(btex $(p_+,\dot\theta_+,q_+)$ etex, point 1 of paths[4]);
putcopyspace;putcopy;
\end{mpostfile}

\begin{mpostcmd}
verbatimtex
\end{document}
etex

end;
\end{mpostcmd}

\mpostdone

\title{Smooth Wilson Loops\texorpdfstring{\\}{ }in
\texorpdfstring{$\mathcal{N}=4$}{N=4} Non-Chiral Superspace
}
\author{Niklas Beisert, Dennis M\"uller, Jan Plefka, Cristian Vergu}

\begin{document}
\pdfbookmark[1]{Title Page}{title}
\thispagestyle{empty}

\begingroup\raggedleft\footnotesize\ttfamily
HU-EP-15/20\\
\arxivlink{1506.07047}\par
\vspace{15mm}
\endgroup

\begin{center}
{\Large\bfseries\thetitle\par}%
\vspace{15mm}

\begingroup\scshape\large
Niklas Beisert${}^{1}$, Dennis M\"uller${}^{2}$, Jan Plefka${}^{1,2}$%
\\
and Cristian Vergu${}^{1,3}$\par
\endgroup
\vspace{5mm}

\textit{${}^{1}$ Institut f\"ur Theoretische Physik,\\ Eidgen\"ossische
  Technische Hochschule
  Z\"urich,\\ Wolfgang-Pauli-Strasse 27, 8093 Z\"urich, Switzerland}\\[0.1cm]
{\small\verb+nbeisert@itp.phys.ethz.ch+} \vspace{5mm}

\textit{${}^{2}$ Institut f\"ur Physik und IRIS Adlershof,
\\ Humboldt-Universit\"at zu Berlin, \phantom{$^\S$}\\
  Zum Gro{\ss}en Windkanal 6, D-12489 Berlin, Germany} \\[0.1cm]
{\small\verb+{dmueller,plefka}@physik.hu-berlin.de+} \vspace{5mm}

\textit{${}^{3}$ Department of Mathematics, King's College London \\
The Strand, WC2R 2LS, London, UK} \\[0.1cm]
{\small\verb+c.vergu@gmail.com+} \vspace{8mm}


\textbf{Abstract}\vspace{5mm}\par
\begin{minipage}{14.7cm}
We consider a supersymmetric Wilson loop operator for 4d $\mathcal{N}=4$ super Yang-Mills theory
which is the natural object dual to the $AdS_{5}\times S^{5}$ superstring in the
AdS/CFT correspondence. It generalizes the traditional bosonic $1/2$ BPS Maldacena-Wilson loop
operator and completes recent constructions in the literature
to smooth (non-light-like) loops in the
full $\mathcal{N}=4$ non-chiral superspace. This Wilson loop operator enjoys global superconformal
and local kappa-symmetry of which a detailed discussion is given.
Moreover, the finiteness of its vacuum expectation value is proven
at leading order in perturbation theory.
We determine the leading vacuum expectation value for general paths both at the component
field level up to quartic order in anti-commuting coordinates
and in the full non-chiral superspace in suitable gauges.
Finally, we discuss loops built from quadric splines joined in such a way
that the path derivatives are continuous at the intersection.
\end{minipage}\par
\end{center}
\newpage

 \renewcommand{\thefootnote}{\arabic{footnote}} \setcounter{footnote}{0}

 \setcounter{tocdepth}{2} \hrule height 0.75pt
\tableofcontents
 \vspace{0.8cm} \hrule height 0.75pt \vspace{1cm}


\section{Introduction}
\label{sec:introduction}

The four-dimensional, maximally supersymmetric Yang-Mills ($\mathcal{N} = 4$ SYM)
theory is a highly symmetric and distinguished model uniting all of the concepts of gauge field theory.
This makes it an ideal quantum field theory laboratory. The model is superconformally invariant
and uniquely specified by a choice of coupling $g$
and gauge group. For the choice of  a $\grp{U}(N)$ gauge group
in the planar large $N$-limit the $\mathcal{N} = 4$ SYM theory
is conjectured to be both,  integrable \cite{Beisert:2010jr} and
dual to the free IIB superstring on an $AdS_{5}\times S^{5}$ background
\cite{Maldacena:1997re}. By now there is overwhelming evidence for these properties.
However, we are still lacking a general theory of integrability in four dimensional
$\mathcal{N} = 4$ SYM theory. So far we have detected integrability
in the study of a variety of gauge invariant observables
in the theory.
The best established examples
are the two-point functions of local gauge invariant operators related to integrable
spin-chain Hamiltonians, as well as the Yangian invariance of scattering amplitudes \cite{Drummond:2009fd}
and their weak-coupling dual description via light-like Wilson loops
\cite{Anastasiou2003, Bern2005, Alday2007a, Bern2008a, Brandhuber2008,
Drummond2008a, Drummond2008b, Drummond:2008aq, Drummond:2008vq, Drummond2010b, Kosower2011}.

The Wilson loop/scattering amplitude duality relates scattering amplitudes to a chiral
light-like polygonal Wilson loop in superspace. These chiral super Wilson loops were defined and studied
from a twistor space \cite{Mason2010} and  from a space-time perspective  \cite{Caron-Huot2011}.
The scattering amplitudes are IR divergent (due to massless particles)
while the polygonal Wilson loops (and also their supersymmetric cousins)
are UV-divergent (due to cusps and light-likeness).
Interestingly, the structure of these duality-related divergences
in dimensional regularization are known to all orders in perturbation
theory~\cite{Bern2005, Polyakov:1980ca, Korchemsky1987}.
Because of these divergences, some generators of the superconformal symmetry algebra
become anomalous preventing the use of the full symmetry in the problem.
Nevertheless, progress has been made in the understanding of the anomalous generators
\cite{Drummond2010b, Bullimore:2011kg, Belitsky:2012nu, Caron-Huot2012, Caron-Huot:2013vda}.
The hidden integrability in the problem manifests itself from studying
how the superconformal symmetries are mapped under the scattering 
amplitude/Wilson loop duality.
 In \cite{Drummond:2008vq} it was
shown that the usual and the dual superconformal algebras are different but partially overlap.
The closure of these transformations is a Yangian algebra \cite{Drummond:2009fd}, which is a
hallmark of integrability. Yet another way in which integrability
has appeared in the study of polygonal light-like Wilson loops
is via the Wilson loop operator product expansion initiated in \cite{Alday:2010ku}
and scrutinized in \cite{Basso:2013vsa,Basso:2013aha,Basso:2014koa,Basso:2014nra}.
 Motivated by an attempt to understand the \(\bar{\gen{Q}}\)-anomaly
of these dual Wilson loops, a non-chiral super Wilson loop was defined in
ref.~\cite{Caron-Huot2011a} to the first order in the  anti-commuting
coordinate \(\bar{\theta}\).
In refs.~\cite{Beisert2012a, Beisert2012} a more thorough study of these non-chiral
super Wilson loops was undertaken, including a study of the Yangian anomalies when using
different regularizations.

Here we turn to a prominent further class of non-local observables in $\mathcal{N}=4$ SYM,
which is in fact almost
as old as the proposal of the AdS/CFT correspondence itself: the Maldacena-Wilson
loop operators \cite{Maldacena:1998im,Rey:1998ik}. These extend the usual Wilson loop
operator, which integrates the gauge field $A_{\mu}$ along a bosonic
closed path $C$ in $\Real^{1,3}$, by an additional coupling to
the six adjoint scalar fields $\phi_{i}$ of the
$\mathcal{N}=4$ SYM theory to a path $\tilde C$ on
$\Real^{1,3}\times S^{5}$ via
\begin{align}
\tilde W= \frac{1}{N} \tr  \pathord \exp{ \left(\oint_{\tilde{C}} \mathrm{d} \tau
\left(  \dot{x}^{\mu} A_{\mu}(x) +  n^i \sqrt{\dot{x}^{\mu}\dot{x}_{\mu}} \, \phi_i (x)\right) \right)}
\qquad\text{with } ~
(n^{i})^2=1 \, . \label{eins}
\end{align}
The coupling is such that from an effective 10d perspective the path is light-like:
$\dot{x}^{\idxtvec{m}}=(\dot{x}^{\mu},n^{i}\sqrt{\dot{x}^{\mu}\dot{x}_{\mu}})$ with
$\dot{x}^{\idxtvec{m}}\, \dot{x}_{\idxtvec{m}}=0$ in mostly minus signature.
This operator is locally $1/2$ BPS and invariant under conformal transformations.
Remarkably, its vacuum expectation value is finite for smooth, non-intersecting loops
\cite{Drukker:1999zq} securing conformal invariance also at the quantum level.
In fact, for a circular loop  configuration (related to an
infinite straight line via inversion)
the exact non-perturbative vacuum expectation value was conjectured in
\cite{Erickson:2000af,Drukker:2000rr} in terms of a simple Bessel function
$I_{1}(\sqrt{\lambda})$ with $\lambda=g^{2} N$. This first exact result in
$\mathcal{N}=4$ SYM at finite coupling in fact predates the
discovery of integrability in the theory. The result was proven
later using localization techniques \cite{Pestun:2007rz}.

 In the AdS/CFT duality, the leading order strong coupling contribution to the Wilson loop
expectation value is computed by the (suitably regularized) minimal string worldsheet
ending on the boundary of $AdS_{5}$ on a contour prescribed
by the Wilson loop path~\cite{Rey:1998ik, Maldacena:1998im}.
For the $AdS_{5}\times S^{5}$  superstring model ref.~\cite{Berkovits2008a}
showed that the transformation which maps the superconformal symmetry
to its dual is an interesting combination of T-dualities,
some of which are along fermionic directions.
The fact that the dual generators correspond to non-local worldsheet generators
of Yangian type was emphasized in refs.~\cite{Beisert2008, Berkovits2008a}.
Such non-local generators in two-dimensional systems
reflect their classical integrability.
In recent work \cite{Munkler:2015gja} the minimal surface problem relevant for
a strong-coupling description of smooth Wilson loops in the gauge theory was
lifted to the full supersymmetric scenario of the $AdS_{5}\times S^{5}$ superstring.
Exploiting the hidden Yangian symmetry of the superstring model it was shown that
the vacuum expectation value of the smooth Wilson loop at strong coupling is
Yangian invariant. Recently smooth bosonic
Wilson loops where studied at strong coupling as a continuum limit of light-like
polygonal boundary curves in \cite{Toledo:2014koa}.

In this work we present the supersymmetric generalization
of the Maldacena-Wilson loop operator \eqref{eins} which is the natural dual object to
the minimal superstring surface in the AdS/CFT correspondence.
This construction was initiated in ref.~\cite{Muller:2013rta}, which also showed
the Yangian invariance of this object at leading order in the coupling
and anti-commuting coordinates $\theta$ and $\bar\theta$. Here we define the loop operator to all orders in anti-commuting coordinates
and study its symmetries.
The proposed smooth super Maldacena-Wilson loop operator in $\mathcal{N}=4$ superspace reads
\begin{align}
\label{ourloop}
\tilde{\mathcal{W}}= \frac{1}{N} \tr  \pathord \exp \biggl( \oint_{\tilde{Z}} \mathrm{d} \tau
\, \Bigr ( & p^{\mu}  \sgauge_{\mu}(x,\theta,\bar\theta)  +
q^{i}\, \Phi_{i}(x,\theta,\bar\theta)\, \nonumber \\
&+
\dot\theta^{a\alpha} \sgauge_{a\alpha}(x,\theta,\bar\theta) +
\dot{\bar\theta}^{\dot\alpha}{}_{a} \, \sgauge^{a}{}_{\dot\alpha}(x,\theta,\bar\theta)
  \Bigr) \, \biggr) \, ,
\end{align}
where
$p^{\mu}=\dot x^{\mu}+ \theta \sigma^{\mu} \dot{\bar\theta} -
\dot{\theta} \sigma^{\mu} \bar{\theta}$ and $q^{i}q^{i}=p^{\mu}p_{\mu}$
depend on the path $\tilde{Z}=\{x,\theta,\bar\theta;q\}$ in non-chiral $\mathcal{N}=4$
superspace and $\sgauge_{\mu}$,
$\sgauge_{a\alpha}$, $\sgauge^{a}{}_{\dot\alpha}$ and $\Phi_{i}$ are
fields on this superspace.
We note that the super-momentum $p^{\mu}$ is real due to the reality convention
 \[
\bar \theta = -i\theta^\dagger = \theta^\ddagger
\]
that we use in this work.
For Gra{\ss}mann variables \(\theta\), \(\psi\), we take \((\theta \psi)^\ddagger =
-\psi^\ddagger \theta^\ddagger\).
Most of the literature on
field theory, instead, uses the reality condition
\(\theta^\dagger = \bar{\theta}\) with \((\theta \psi)^\dagger =
\psi^\dagger \theta^\dagger\).
See \appref{sec:conventions} for a complete discussion of our conventions.
In fact, a related supersymmetrized Maldacena-Wilson loop as $\tilde{\swilson}$ in
\eqref{ourloop} was introduced some 15 years ago in \cite{Ooguri:2000ps}.

The smooth non-chiral super Wilson loop operator \eqref{ourloop} differs
from the chiral polygonal super Wilson loops in several ways.
First, it is non-chiral, second, it contains a coupling to scalar
and  fermions as well,
and third, it is defined on a smooth curve in superspace.
The super Maldacena-Wilson loop may be also thought of as the generalization of the non-chiral polygonal light-like
super-loop operator introduced in \cite{Beisert2012a, Beisert2012}.
It should be stressed that the definition \eqn{ourloop} necessarily uses an on-shell 
superspace with gauge fields  $\sgauge_{\mu}(x,\theta,\bar\theta)$,
$\sgauge_{a\alpha}(x,\theta,\bar\theta)$ and
$\sgauge^{a}{}_{\dot\alpha}(x,\theta,\bar\theta)$.  These need to be constrained
in order to eliminate the non-physical components of
these superfields.  We discuss them in more detail in
eqs.\ \eqref{eqn:first_relation}--\eqref{eqn:forth_relation}
and~\eqref{eq:constraints-4d-language}. In fact these constraints are equivalent to
the equations of motion as shown in \cite{Harnad:1984vk,Witten:1985nt,Harnad:1985bc}. 
In our work we shall not go beyond the
one-loop order in perturbation theory. 
To this order we will show that smooth, non-selfintersecting loops are finite. 
Therefore there are no ambiguities arising from the
on-shell nature of \eqn{ourloop}. It remains to be seen what happens at higher loops,
this is, however, left for future work. Hence, all our claims towards finiteness and
symmetries are only secured at leading order pertubation theory. Nevertheless, we are
convinced that they may be extended to all loops.

Our paper is organized as follows.
The smooth, non-chiral super Wilson loop that we present is intimately related to the superspace
formulation of $\mathcal{N}=1$ SYM in ten dimensions \cite{Witten:1985nt,Harnad:1985bc}
which we discuss in \secref*{sect:2}.
The dimensional reduction to 4d within this formalism provides the guideline for the
supersymmetric generalization of \eqref{eins}. In fact, we shall show that
the $1/2$ BPS property of the bosonic Maldacena-Wilson loop is the consequence
of the local fermionic kappa-symmetry of
the super Wilson loop inherited from the 10d superparticle in a suitably deformed fashion.
Moreover, the superconformal symmetry of our proposed super Wilson loop is proven at the operator level.
In \secref*{sect:1loopvevtheta} we determine the component field form of the super Wilson loop operator
upon expanding in the anti-commuting coordinates $\theta$ and $\bar\theta$.
Picking a specific
convenient fermionic gauge due to \cite{Harnad:1985bc} we
provide the full expansion for the linear or abelian theory of the super-loop operator and
compute the one-loop vacuum expectation value through fourth order in Gra{\ss}mann odd coordinates.
Within this expansion we moreover show the equivalence of the super Maldacena-Wilson loop operator
\eqref{ourloop} to the construction
of \cite{Muller:2013rta} at leading order in $\theta$ and $\bar\theta$.
Finiteness and superconformal symmetry of the one-loop
vacuum expectation value is shown at the component field level.
In \secref*{sect:4} we lift our analysis to a one-loop study in the full $\mathcal{N}=4$ non-chiral superspace.
For this the superspace propagators are established. Using these we prove the finiteness of
the loop expectation value to all orders in the theta expansion and display the superconformal
symmetry.
Finally, in \secref*{sec:conclusions} we conclude pointing to an upcoming companion paper
\cite{usyangian}
in which the Yangian symmetry of our super Maldacena-Wilson loop operator is investigated in the full superspace.

\section{Definition, Geometry and Symmetries of the Super Wilson Loop}
\label{sect:2}

\subsection{Superspace and Gauge Theory}
\label{sec:supersp-gauge-theory}

We begin by describing gauge theory in superspace.  This material
is standard (see ref.~\cite{Harnad:1985bc} for a good presentation),
but we use this opportunity to introduce our conventions.

We start the presentation with the \(\mathcal{N}=1\) super Yang-Mills theory
in ten dimensions, which will then be dimensionally reduced to four
dimensions.  In ten dimensions the superspace is parametrized by
\((x^{\idxtvec{m}}, \theta^\idxtspin{A})\), with \(\idxtvec{m}=0, \dotsc, 9\) and \(\idxtspin{A}=1, \dotsc,
16\).  The odd variables \(\theta\) form a Majorana-Weyl spinor
transforming in the \(\mathbf{16}\) of the
\(\grp{Spin}(1,9)\) group.
Quantities with a lower spinor index transform in the \(\mathbf{16}'\).

The supersymmetry covariant derivatives are
\begin{equation}\label{eq:coder}
  \partial_{\idxtvec{m}}, \qquad
  \sdel_\idxtspin{A} = \partial_\idxtspin{A} + (\theta \Gamma^{\idxtvec{m}})_\idxtspin{A} \partial_{\idxtvec{m}},
\end{equation}
where we have used the ten-dimensional generalization of the
Pauli matrices \(\Gamma_{\idxtspin{A}\idxtspin{B}}^{\idxtvec{m}}\).
The matrices \(\Gamma_{\idxtspin{A}\idxtspin{B}}^{\idxtvec{m}}\) and \((\tilde{\Gamma}^{\idxtvec{m}})^{\idxtspin{A}\idxtspin{B}}\)
are \(16 \times 16\) matrices which satisfy the algebra
\(\Gamma^{\idxtvec{m}} \tilde{\Gamma}^{\idxtvec{n}} + \Gamma^{\idxtvec{n}} \tilde{\Gamma}^{\idxtvec{m}}
= 2 \eta^{\idxtvec{m} \idxtvec{n}}\) and \(\tilde{\Gamma}^{\idxtvec{m}} \Gamma^{\idxtvec{n}}
+ \tilde{\Gamma}^{\idxtvec{n}} \Gamma^{\idxtvec{m}} = 2 \eta^{\idxtvec{m} \idxtvec{n}}\).
See \appref{sec:gamma-matrices} for our conventions on \(\Gamma\) matrices.
These Pauli matrices are symmetric in the spinor indices.
The fermionic covariant derivatives obey the
anti-commutation relations
\(\{\sdel_\idxtspin{A}, \sdel_\idxtspin{B}\} = 2
\Gamma_{\idxtspin{A}\idxtspin{B}}^{\idxtvec{m}} \partial_{\idxtvec{m}}\)
representing torsion of superspace.

For each of these covariant derivatives we introduce a connection
\(\sgauge\) which is a function of the superspace coordinates
\begin{gather}
  \scdelB_{\idxtvec{m}} = \partial_{\idxtvec{m}} + \sgauge_{\idxtvec{m}}(x,\theta), \qquad
  \scdelF_\idxtspin{A} = \partial_\idxtspin{A}
+ (\theta \Gamma^{\idxtvec{m}})_\idxtspin{A} \partial_{\idxtvec{m}} + \sgauge_\idxtspin{A}(x, \theta).
\end{gather}
These connections are defined up to
gauge transformations
\begin{gather}
  \delta \sgauge_{\idxtvec{m}} = [\scdelB_{\idxtvec{m}}, \chi], \qquad
  \delta \sgauge_\idxtspin{A} = [\scdelF_\idxtspin{A}, \chi].
\end{gather}

Even after identifying gauge connections which differ by a gauge
transformation, the Gra{\ss}mann expansion of these connections
produces many more fields than appropriate.  To reproduce the expected
spectrum, we need to impose some constraints on them.
In order for the constraints to be gauge invariant
they must be expressed in terms of the field strengths.
The field strengths are computed in
the usual way
\begin{gather}
  \sfstr_{\idxtspin{A}\idxtspin{B}} = \{\scdelF_\idxtspin{A}, \scdelF_\idxtspin{B}\}
- 2 \Gamma_{\idxtspin{A}\idxtspin{B}}^{\idxtvec{m}} \scdelB_{\idxtvec{m}}, \qquad
  \sfstr_{\idxtvec{m} \idxtspin{A} } = [ \scdelB_{\idxtvec{m}},\scdelF_\idxtspin{A}], \qquad
  \sfstr_{\idxtvec{m} \idxtvec{n}} = [\scdelB_{\idxtvec{m}}, \scdelB_{\idxtvec{n}}],
\end{gather}
where for \(\sfstr_{\idxtspin{A}\idxtspin{B}}\) we had to subtract a piece which
arises due to the superspace torsion.  We impose the constraints that
\[\label{eq:superspaceconstraint}
\sfstr_{\idxtspin{A}\idxtspin{B}} = 0.\]
If \(\sfstr_{\idxtspin{A}\idxtspin{B}}\)
were non-vanishing, then it would
correspond to a dimension-1 field transforming in the symmetric
product of two \(\mathbf{16}\) representations.%
\footnote{This
  representation is reducible and decomposes into a vector and a
  selfdual five-form.  But no fields of dimension \(1\) with these
  transformation properties exist in this theory.  The bosonic gauge
  field is a vector of dimension \(1\), but it is not gauge
  covariant.}

Bianchi identities generate further constraints.  For example,
\begin{equation}
\bigcomm{\scdelF_\idxtspin{A}}{\{\scdelF_\idxtspin{B}, \scdelF_\idxtspin{C}\}}
+ \text{cycle}(\idxtspin{A},\idxtspin{B},\idxtspin{C}) = 0,
\end{equation}
implies that \(\Gamma_{(\idxtspin{A}\idxtspin{B}}^{\idxtvec{m}} \sfstr_{\idxtspin{C}),{\idxtvec{m}}} = 0\), where the
parentheses indicate total symmetrization of the enclosed indices.
These constraints can be solved by \(\sfstr_{\idxtvec{m} \idxtspin{A}} = \Gamma_{\idxtspin{A}\idxtspin{B}, \idxtvec{m}}
\Psi^\idxtspin{B}\) which allows us to identify the field strength \(\sfstr_{\idxtvec{m}\idxtspin{A}}\)
with the fermionic superfield \(\Psi\).  This is possible thanks to
the ``magic identity'' \(\Gamma_{(\idxtspin{A}\idxtspin{B}}^{\idxtvec{m}} \Gamma_{\idxtspin{C})\idxtspin{D}, \idxtvec{m}} = 0\).

Continuing the analysis of the constraints generated by Bianchi
identities, one finds that they can only be realized on-shell \cite{Witten:1985nt}.
We can define the vielbeine, which are dual to the supersymmetry
covariant derivatives%
\footnote{In our conventions the differential forms
in superspace have two gradings: a form degree and a Gra{\ss}mann degree.
For example, \(\der x\) has form degree one and Gra{\ss}mann degree zero while \(\der \theta\)
has form degree one and Gra{\ss}mann degree one.
When commuting two differential forms, we pick up two contributions,
one of which is the usual sign from commuting differential forms
and the other is the sign from commuting Gra{\ss}mann variables.
See \appref{sec:supersp-geom} for a more in-depth discussion.}
\begin{equation}
  \label{eq:vielbeine10D}
  \svielB^{\idxtvec{m}} = \der x^{\idxtvec{m}} + \theta \Gamma^{\idxtvec{m}} \der \theta, \qquad
  \sviel^\idxtspin{A} = \der \theta^\idxtspin{A},
\end{equation}
such that we have \(\der = \svielB^{\idxtvec{m}} \partial_{\idxtvec{m}} + \svielF^\idxtspin{A} \sdel_\idxtspin{A}\).  If we
introduce \(\sgauge = \svielB^{\idxtvec{m}} \sgauge_{\idxtvec{m}} + \svielF^\idxtspin{A} \sgauge_\idxtspin{A}\), we have
\(\der + \sgauge = \svielB^{\idxtvec{m}} \scdelB_{\idxtvec{m}} + \svielF^\idxtspin{A} \scdelF_\idxtspin{A}\).
The field strength defined by \(\sfstr = \der \sgauge + \sgauge \wedge \sgauge\)
transforms covariantly under gauge transformations.%

Dimensionally reducing to four dimensions using the formulas and convention in
\appref{sec:gamma-matrices} the 10d superspace and superfields decompose into 4d
quantities as
\begin{align}
\{ x^{\idxtvec{m}}, \theta^{\idxtspin{A}}\, \} &\to \{x^{\dot\alpha\alpha},\theta^{a\alpha}, \bar\theta^{\dot\alpha}{}_{a}\}\, , \nln
\sgauge_{\idxtvec{m}}(x,\theta) &\to \{\sgauge_{\alpha\dot\alpha}(x,\theta, \bar\theta) ,
\Phi_{ab}(x,\theta, \bar\theta) \}\,  , \nln
\sgauge_{\idxtspin{A}}(x,\theta) &\to  \{\sgauge_{a\alpha}(x,\theta, \bar\theta) ,
\sgauge^{a}{}_{\dot\alpha} (x,\theta, \bar\theta)\} \, .
\end{align}
Here we recognize the 4d gauge superfield $\sgauge_{\alpha\dot\alpha}$ and a 4d scalar superfield
$\Phi_{ab}$ whose leading component is the adjoint scalar field $\phi_{ab}(x)$ of $\mathcal{N}=4$ SYM.

\subsection{Super Wilson Loop Operator}

The standard bosonic Wilson loop operator is a functional of the closed path $C$
defined as
\[
W= \frac{1}{N} \tr \pathord \exp \left ( \oint_C \mathrm{d}\tau\, \dot x^{\mu}
A_{\mu}(x)\right )\, ,
\]
where we consider an $\grp{SU}(N)$ or $\grp{U}(N)$ gauge group and $\pathord$ stands for path ordering
(with increasing path variable $\tau$ from left to right).
It is gauge and reparametrization invariant.
Given the definitions of the super-connection and the super-vielbeine it is
straight-forward to define the super Wilson loop operator for
\(\mathcal{N}=1\) super Yang-Mills theory in ten dimensions. We now consider a
closed path in superspace $Z=\{x^{\idxtvec{m}}(\tau), \theta^{\idxtspin{A}}(\tau)\}$ and
define the super Wilson loop operator
\[
\label{eq:10dSWL}
\swilson= \frac{1}{N} \tr \pathord \exp  \left ( \oint_Z \sgauge\right )=
\frac{1}{N} \tr \pathord \exp  \left ( \oint_Z \mathrm{d} \tau \left(
p^{\idxtvec{m}} \sgauge_{\idxtvec{m}} + \dot\theta^{\idxtspin{A}} \sgauge_{\idxtspin{A}} \right)
\right)
\, ,
\]
where we have introduced the super-momentum
\[
\label{eq:supermomentum}
p^{\idxtvec{m}}= \dot x^{\idxtvec{m}} + \theta \Gamma^{\idxtvec{m}} \dot \theta \, .
\]

This operator is gauge and reparametrization invariant as well as supersymmetric by construction.
The supersymmetry variations follow from the supersymmetry transformation of the contour.
The smooth super Wilson loop of the
$\mathcal{N}=4$ super Yang-Mills theory in four dimensions of interest to us is
not simply obtained through a dimensional reduction of the 10d operator
\eqref{eq:10dSWL}. The relevant bosonic Wilson loop was introduced by Maldacena
\cite{Maldacena:1998im} and also couples to the six scalar fields $\phi_i (x)$ in the
adjoint of $\grp{U}(N)$ via a path $n^{i}(\tau)$ on an $S^{5}$.
Given a path \(\tilde{C}\) specified by \((x^\mu(\tau), n^i(\tau))\), we define
\[
\label{eq:MaldaWil}
\tilde{W}= \frac{1}{N} \tr  \pathord \exp \left( \oint_{\tilde{C}} \mathrm{d} \tau
\left ( \dot{x}^{\mu}  A_{\mu}(x)  + n^i \sqrt{\dot{x}^{\mu}\dot{x}_{\mu} } \,
 \phi_i (x)\right) \right)\qquad\text{with } ~(n^{i})^2=1 \, .
\]
This coupling was motivated in \cite{Maldacena:1998im,Drukker:1999zq} through
the induced coupling of the gauge field to a massive $W$ boson obtained by spontaneous
symmetry breaking of $\grp{U}(N+1)\to \grp{U}(N)\times \grp{U}(1)$. It may also be understood as arising
from a 10d light-like path $x^{\idxtvec{m}}=\{x^{\mu},\sqrt{\dot{x}^{2}}\,  n^i\}$ giving
rise to a $1/2$ BPS property to be discussed in \secref{sect:kappasym}.
This coupling also guarantees the finiteness of the vacuum expectation value for
smooth, non-intersecting curves.

How does one supersymmetrize this operator? As in the ordinary case discussed above the
guiding principle is the invariance under supersymmetry transformations.
Here we propose the following supersymmetric generalization of the Maldacena-Wilson
loop operator coupling to the superpath $\tilde{Z}=\{ x^{\mu},\theta^{\idxtspin{A}}; q^{i} \}$
\[
\label{eq:superWL}
\tilde{\swilson}= \frac{1}{N} \tr  \pathord \exp \left( \oint_{\tilde{Z}} \mathrm{d} \tau
\left ( p^{\mu}  \sgauge_{\mu} + \dot\theta^{\idxtspin{A}} \sgauge_{\idxtspin{A}} +
q^{i}\, \Phi_{i}\,
 \right ) \, \right) \, ,
\]
where $q^i$ is constrained by
\[
q^{2}=q^{i}q^{i}=p^{\mu}p_{\mu} \, ,
\]
and thus $q^i=\sqrt{p^{\mu}p_{\mu}}\, n^i$ with $n^i$ being a unit vector as in \eqn{eq:MaldaWil}.
This operator is manifestly gauge and reparametrization invariant and
limits to the bosonic predecessor \eqref{eq:MaldaWil}.
Its supersymmetry transformations can again be obtained by transforming the contour.
We shall now show that it
is also invariant under kappa-symmetry as a generalization of the $1/2$ BPS property,
as well as under superconformal transformations. 

A similar super Wilson loop operator was
in fact proposed in \cite{Ooguri:2000ps}.
The explicit construction of this super Maldacena-Wilson loop operator in terms of component
fields in a particular gauge  to be discussed in \secref{sect:1loopvevtheta},
shows that the proposed operator \eqref{eq:superWL} exactly reproduces the
explicit construction of \cite{Muller:2013rta} to the first two orders
in the theta expansion.
Finally, for $p^{\mu}p_{\mu}=0$ it turns into the light-like
super Wilson loop of \cite{Beisert:2012gb,Beisert2012}. 

\subsection{Kappa-Symmetry}
\label{sect:kappasym}

In this section we discuss the kappa-symmetry of the super Wilson
loops.  The kappa-symmetry is closely related to the BPS property
for local operators so we first discuss this simpler case.  Then we
discuss the kappa-symmetry for ten-dimensional super Wilson loops
and finally give the kappa-symmetry transformations for the
four-dimensional super Wilson loop with a scalar coupling.

\paragraph{BPS property of local operators.}

Let \(O(x)\) be a local gauge-invariant operator in a supersymmetric
theory.  We can formally construct its supersymmetric completion by
defining%
\begin{equation}
  \mathcal{O}(x,\theta) = e^{\theta \genfield{Q}} O(x) e^{-\theta \genfield{Q}},
\end{equation} where \(\genfield{Q}\) are the supercharges acting on the
fields.  This is equivalent
to the familiar translation property
\begin{equation}
  O(x) = e^{x \genfield{P}} O(0) e^{-x \genfield{P}},
\end{equation} with the difference that the translations in superspace
do not commute in general.  Using this we can define the operator at
an arbitrary point in superspace by
\begin{equation}
  \mathcal{O}(x,\theta) = e^{x \genfield{P} + \theta \genfield{Q}} O(0) e^{-x
  \genfield{P}-\theta \genfield{Q}}.
\end{equation}

The supersymmetry transformations of a local gauge-invariant operator
are given by
\begin{equation}
  e^{\epsilon \genfield{Q}} \mathcal{O}(x,\theta) e^{-\epsilon \genfield{Q}} =
  \mathcal{O}(x-\epsilon \Gamma \theta, \theta+\epsilon).
\end{equation}  Looking at the linear terms in \(\epsilon\) we find
that\footnote{The minus sign in the equation below is crucial.  It
  ensures the invariance of a superfield under the combined field and coordinate transformations.}
\begin{equation}
  [\genfield{Q}_{\idxtspin{A}}, \mathcal{O}(x,\theta)] = -\gen{Q}_{\idxtspin{A}} \mathcal{O}(x,\theta),
\end{equation} where \(\gen{Q}_{\idxtspin{A}}\) is the representation of the supercharge
acting on the coordinates.  Explicitly, we have
\(\gen{Q}_{\idxtspin{A}} = - \partial_{\idxtspin{A}}+ (\Gamma^{\idxtvec{m}} \theta)_{\idxtspin{A}} \partial_{\idxtvec{m}}\).
Similarly, we obtain that
\(\gen{P}_{\idxtvec{m}} = -\partial_{\idxtvec{m}}\) and \([\genfield{P}_{\idxtvec{m}}, O(x)] = -\gen{P}_{\idxtvec{m}}
O(x) = \partial_{\idxtvec{m}} O(x)\).  The representation on the coordinates of course satisfies the
same graded commutation relations as the representation on the fields, i.e.\ \(\{\gen{Q}_{\idxtspin{A}},
\gen{Q}_{\idxtspin{B}}\} = 2 \Gamma_{\idxtspin{A} \idxtspin{B}}^{\idxtvec{m}} \gen{P}_{\idxtvec{m}}\).

It may happen that a particular combination of Poincar\'e supercharges annihilates
an operator \([\epsilon \genfield{Q}, O(x)] = 0\) for some spinor \(\epsilon\).  Then the
supersymmetric version \(\mathcal{O}(x,\theta)\) depends only on some
\(\theta\)'s and therefore the Gra{\ss}mann expansion is shorter.
These are BPS operators.  The condition \([\epsilon \genfield{Q}, O(x)] = 0\) on the
bottom component gets translated into another condition on its
supersymmetric version \(\mathcal{O}(x, \theta)\).  Working to first
order in \(\epsilon\), we find
\begin{align}
  \mathcal{O}(x, \theta) &= e^{\theta \genfield{Q}} e^{\epsilon \genfield{Q}} O(x)
  e^{-\epsilon \genfield{Q}} e^{-\theta \genfield{Q}} = e^{(\theta + \epsilon) \genfield{Q}
  + (\epsilon \Gamma^{\idxtvec{m}} \theta) \genfield{P}_{\idxtvec{m}}}
  O(x) e^{-(\theta + \epsilon) \genfield{Q} - (\epsilon \Gamma^{\idxtvec{m}} \theta)
    \genfield{P}_{\idxtvec{m}}}
\nonumber\\
&=\mathcal{O}(x + \epsilon \Gamma \theta,
  \theta+\epsilon) = \mathcal{O}(x, \theta) + (\epsilon \sdel)
  \mathcal{O}(x, \theta) + \mathcal{O}(\epsilon^{2}).
\end{align}
  Here \(\sdel_{\idxtspin{A}}\) is the supersymmetry covariant
derivative \eqref{eq:coder}.  The conclusion of this computation is
that \([\genfield{Q}_{\idxtspin{A}}, O(x)] = 0\) implies \(\sdel_{\idxtspin{A}} \mathcal{O}(x,\theta) = 0\).
This concludes the discussion of local operators.

\paragraph{Kappa-symmetry.}

Let us now consider
a Wilson loop on a light-like path in ten dimensions.  The connection
can be written as \(A(\tau) = \der \tau\,  \dot{x}^{\idxtvec{m}} A_{\idxtvec{m}}\).
Using the infinitesimal action of the supercharges on the fundamental fields
\(A\) and \(\psi\) yields
\begin{align}
  \delta_{\epsilon} A_{\idxtvec{m}}(x) &= [\epsilon \genfield{Q}, A_{\idxtvec{m}}(x)] = - \epsilon
  \Gamma_{\idxtvec{m}} \psi(x),
\\
  \delta_{\epsilon} \psi(x) &= [\epsilon \genfield{Q}, \psi(x)] = \half (\tilde{\Gamma}^{\idxtvec{m}
    \idxtvec{n}} \epsilon) F_{\idxtvec{m} \idxtvec{n}}(x),
\end{align}
and we find
\begin{equation}
  [\epsilon \genfield{Q}, A] = - \der \tau (\epsilon \dot{x}^{\idxtvec{m}} \Gamma_{\idxtvec{m}} \psi).
\end{equation}
If \(\dot{x}\) is a null vector, this variation vanishes if \(\epsilon
= \kappa \dot{x}^{\idxtvec{m}} \tilde{\Gamma}_{\idxtvec{m}}\), where \(\kappa\) is a
spinor and \(\tilde{\Gamma}\) are the ten-dimensional Pauli matrices
with upper indices (see \appref{sec:gamma-matrices}).
This is a local condition since \(\dot{x}\) is in
general not constant, so different supersymmetries are preserved at
different points around the Wilson loop contour. It is a $1/2$ BPS condition, as
the spinor $\kappa_{\idxtspin{A}}$ has 8 degrees of freedom as
$\dot{x}^{\idxtvec{m}} \tilde{\Gamma}_{\idxtvec{m}}$ for null $\dot x^{\idxtvec{m}}$ has an 8 dimensional
kernel.

Just like for local operators, the invariance of the bosonic Wilson
loop under some local worldline supersymmetry transformations lifts to the
invariance of the super Wilson loop under some local
super-diffeomorphisms.  These super-diffeomorphisms are obtained by
replacing
\(\epsilon \to \kappa p^{\idxtvec{m}} \tilde{\Gamma}_{\idxtvec{m}}\) and \(\genfield{Q}
\to \sdel\),
where
\(p^{\idxtvec{m}} = \dot{x}^{\idxtvec{m}} + \theta \Gamma^{\idxtvec{m}}\dot{\theta}\)
is the super-momentum, a supersymmetrization of \(\dot{x}^{{\idxtvec{m}}}\).

The kappa-symmetry transformations are just the transformations
of the superspace coordinates under the super-diffeomorphism
\(\kappa_\idxtspin{A} p^{\idxtvec{m}} \tilde{\Gamma}_{\idxtvec{m}}^{\idxtspin{A}\idxtspin{B}} \sdel_\idxtspin{B} \equiv
\epsilon(\kappa)^\idxtspin{A} \sdel_\idxtspin{A}\).  For example, we find
\begin{equation}
  \delta_\kappa p^{\idxtvec{m}} = [\epsilon(\kappa)^\idxtspin{A} \sdel_\idxtspin{A}, p^{\idxtvec{m}}]
= 2 \epsilon(\kappa) \Gamma^{\idxtvec{m}} \dot{\theta}, \qquad
  \delta_\kappa \theta^\idxtspin{A} = [\epsilon(\kappa)^\idxtspin{B} \sdel_\idxtspin{B}, \theta^\idxtspin{A}]
= \epsilon(\kappa)^\idxtspin{A}.
\end{equation}
We should note that out of the \(16\) components of \(\kappa\), only
\(8\) appear in the transformations, while the other \(8\) are
projected out when acted upon by the matrix \(p^{\idxtvec{m}}
\tilde{\Gamma}_{\idxtvec{m}}\).
Using the light-like condition on \(p\) it is easy to show that
there is an equivalence relation \(\kappa \sim \kappa + p \cdot \Gamma \xi\)
on the spinors \(\kappa\).  That is, \(\kappa\) and \(\kappa + p \cdot \Gamma \xi\)
produce the same kappa-transformation.

Let us now show that the full 10d super Wilson loop \eqref{eq:10dSWL} is
invariant under the kappa-symmetry transformations.
The kappa-transformations can, as explained above, be seen as diffeomorphisms in superspace generated by the vector field
\[
\delta_{\kappa}\theta^{\idxtspin{A}} \del_{\idxtspin{A}}
+ \brk{\delta_{\kappa}\theta \Gamma^{\idxtvec{m}} \theta} \del_{\idxtvec{m}}
= \delta_{\kappa}\theta^{\idxtspin{A}} \sdel_{\idxtspin{A}}
=: \zeta^{\idxall{A}} \sdel_{\idxall{A}}
\qquad
\text{with}\qquad
\zeta^{\idxall{A}}=(0,\delta_{\kappa}\theta^{\hat\alpha}) \, ,
\]
where $\idxall{A}=(\idxtvec{m},\idxtspin{A})$ is a collective superspace index.
Under a general diffeomorphism $\zeta^{\idxall{A}} \sdel_{\idxall{A}}$
the super-connection transforms as
\[
\delta_{\zeta} \oint \idxall{A}=
\left(\zeta^{\idxall{A}} \sdel_{\idxall{A}} \right)\oint \der \tau\, p^{\idxall{B}} \sgauge_{\idxall{B}}
= \oint \der \tau \Bigl( p^{\idxall{B}} \zeta^{\idxall{A}}  \sfstr_{\idxall{A}\idxall{B}}
+ \left[p^\idxall{A} \scdelF_\idxall{A}, \zeta^\idxall{B} \sgauge_\idxall{B}\right] \Bigr) \, ,
\label{eqn:Avariation}
\]
where we have introduced the abbreviation $p^{\idxall{A}}=(p^{\idxtvec{m}},\dot\theta^{\idxtspin{A}})$.
Note that the latter term in equation \eqref{eqn:Avariation} represents an infinitesimal,
field-dependent gauge transformation and can therefore be neglected if $\delta_{\zeta}$
is applied to a gauge invariant operator. By applying equation \eqref{eqn:Avariation}
to the case of kappa-transformations one finds the following expression
for the variation of the super Wilson loop operator
\begin{align}
\delta_{\kappa} \swilson&=
\frac{1}{N}
\tr \pathord \left\{ \exp \biggl(\oint_Z \sgauge \biggr) \oint_Z \delta \sgauge \right\}
\nonumber \\&=
\frac{1}{N}
\tr \pathord \left\{ \exp \biggl(\oint_Z \sgauge \biggr) \oint_Z \der \tau
\Bigl( p^{\idxtvec{m}}\, \delta_{\kappa}\theta^{\idxtspin{A}}\, \sfstr_{\idxtspin{A}\idxtvec{m}} +
\dot\theta^\idxtspin{B}\, \delta_{\kappa}\theta^{\idxtspin{A}}\, \sfstr_{\idxtspin{A}\idxtspin{B}}
\Bigr)  \right\} \, .
\end{align}
With the constraints $\sfstr_{\idxtspin{A}\idxtspin{B}}=0$ \eqref{eq:superspaceconstraint}
and \(\sfstr_{\idxtvec{m} \idxtspin{A}} = \Gamma_{\idxtspin{A}\idxtspin{B}, \idxtvec{m}}
\Psi^\idxtspin{B}\) as discussed in \secref{sec:supersp-gauge-theory} we see that the above
vanishes for $p^{\idxtvec{m}}\, p_{\idxtvec{m}}=0$ as
\[\delta_{\kappa}\theta^{\idxtspin{A}}\, p^{\idxtvec{m}}\, \Gamma_{\idxtspin{A}\idxtspin{B}, \idxtvec{m}}
\Psi^\idxtspin{B} = (\kappa \tilde \Gamma_{\idxtvec{n}}\Gamma_{\idxtvec{m}}\Psi)\, p^{\idxtvec{n}} \, p^{\idxtvec{m}}
\propto  p^{\idxtvec{m}} \, p_{\idxtvec{m}}=0.\]

A closely related transformation is loop reparametrization under which the
super Wilson loop is invariant.  On the superspace coordinates the
infinitesimal reparametrization transformations acts as the diffeomorphism
\[
  \delta_\sigma \theta(\tau) = \sigma \dot{\theta}(\tau), \qquad
  \delta_\sigma x^{\idxtvec{m}}(\tau) = \sigma \dot{x}^{\idxtvec{m}}(\tau).
\]
Consistency requires that the
kappa-transformations and the reparametrization transformations
form a closed algebra.
Indeed, we find (see \appref{sec:cosets} for
more details)
\begin{align}
  [\delta_{\kappa_1}, \delta_{\kappa_2}] &= \delta_{\kappa(\kappa_1, \kappa_2)} + \delta_{\sigma(\kappa_1, \kappa_2)},\\
  [\delta_{\sigma_1}, \delta_{\sigma_2}] &= \delta_{\sigma(\sigma_1, \sigma_2)},\\
  [\delta_{\sigma_1}, \delta_{\kappa_2}] &= \delta_{\kappa(\sigma_1,\kappa_2)},
\end{align}
where
\begin{align}
  \kappa(\kappa_1, \kappa_2) &= 4 (\dot{\theta} \kappa_2) \kappa_1 -
  4 (\dot{\theta} \kappa_1) \kappa_2 +
  2 (\kappa_1 \tilde{\Gamma}_{\idxtvec{m}} \kappa_2) (\Gamma^{\idxtvec{m}} \dot{\theta}),
&
  \kappa(\sigma_1, \kappa_2) &= \dot{\sigma}_1 \kappa_2 - \sigma_1 \dot{\kappa}_2,
\\
  \sigma(\kappa_1, \kappa_2) &= - 4 (\kappa_1 p^{\idxtvec{m}} \tilde{\Gamma}_{\idxtvec{m}} \kappa_2),
&
  \sigma(\sigma_1, \sigma_2) &= \sigma_2 \dot{\sigma}_1 - \sigma_1 \dot{\sigma}_2.
\end{align}

Let us now use the same ideas for a 4d bosonic Maldacena-Wilson loop with a scalar
coupling.  We find, for the supersymmetry transformations
\begin{equation}
  \bigcomm{\epsilon \genfield{Q}}{\dot{x}^{\mu} A_{\mu}(x) +
  \sqrt{\dot{x}^{2}} \phi_i(x) n^i}
= -\epsilon \bigbrk{\dot{x}^{\mu}
  \Gamma_{\mu} + \sqrt{\dot{x}^{2}} \Gamma_i n^i} \psi(x),
\end{equation} where we have used the dimensionally reduced
supersymmetry transformations.  Note that the matrix \(\dot{x}^{\mu}
\Gamma_{\mu} + \sqrt{\dot{x}^{2}} \Gamma_i n^i\) has an
eight-dimensional kernel.  Therefore, there is an eight-dimensional
space of operators \(\epsilon \genfield{Q}\) which
locally annihilate the exponent of the
Wilson loop.  The spinors \(\epsilon\) for which this holds are of type%
\footnote{Here one should keep in mind that
\(\Gamma_i \tilde{\Gamma}_j + \Gamma_j \tilde{\Gamma}_i = 2 \eta_{ij} = -2 \delta_{ij}\)
since our metric signature is mostly minus.}
\(\epsilon = \kappa (\dot{x}^{\mu} \tilde{\Gamma}_{\mu} +
\sqrt{\dot{x}^{2}} \tilde{\Gamma}_i n^i)\).  Just like for local operators, this
translates on the super Wilson loop to a condition involving
supersymmetry covariant derivatives.  These conditions are precisely
the conditions of invariance under kappa-symmetry
transformations. Hence the $1/2$ BPS property of the Maldacena-Wilson loop is
the kappa-symmetry of the super Wilson loop in disguise.

In the super Wilson loop case, if we follow the same recipe as in ten dimensions and
replace \(\dot{x}^{\mu} \to p^{\mu}\)
and \(\sqrt{\dot{x}^2}n^i\to q^i\), we find that the
kappa-symmetry transformations are generated by
\begin{align}
  \kappa_\idxtspin{A} (p^{\mu} \tilde{\Gamma}_{\mu} +
   \tilde{\Gamma}_i q^i)^{\idxtspin{A}\idxtspin{B}} \sdel_\idxtspin{B} &=
  (\bar{\kappa}^a{}_{\dot{\alpha}} p^{\dot{\alpha} \alpha}
+  q^{a b} \epsilon^{\alpha \beta} \kappa_{\beta b}) \sdel_{a \alpha}
+
  (\kappa_{\alpha a} p^{\dot{\alpha} \alpha}
+ \bar{q}_{a b} \epsilon^{\dot{\alpha} \dot{\beta}} \bar{\kappa}^b{}_{\dot{\beta}}) \bar{\sdel}^a{}_{\dot{\alpha}},
\end{align} where we have used the formulas in
\appref{sec:gamma-matrices} to reduce ten-dimensional gamma-matrices
to four dimensions.  Also, \(q^{a b}\) is the vector \(q^i\) written
in spinor form and \(\bar{q}_{a b} = \tfrac 1 2 \epsilon_{a b c d}
q^{c d}\).

The transformations of the odd variables follow immediately,
\[
\delta_{\kappa}\theta^{\idxtspin{A}}= \kappa_{\idxtspin{B}}(p^{\mu} \tilde{\Gamma}_{\mu} +
 q^{i} \tilde{\Gamma}_i )^{\idxtspin{B}\idxtspin{A}},
  \]
or in four-dimensional language
\begin{align}
  \delta_\kappa \theta
&= \bar{\kappa} p
- q \kappa^\trans \epsilon,
\\
  \delta_\kappa \bar{\theta} &=
  p \kappa
- \epsilon \bar{\kappa}^\trans \bar{q},
\end{align}
where $p=\dot x+2\bar\theta\dot\theta-2\dot{\bar\theta}\theta$.
In matrix notation we take \(\kappa\) to be a \(2 \times 4\) matrix
and \(\bar{\kappa}\) to be a \(4 \times 2\) matrix. The
kappa-transformations of the remaining quantities of interest are in vector language
\begin{gather}
\delta_{\kappa}x^{\mu}= -\theta\Gamma^{\mu}\delta_{\kappa}\theta , \qquad
\delta_{\kappa}p^{\mu}= -2\dot{\theta}\Gamma^{\mu}\delta_{\kappa}\theta  , \qquad
\delta_{\kappa}q^{i}= -2\dot{\theta}\Gamma^{i}\delta_{\kappa}\theta  ,
\end{gather}
or in the compact spinor matrix notation
\begin{align}
  \delta_\kappa x &= 2 (p \kappa - \epsilon \bar{\kappa}^\trans \bar{q}) \theta
     -2 \bar{\theta} (\bar{\kappa} p - q \kappa^\trans \epsilon),\\
  \delta_\kappa p &= 4 (p \kappa - \epsilon \bar{\kappa}^\trans \bar{q}) \dot{\theta}
    -4 \dot{\bar{\theta}} (\bar{\kappa} p - q \kappa^\trans \epsilon),\\
   \delta_\kappa q^{ab} &= -8 (\dot{\theta} \epsilon (\delta_\kappa \theta)^\trans)^{[ab]}
    -4 \epsilon^{a b c d} (\dot{\bar{\theta}}^\trans \epsilon (\delta_\kappa \bar{\theta}))_{c d}.
\end{align}
Here we have taken the transformations for \(q\) to be the dimensional
reduction of the kappa-transformations for the extra components
of ten-dimensional momentum.

\begin{figure}
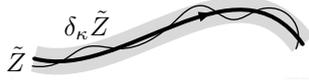
\centering
\includegraphicsbox{FigKappa.mps}
\caption{$1|8$-dimensional submanifold of superspace
describing Wilson lines equivalent by kappa-symmetry.}
\label{fig:kappa}
\end{figure}

The kappa-transformations and reparametrizations
can be thought of as transformations mapping points along some path to
nearby points in superspace.
Since these transformations close under the Lie bracket,
they locally generate some manifold of dimension \(1 \vert 8\),
see \figref{fig:kappa}.
Any two paths within this manifold are therefore
related by kappa-transformations and reparametrizations,
and the corresponding Wilson loops are equivalent.

\subsection{Geometry and Superconformal Cosets}
\label{sect:geometrycosets}

Before we proceed, we would like to understand the geometric
foundations of the Wilson line.
An ordinary Wilson line is defined by a path
in space-time on which the gauge theory is formulated.
In a supersymmetric gauge theory formulated on superspace,
the generalization of the Wilson line is straight-forward
by lifting the path to the corresponding superspace.
However, the above Wilson line has additional couplings
to the scalar field (on superspace).
From the ten-dimensional perspective this coupling is natural
upon dimensional reduction.
However, from the four-dimensional point of view
it appears somewhat ad-hoc.%
\footnote{For example, one might contemplate
similar direct couplings to the spinor fields.}
What is the geometric meaning of the scalar couplings
since they have no immediate connection to space-time or superspace?
Is there an extension of superspace to accommodate for the scalar couplings?

\paragraph{Superconformal symmetry.}

In particular, we are interested in symmetry properties of the Wilson line.
Therefore we should ask whether it is possible to define an action of
the superconformal group on such an extended superspace.
Depending on whether we admit only unit vectors $n^i$ for the coupling
to the scalars or more general vectors of unrestricted length,
this space-time will have 5 or 6 additional bosonic coordinates.%
\footnote{This could be seen as an obstacle to superconformal
symmetry which is known to exist only up to and including 6 dimensions.
However, we will not require Lorentz symmetry in $4+6$ dimensions,
but merely Lorentz symmetry in 4 dimensions and rotational
symmetry in the 6 internal directions.}

In fact, the requirement of superconformal symmetry
resolves the geometry because it can only act consistently
on its coset spaces or collections thereof.
Ordinary superspace is the coset space of the superconformal group
by the super Poincar\'e group, dilatations and internal rotations
\[
\Real^{3,1|16} = \frac{\grp{PSU}(2,2|4)}{\bigbrk{\grp{SO}(6)\times\grp{SO}(1,1)}\ltimes \grp{SPoin}^*}\,.
\]
Here the \(\grp{SPoin}^*\) group is not the usual super Poincar\'e group,
but the dual one generated by \((\gen{K}, \gen{M}, \gen{S}, \bar{\gen{S}})\).
The dimension of the superconformal group is $(15+15|32)$ while the denominator group
has dimension $(15+1+10|16)$;
the difference of dimensions agrees with superspace.
Now, to add a unit vector $n^i\in S^5$ to this space is not difficult
\cite{Ooguri:2000ps}:
\[
\Real^{3,1|16}\times S^5 = \frac{\grp{PSU}(2,2|4)}{\bigbrk{\grp{SO}(5)\times\grp{SO}(1,1)}\ltimes \grp{SPoin}^*}\,.
\]
This is because $S^5$ is the coset space $\grp{SO}(6)/\grp{SO}(5)$.
The path for the extended Wilson line is therefore defined on the
above coset space which extends superspace by an $S^5$.
The fields of $\superN=4$ SYM, however, are defined on ordinary superspace.
These two spaces are canonically connected by projection from $S^5$ to a point.

The extension by $S^5$ is therefore very natural. However, it is also possible
to extend the space for the paths to $\Real^6$ and thus to relax the
unit vector constraint for $n^i$.
To that end, one can view $\Real^6$ as the union of
spheres with different radii. Therefore, it is possible to extend the
superconformal action to this space as well.
The transformations will never change the modulus $|n|$,
i.e.\ the modulus can be viewed as a coordinate on $\Real^+$
unrelated to superconformal symmetry.

The above discussion shows that a superconformal action on the path of the extended Wilson line
can be established. In the following we will develop the various transformation
properties more explicitly.

\paragraph{Non-null curved lines.}

In view of the relevance of null polygonal Wilson loops to scattering amplitudes
\cite{Alday2007a,Brandhuber2008,Drummond2008a},
it is natural to ask whether these objects can be generalized to our Wilson loops.
The following is somewhat outside the scope of the present paper, and it may be skipped
without penalty for the understanding of the following sections.

First, let us briefly review the case of null lines, before discussing the non-null case.
Straight null lines are parametrized by ambi-twistors
(see for example refs.\ \cite{Howe1995,Mason2006,Beisert:2012gb}), which again form a coset
space of the superconformal group.  An ambi-twistor is a pair \((\mathscr{Z}, \mathscr{W}) \in \mathbb{CP}^3 \times \mathbb{CP}^3\)
and such that \(\mathscr{Z} \cdot \mathscr{W} = 0\).
The pair \((\mathscr{Z}, \mathscr{W})\)
has six degrees of freedom from which we subtract
the \(\mathscr{Z} \cdot \mathscr{W} = 0\) constraint to obtain five degrees of freedom.  The conformal group acts by the natural \(\grp{SL}(4)\) action\footnote{This natural action is just the matrix multiplication on the homogeneous coordinates.} on the first \(\mathbb{CP}^3\) and the inverse of this action on the second \(\mathbb{CP}^3\).
The stability group of a light-like ray can be found as follows.
Let us take a light-like ray in the \(x^{(+)} \equiv x^0 + x^1\) direction.
The broken generators (which do not preserve this light-like direction)
are \(\gen{P}_{-}\), \(\gen{P}_i\) and \(\gen{M}_{i-}\) for \(i=2,3\).
So the space of light-like lines is five-dimensional.
This is also what we obtained from the ambi-twistor description.
The unbroken generators are the rest of the generators,
\(\gen{P}_{+}\), \(\gen{K}_{+}\), \(\gen{K}_{-}\), \(\gen{K}_i\), \(\gen{M}_{+-}\),
\(\gen{D}\), \(\gen{M}_{+i}\), \(\gen{M}_{23}\)
and they generate the stability group by which we need to quotient in the coset.

Next, we should extend the notion of a straight null line to a non-null line.
We are most interested in preserving conformal symmetry, and it is known that
non-null straight lines are typically transformed to circles and/or hyperbolae
under conformal transformations.
More precisely, the class of non-null straight lines, circles and/or hyperbolae
closes under conformal transformations,
and hence these shapes are natural generalizations of straight null lines.

The moduli space for hyperbolae
(including straight time-like lines)
in bosonic space-time is given by the coset
(up to global issues)
\[
\frac{\grp{SO}(2,4)}{\grp{SO}(3)\times \grp{SO}(2,1)}\,.
\]
This is most easily seen by considering the stabilizer of a
straight time-like line.
The time-like direction \(x^0\) is preserved by the transformations \((\gen{K}_0, \gen{D}, \gen{P}_0)\).
These generators form the conformal algebra in one dimension with signature \((1,0)\)
which generates the group \(\grp{SO}(2,1)\).  It is also preserved by \(\grp{SO}(3)\)
rotations in the transverse space \((x^1, x^2, x^3)\).
Conversely, circles (including straight space-like lines)
are parametrized by the coset
\[
\frac{\grp{SO}(2,4)}{\grp{SO}(1,2)\times \grp{SO}(1,2)}\,.
\]
This time we need to find the stability group of a space-like axis, \(x^3\).
The \(x^3\) axis is preserved by the transformations \((\gen{K}_3, \gen{D}, \gen{P}_3)\).
These generators form the conformal algebra in one dimension with signature \((0,1)\)
which generates the group \(\grp{SO}(1,2)\).  It is also preserved by \(\grp{SO}(1,2)\)
rotations in the transverse space \((x^0, x^1, x^2)\).
Both cosets have a dimension of $15-3-3=9$, i.e.\
the curves are specified by 9 moduli. These correspond to:
\begin{itemize}
\item
4 parameters to describe a specific point on the curve;
\item
minus 1 parameter to choose this point on the curve;
\item
3 parameters to specify the direction of the curve at this point;
\item
3 parameters to specify the curvature radius and orientation
(orthogonal to the direction).
\end{itemize}
Alternatively, one could count the parameters as follows:
\begin{itemize}
\item
4 parameters to specify the center of the circle;
\item
1 parameter to specify the radius;
\item
4 parameters to describe the orientation of the great circle.
\end{itemize}

The generalization to superspace exists only for the time-like case
\[
\frac{\grp{PSU}(2,2|4)}{\grp{OSp}(2|2,\Quat)}\,.
\]
Here, the bosonic component $\grp{Sp}(2)=\grp{Sp}(2,\Quat)$ is
the spin group for $\grp{SO}(5)$,
and $\grp{SO}(2,\Quat)$ is locally equivalent to $\grp{SO}(3)\times\grp{SO}(2,1)$.
The space-like case does not correspond to a real coset space.
However, one can achieve such curves by allowing
some coordinates to become imaginary;
ordinarily, the scalar couplings $n^i$ are accompanied by a factor of the imaginary unit $i$.
This case can be embedded into the complexified space $\grp{PSL}(4|4,\Comp)/\grp{OSp}(4|4,\Comp)$.
See \cite{Drukker:2006xg,Drukker:2007qr} for related discussions.
The coset has dimension $(15|32|15)-(6|16|10)=(9|16|5)$,
where the middle number counts the fermionic degrees of freedom and
the latter one the bosonic degrees of freedom related to the scalars.
The additional 5 bosonic degrees of freedom specify the point $n^i$ on the sphere $S^5$
whereas the 16 fermionic degrees of freedom
describe the orientation of the curve along the odd directions of superspace.

To be slightly more concrete, let us express a straight line in superspace
which is described by a particular class of points in the above moduli space.
Its direction is determined
by a time-like vector $p=v$, $v^2=1$,
and a unit internal vector $q=n$, $n^2=1$;
Note that the fermionic directions are completely determined by these moduli
as well though the orthogonality conditions
$\der\theta \epsilon v^\trans = n\der\bar\theta^\trans$
and $\bar n \der\theta = \der\bar\theta^\trans \epsilon v$.
The location of the line in space-time is further specified by
a vector $b$ and a spinor $(\beta,\bar\beta)$.
The (fat) line is a manifold of dimension $1|8$
given by
\[
x = v \tau+b  -2\bar\theta \beta + 2\bar\beta \theta,
\qquad
\theta =  \bar\lambda v - n  \lambda^\trans\epsilon + \beta,
\qquad
\bar\theta =  v\lambda - \epsilon \bar\lambda^\trans \bar n + \bar\beta.
\]
Here, $\tau$ is a bosonic coordinate of the manifold, and $\lambda,\bar\lambda$
are 16 fermionic coordinates out of which $8$ are projected out
in the above equation.
Note that the manifold is straight in the sense
that the dependence on the coordinates $(\tau,\lambda,\bar\lambda)$ is linear.
To obtain a more conventional straight one-dimensional (thin) line,
one may restrict the submanifold by assuming the fermionic coordinates
to be linear functions of $\tau$,
e.g.\ $\lambda=\omega \tau$, $\bar\lambda=\bar\omega \tau$;
however, due to kappa-symmetry, such a restriction
along with the value of $\omega$ has no physical significance.

By changing the coordinates $\tau,\lambda,\bar\lambda$
one can see that only $3|8$ of the $4|16$ components of
$b,\beta,\bar\beta$ are physical.
Together with the $3+5$ directions, there are $6|8|5$ moduli for straight lines.
The more general form of hyperbolae is obtained by
conformal inversion and translation in superspace amounting to
$3|8$ additional degrees of freedom and $9|16|5$ moduli altogether.
One can convince oneself that all of these $1|8$-dimensional sub-manifolds of superspace
are invariant under some $\grp{OSp}(2|2,\Quat)$ subgroup of $\grp{PSU}(2,2|4)$.
We conclude that their moduli space is indeed the above coset.

\paragraph{Polygons and splines.}

\begin{figure}
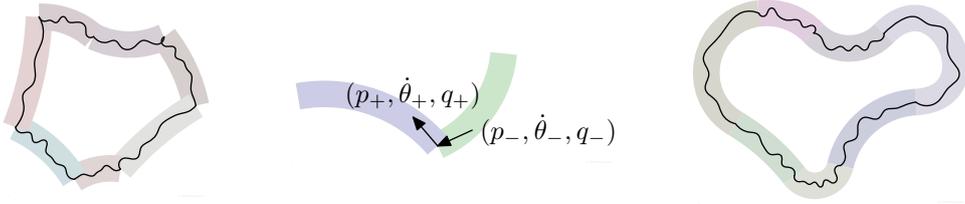
\centering
\includegraphicsbox{FigPoly.mps}
\qquad
\includegraphicsbox{FigCorner.mps}
\qquad
\includegraphicsbox{FigSpline.mps}
\caption{A polygon (left) and a spline (right)
composed from curved line segments
with an inscribed Wilson loop.
Two lines segments meet at a corner (middle)
which can act as a source for a UV-singularity.}
\label{fig:polygons}
\end{figure}

Closed curves can be constructed from segments of the above curves.
Let us first understand how to make two curves meet in a point.
The codimension of a $(1|8)$-dimensional curve
in $(4|16)$-dimensional superspace is $(3|8)$.
Two curves thus imply $(6|16)$ equations on the
$(4|16)$ coordinates of the intersection, i.e.\ generically
the curves will not intersect. Only after imposing $(2|0)$ suitable constraints
on the moduli of the curves, they will generically intersect at a single point
in superspace.
A closed curve made from $n$ segments thus has $n\cdot (7|16|5)$ degrees of freedom,
see \figref{fig:polygons}.

In a Wilson loop expectation value, the segments of the curve are expected to be finite
whereas the corners may introduce UV-divergences.
Since kappa-symmetry appears to imply finiteness,
one should make sure that the corners preserve this symmetry as well.
Kappa-symmetry requires $p^2=q^2$ for all points on the curve.
At the corners $p$ and $q$ are discontinuous, and a reasonable regularization
of the constraint is to demand $p_-\cdot p_+=q_-\cdot q_+$,
where $p_\pm$ and $q_\pm$ describe the values of $p$ and $q$ just before and after
the discontinuity, cf.\ \figref{fig:polygons}.%
\footnote{Using the finiteness derivations in later sections,
one can convince oneself that this constraint is sufficient for one-loop UV-finiteness
of the corners.}
\footnote{The only solution of this constraint
with all parameters real (time-like curves) is the trivial one
where $p$ and $q$ are continuous.
This can be seen by writing the constraint as
$p^2\cosh\xi=q^2\cos\theta$ where $\theta$ is the angle between $q_\pm$
and $\xi$ is the corresponding hyperbolic angle between $p_\pm$.}
This amounts to one bosonic constraint and consequently,
a composite curve of $n$ segments has $n\cdot (6|16|5)$ degrees of freedom.

In a more conservative approach, the transition
between the segments can be smoothed out by one order in the derivatives.
In other words, the direction of the path can be made continuous
while its derivative must be allowed to have jumps, see \figref{fig:polygons}.
The resulting shape can be called a quadric \emph{spline}.%
\footnote{A spline is a piecewise polynomial function
which is smooth to a certain degree.
They are often used to approximate other functions, e.g.\
in computer graphics and modeling.
A polygon is a spline of lowest degree,
here we require one extra degree of smoothness.
The segments of our splines are not polynomials,
but rather circles or hyperbolae.}
For this purpose, the variable $q^i$ should be considered on equal footing as
the path derivative $p^\mu\approx \dot x^\mu$.
Matching of the directions of $p_\pm$ amounts to 3 constraints
(the magnitude may jump due to a discontinuity in parametrization)
Similarly, the $q_\pm$ are aligned by 5 constraints
(the magnitude is linked to $p$).
Finally, only half of the fermionic directions $\dot\theta$ are physical
while the other ones can be adjusted by kappa-symmetry.
Altogether, continuity imposes $(3|8|5)$ constraints and
leaves $n\cdot(4|8|0)$ degrees of freedom for a spline with $n$ segments.
Curiously, this is merely one additional bosonic degree of freedom
per segment compared to the null polygon in full superspace \cite{Beisert2012a}.
As such one might consider it as a massive generalization of these objects,
and wonder whether there might be corresponding or dual objects
in the AdS/CFT dictionary. Could the 4 bosonic degrees of freedom
correspond to a massive particle momentum? Are the corresponding string
configurations or their T-duals distinguished in some way?
What is the twistor space description?%
\footnote{See \cite{Bullimore:2011ni} for a similar construction
of higher degree curves in twistor space,
which may or may not be related.}

Unfortunately, these polygons or splines are not very
amenable to concrete calculations
because the moduli space for each segment is already very large:
There are $(9|16|5)$ moduli to specify the curved line plus
2 times $(1|8)$ parameters to specify the end points of the segment.
In practice, for the one-loop expectation value
one would need to compute the correlator of two segments.
However, the combination of two segments already
breaks all of conformal symmetry and most of the internal symmetry.
Therefore, this correlator will be a function of several cross ratios
\cite{Plefka:2001bu},
and conformal symmetry cannot be used to constrain this function fully.
Supposing that the residual internal symmetry
is $\grp{SO}(4)\subset\grp{SO}(6)$ of dimension $(0|0|6)$,
there would be $2\cdot(9|16|5)+(0|0|6)-(15|32|15)=(3|0|1)$ superconformal cross ratios
in this configuration. Furthermore, the correlator of two segments would
depend on the four end points specified by $(1|8)$ coordinates each.
Nevertheless, Yangian symmetry may provide some clues towards their calculation.
This will be investigated in \cite{usyangian}.
For instance, some modification of the TBA-related methods found in \cite{Alday:2010vh}
could be used to describe the moduli of the spline along with
the Wilson loop expectation values. The advantage of this shape over
the null polygons would be the manifest absence of singularities.
The latter should be approachable by taking a limit of splines
and observing how the divergences develop.
Finally, some special kinematical configurations might further simplify
the analysis so that concrete results could be computed, with or without integrability.

\subsection{Conformal Symmetry}
\label{sec:conformal-symmetry}

After dimensional reduction to four dimensions, the symmetry is
enhanced with respect to what one might naively expect.  Some extra
generators arise, like conformal boosts and conformal supersymmetry.

Before moving on to the supersymmetric case let us study the bosonic
Maldacena-Wilson loop whose exponential is
\begin{equation}
  \der x^{\mu} A_{\mu}(x) + \der \tau \sqrt{\dot{x}^{2}} \phi_{i}(x) n^{i},
\end{equation}
where \(n^{i}\) is some (possibly \(\tau\)-dependent) six-dimensional unit
vector.  Obviously this is invariant under Poincar\'e transformations
and six-dimensional rotations.  If we show that it is invariant under
inversion, then conformal invariance follows.  Using the
transformations of the fields under inversion, this is easy to
establish once we use the fact that%
\footnote{Recall that under inversions $\phi'(x')=x^{2}\, \phi(x)$. There is a potential sign
  issue when taking the square root if \(x^{2}(\tau)\) changes sign.}
\begin{equation}
  \der x^{\mu} A_{\mu}(x) = \der {x'}^{\mu} A'_{\mu}(x'), \qquad
  \dot{x}'^{2} = \frac {\dot{x}^{2}}{(x^{2})^{2}}\,.
\end{equation}

Let us now work out the action of the extra generators on superfields.
Consider the following transformations, which will turn out to
correspond to inversions
\begin{align}
  I(x^{\pm}) &= \epsilon (x^{\mp})^{\trans,-1} \epsilon,
\\
  I(\theta) &= -M \bar{\theta}^{\trans} (x^{-})^{\trans,-1} \epsilon,
\\
  I(\bar{\theta}) &= \epsilon (x^{+})^{\trans,-1} \theta^{\trans} M^{-1}.
\end{align}
Here we use the notation \(x^\pm = x \mp 2 \bar{\theta} \theta\) and we have introduced a constant (i.e.\ \(x\)-independent) \(4
\times 4\) matrix \(M\) which acts on the R-symmetry indices.
The matrix \(M\) is such that \(M^{\trans} = M\), \(M^{\dagger} = M^{-1}\).
These properties are necessary to ensure that \(I^{2} =
\operatorname{id}\).  It can be checked that the constraint
\(x^{-}-x^{+}=4 \bar{\theta} \theta\) is preserved by inversion.
The reality conditions \(\theta^{\ddagger} = \bar\theta\)
and
\((x^{+})^{\ddagger} = x^{-}\) are also preserved.  The inversion for
\(x\) can be obtained from \(x = \tfrac 1 2 (x^{+} + x^{-})\)
\begin{equation}
  I(x) = \epsilon (x^{-})^{\trans, -1} x^{\trans} (x^{+})^{\trans, -1}
  \epsilon.
\end{equation}

In order to find the transformation under inversion for the
superfields corresponding to the fundamental fields of
\(\mathcal{N}=4\) super Yang-Mills, we proceed as follows.  First,
consider the total differential
\begin{equation}
  \der = \svielF^{a \alpha} \sdel_{a \alpha} +
  \svielC{}^{\dot{\alpha}}{}_a
  \bar{\sdel}^{a}{}_{\dot{\alpha}} + \svielB^{\mu} \partial_{\mu},
\end{equation}
where
\begin{align}
  \sdel_{a \alpha} &= \partial_{a \alpha} +
  \bar{\theta}^{\dot{\alpha}}{}_a \partial_{\alpha \dot{\alpha}},
\\
  \bar{\sdel}^{a}{}_{\dot{\alpha}} &= \bar{\partial}^{a}{}_{\dot{\alpha}} +
  \theta^{a \alpha} \partial_{\alpha \dot{\alpha}},
\\
  \svielB^{\mu} &= \der x^{\mu} - \der \theta^{a \alpha} \sigma_{\alpha
    \dot{\alpha}}^{\mu}
  \bar{\theta}^{\dot{\alpha}}{}_a + \theta^{a
    \alpha} \sigma_{\alpha \dot{\alpha}}^{\mu}
\der \bar{\theta}^{\dot{\alpha}}{}_a.
\end{align}

Using the inversion formulas we can compute new vielbeine in terms of
the old.  For our purposes it is more useful to present the old
vielbeine in terms of the new.  We find, using the matrix form for
these quantities
\begin{align}
  \svielB &= - x^{-} \epsilon \svielB^{\prime \trans} \epsilon x^{+},
\\
  \svielF &= (1-4 \theta x^{-,-1} \bar{\theta}) M \svielC^{\prime
    \trans} \epsilon x^{+} - \theta \epsilon \svielB^{\prime \trans}
  \epsilon x^{+},
\\
  \svielC &= -x^{-} \epsilon \svielF^{\prime \trans} M^{-1} (1 + 4
  \theta x^{+,-1} \bar{\theta}) - x^{-} \epsilon \svielB^{\prime \trans}
  \epsilon \bar{\theta}.
\end{align}
  Plugging these equations in the formula for the total
differential $\der$ we find
\begin{align}
  \sdel'_{a \alpha} &= - (x^{-}
  \epsilon)^{\dot{\alpha}}_{\hphantom{\dot{\alpha}} \alpha} (M^{-1} (1
  + 4 \theta x^{+,-1} \bar{\theta}))_{a b}
  \bar{\sdel}^{b}{}_{\dot{\alpha}},
\\
  \bar{\sdel}^{\prime a}{}_{\dot{\alpha}} &= (\epsilon
  x^{+})_{\dot{\alpha}}^{\hphantom{\dot{\alpha}} \alpha} ((1 - 4
  \theta x^{-,-1} \bar{\theta}) M)^{b a} \sdel_{b \alpha},
\\
  \partial_{\alpha \dot{\alpha}}' &= -(x^{-}
  \epsilon)^{\dot{\beta}}_{\hphantom{\dot{\beta}} \alpha} (\epsilon
  x^{+})_{\dot{\alpha}}^{\hphantom{\dot{\alpha}} \beta} \partial_{\beta \dot{\beta}} - 2 (\epsilon
  x^{+})_{\dot{\alpha}}^{\hphantom{\dot{\alpha}} \beta} (\theta
    \epsilon)^{a}_{\hphantom{a} \alpha} \sdel_{a \beta} - 2 (x^{-}
  \epsilon)^{\dot{\beta}}_{\hphantom{\dot{\beta}} \alpha} (\epsilon
  \bar{\theta})_{\dot{\alpha} a} \bar{\sdel}^{a}{}_{\dot{\beta}}.
\end{align}

These supersymmetry covariant derivatives get promoted to
supersymmetry and gauge covariant derivatives, by defining
\begin{gather}
  \scdelF_{a \alpha} = \sdel_{a \alpha} + \sgauge_{a \alpha}, \qquad
  \scdelC^{a}{}_{\dot{\alpha}} = \bar{\sdel}^{a}{}_{\dot{\alpha}} +
  \sgauge^{a}{}_{\dot{\alpha}}, \qquad
  \scdelB_{\alpha \dot{\alpha}} = \partial_{\alpha \dot{\alpha}} +
  \sgauge_{\alpha \dot{\alpha}}.
\end{gather} Then, if inversion is to act as a symmetry, we need to
take the transformations of the gauge connections to be the same as
the transformations of the gauge covariant derivatives.

It remains to show that these inversion transformations are indeed a
symmetry of the theory.  For this it is sufficient to check that they
preserve the supersymmetry constraints, which, in turn, imply the
equations of motion.  As a by-product we also obtain the
transformations under inversion of the scalar superfields.

In dimensionally reduced form and with the conventions in
\appref{sec:gamma-matrices} for gamma-matrices, the constraints read
\begin{align}
  \{\scdelF_{a \alpha}, \scdelF_{b \beta}\} &= -4
  \epsilon_{\alpha \beta} \Phi_{a b},
\\
  \{\scdelC^{a}{}_{\dot{\alpha}},
  \scdelC^{b}{}_{\dot{\beta}}\} &= -2 \epsilon_{\dot{\alpha}
    \dot{\beta}} \epsilon^{a b c d} \Phi_{c d},
\\
  \{\scdelF_{a \alpha}, \scdelC^{b}{}_{\dot{\beta}}\} &= 2
  \delta_{a}^{b} \scdelB_{\alpha \dot{\alpha}}.
\end{align}
  Plugging in the expressions for the transformed
covariant derivatives, we find
\begin{equation}
  \label{eq:scalar-inv}
  \Phi'_{a b}\bigbrk{I(x), I(\theta), I(\bar{\theta})}
= (\det x^{-}) \bigbrk{ M (1+4 \theta x^{+,-1} \bar{\theta})}_{a c}
\bigbrk{M (1+4 \theta x^{+,-1} \bar{\theta})}_{b d}
\bar{\Phi}^{c d}(x, \theta, \bar{\theta}),
\end{equation}
where \(\bar{\Phi}^{ab} = \tfrac 1 2
\epsilon^{a b c d} \Phi_{c d}\).

It is useful to compute the determinant \(\det (1+4 \theta x^{+,-1}
\bar{\theta})\).  To do this, we use the fact that \(\det (1+A B) =
\det (1+B A)^{-1}\) where \(A\) is an odd \(n \times m\) matrix and
\(B\) is an odd \(m \times n\) matrix.  We can show this identity by
computing the super-determinant of
\(\left(\begin{smallmatrix} 1 & -A\\ B & 1\end{smallmatrix}\right)\)
in two different ways.%
\footnote{The bosonic counterpart of this identity is \(\det (1+
  A B) = \det (1 + B A)\), for \(A\) an \(n \times m\) even matrix and
  \(B\) an \(m \times n\) even matrix.  This identity can be shown by
  noticing that it holds if \(A\) and \(B\) are invertible square
  matrices and then using a standard deformation argument.}
Using this identity, we have
\begin{equation}
  \det \bigbrk{1+4 \theta x^{+,-1} \bar{\theta}}
= \det \bigbrk{1+4 x^{+,-1}  \bar{\theta} \theta}^{-1}
= \det \bigbrk{1+x^{+,-1}(x^{-}-x^{+})}^{-1}
= \frac {\det x^{+}} {\det x^{-}}
\,.
\end{equation}

Using inversion we can compute the infinitesimal transformations of
\((x, \theta, \bar{\theta})\) under superconformal boosts (see
\appref{sec:supersp-geom} for the explicit expressions).  Then, we
can find how the supersymmetry covariant derivatives transform.  We
have
\begin{equation}
  \sdel_{a \alpha} = (1 + 4 \rho^\trans \theta^\trans +
  4 \bar{\theta}^\trans \bar{\rho}^\trans)_a^{\hphantom{a} b}
  (1 - 4 \rho \theta)_\alpha^{\hphantom{\alpha} \beta} \sdel'_{b \beta}\, ,
\end{equation}
where $\rho$ and $\bar\rho$ are the parameters of the superconformal transformations.
The supersymmetry and gauge covariant derivatives
\(\scdelF_{a \alpha}\) transform in the same way.  Then, the supersymmetry
constraints can be used to compute how the scalar superfield \(\Phi_{a  b}\) transforms.
We find
\begin{equation}
  \Phi_{a b}'(x',\theta',\bar{\theta}') = \Phi_{a b}(x,\theta,\bar{\theta})
  +4 \tr(\rho \theta) \Phi_{a b} -
  4 (\rho^\trans \theta^\trans + \bar{\theta}^\trans \bar{\rho}^\trans)_a^{\hphantom{a} a'} \Phi_{a' b}+
  4 \Phi_{a b'} (\theta \rho + \bar{\rho} \bar{\theta})^{b'}_{\hphantom{b'} b}.
\end{equation}
This transformation law makes it clear that \(\Phi\) is a
superconformal primary: when placed at the origin of superspace it is
invariant under \(\gen{S}\) and \(\bar{\gen{S}}\) transformations.

If \(\Phi_{ab}\) is contracted with \(q^{a b}\) and we want this
contraction to be invariant, then we need to impose that \(q^{a b}\)
transforms in the opposite way.  That is, using matrix notation,
\begin{equation}
  \delta q = -4 (\theta \rho + \bar{\rho} \bar{\theta}) q +
  4 q (\rho^\trans \theta^\trans + \bar{\theta}^\trans \bar{\rho}^\trans) -
  4 \tr (\rho \theta) q.
  \label{eqn:s-sbar-q-transformation}
\end{equation}
The same result can be obtained by studying the transformation of
\(\svielF \epsilon \svielF^\trans\).  More precisely, \(\svielF \epsilon
\svielF^\trans\) is related to \(\svielF' \epsilon \svielF^{\prime \trans}\) in
the same way as \(\der q\) is related to \(\der q'\).

The transformation law for \(\bar{q}\) can be found by using hermitian
conjugation \(\bar{q} = q^\dagger\).
Complex conjugation of odd
variables preserves the order of terms and does not introduce extra
signs.
The reality conditions for the odd variables are \(\theta^\ast
= \bar{\theta}^\trans\), \(\bar{\theta}^\ast = \theta^\trans\), and
similarly for \(\rho\).  Using this we obtain
\begin{equation}
  \delta \bar{q} = -4 (\bar{\theta}^\trans \bar{\rho}^\trans + \rho^\trans \theta^\trans) \bar{q} +
  4 \bar{q} (\bar{\rho} \bar{\theta} + \theta \rho) +
  4 \tr (\bar{\theta} \bar{\rho}) \bar{q}.
\end{equation}
Finally, using a Schouten identity in five indices, the duality
relation \(\bar{q}_{ab} = (\dual q)_{ab} \equiv \tfrac 1 2 \epsilon_{abcd}
q^{cd}\) can be shown to be preserved by these transformations.

We conclude that this super Wilson loop transforms in a simple way
under superconformal transformations.  In fact, both the gauge and the
scalar part have simple transformations by themselves.

One last thing which is not obvious is whether the ten-dimensional
light-like constraint is preserved by superconformal transformations.
It is clearly preserved by Lorentz transformations and R-symmetry
transformations, which descend from ten-dimensional \(\grp{SO}(1,9)\)
Lorentz transformations.  It is less obvious that the light-like
constraint is also preserved by conformal boosts.  These conformal
boosts can be thought of as descending from ten-dimensional conformal
group \(\grp{SO}(2,10)\) which preserves the ten-dimensional
light-like intervals.  So it remains to check if the light-like
constraint is preserved by \(\gen{S}\)-supersymmetries.  We find
\begin{equation}
  \bigcomm{\rho \gen{S} + \bar{\rho} \bar{\gen{S}}}{ (p^2 - q^2)} =
  -4 (p^2 - q^2) \bigbrk{\tr (\rho \theta) + \tr (\bar{\rho} \bar{\theta})},
\end{equation}
where
\(q^2 = \frac 1 8 \epsilon_{a b c d} q^{a b} q^{c d}
= -\tfrac 1 4 \tr(q \bar{q})\)
and
\(p^2 = -\frac 1 2 \tr (p \epsilon p^\trans \epsilon)\).

Let us now study how the superconformal transformations and
kappa-transformations fit together.  We do not need to make any
checks for the transformations which descend from ten-dimensional
super Poincar\'e transformations, since those can be represented as a
left-action in the coset language and their algebra with the right
action of the kappa-transformations can be found from general
arguments.

For the \(\gen{S}\) and \(\bar{\gen{S}}\) supersymmetries we will check the
algebra with kappa-transformations by direct computation.  The
action on \(\theta\) and \(p\) is simple and it yields
\begin{equation}
  [\rho \gen{S} + \bar{\rho} \bar{\gen{S}}, \delta_{\kappa}] = \delta_{\kappa'},
\end{equation}
where \(\kappa' = -4 \kappa \bar{\rho} \bar{\theta} - 4 \kappa \theta
\rho - 4 \rho \theta \kappa\).  However, checking the action of this
commutator on \(q\) is much more involved.

In order to show that the action of this commutator on \(q\) is the
same as on \(\theta\) and \(p\), we need a number of identities
involving the dual of a rank two tensor
\begin{align}
  q^2 (\dual M) &= -\sfrac 1 2 q M^\trans q + \sfrac 1 2 q M q - \sfrac 1 2 q \tr (M q),
\\
\dual (M \bar{q} N) &=
  -\sfrac 1 2 q \tr (M N^\trans ) + \sfrac 1 2 M^\trans q N^\trans
  -\sfrac 1 2 \tr(N) M^\trans q + \sfrac 1 2 N M^\trans q
\\&\qquad
  +\sfrac 1 2 M^\trans N q - \sfrac 1 2 \tr(M) q N^\trans+
  \sfrac 1 2 q M N^\trans + \sfrac 1 2 q N^\trans M
\\ &\qquad
  -\sfrac 1 2 \tr(M) N q + \sfrac 1 2 \tr(N) q^\trans M+
  \sfrac 1 2 N q M + \sfrac 1 2 q \tr(M) \tr(N),
\\
  (\dual M) \tr(N) &= \dual (M N) - \dual (M^\trans N) + N (\dual M) + (\dual M) N^\trans,
\end{align}
where \(M\), \(N\) are \(4 \times 4\) even matrices with zero form
grading.  The last identity can in fact be used to rewrite in a more
compact way the transformation of \(q\) under \(\gen{S}\) and \(\bar{\gen{S}}\)
transformations
\begin{equation}
  \delta q = -4 (\bar{\rho} \bar{\theta} q) +
  4 (\bar{\rho} \bar{\theta} q)^\trans + 8 \dual(\bar{q} \theta \rho) =
  -8 \dual\dual(\bar{\rho} \bar{\theta} q) + 8 \dual(\bar{q} \theta \rho),
\end{equation}
where we have used the fact that \(\dual\dual M = \tfrac 1 2 M - \tfrac 1 2
M^\trans\) (for an anti-symmetric matrix we have the more familiar
relation \(\dual\dual M = M\)).

When acting with \([\rho \gen{S} + \bar{\rho} \bar{\gen{S}}, \delta_{\kappa}] -
\delta_{\kappa'}\) on \(q\), we find
\begin{equation}
    4 \bigbrk{2 \dual (\kappa^\trans \epsilon \rho) +
      \bar{\kappa} \epsilon \bar{\rho}^\trans -
      \bar{\rho} \epsilon \bar{\kappa}^\trans} (p^2 - q^2),
\end{equation}
which vanishes for \(p^2 - q^2 = 0\).

\section{The Super Wilson Loop in Harnad-Shnider Gauge}

\label{sect:1loopvevtheta}
In the previous section we introduced the supersymmetric analog of the
Maldacena-Wilson loop operator and studied its symmetries.
We discussed that our super Maldacena-Wilson loop operator is invariant under superconformal
transformations and moreover enjoys kappa-symmetry.
In this section we will compute the one-loop expectation value through fourth order
in an expansion in the anti-commuting variables and investigate the question of finiteness.
Moreover, we will explicitly check the superconformal invariance
of the one-loop expectation value at the first non-trivial order
in the fermionic coordinates.

\subsection{Component Expansion of the Superfields}

We start by expressing the super-connection in terms of the component fields of $\mathcal{N}=1$ SYM theory in 10d.
The procedure that we will use to do so was introduced
in \cite{Harnad:1984vk} and \cite{Harnad:1985bc}.
By combining the constraint $\sfstr_{\idxtspin{A}\idxtspin{B}}=0$
\eqref{eq:superspaceconstraint}
with the Bianchi identities
for the gauge covariant derivatives one can derive the following set of equations \cite{Witten:1985nt}
\begin{align}
\sfstr_{\idxtspin{A}\idxtspin{B}}&=0  \label{eqn:first_relation} \, ,
\\
\sfstr_{\idxtvec{m} \idxtspin{A}}&=\Gamma_{\idxtvec{m}, \idxtspin{A}\idxtspin{B}} \, \Psi^\idxtspin{B}  \, ,
\\
\bigl\{\scdelF_\idxtspin{A}, \Psi^\idxtspin{B} \bigr\}
&=-\sfrac{1}{2} \, \bigl(\Gamma^{\idxtvec{m} \idxtvec{n}} \bigr){}_\idxtspin{A}{}^\idxtspin{B} \sfstr_{\idxtvec{m} \idxtvec{n}}
\label{eqn:third_relation} \, ,\\
\left[\scdelF_\idxtspin{A}, \sfstr_{\idxtvec{m} \idxtvec{n}} \right]
&=\Gamma_{\idxtvec{m}, \idxtspin{A}\idxtspin{B}} \bigl[\scdelB_{\idxtvec{n}}, \Psi^\idxtspin{B} \bigr]
-\Gamma_{\idxtvec{n}, \idxtspin{A}\idxtspin{B}} \bigl[\scdelB_{\idxtvec{m}}, \Psi^\idxtspin{B} \bigr]
\label{eqn:forth_relation} \, .
\end{align}
The first two of these equations have already been discussed in \secref{sec:supersp-gauge-theory}.
While the fourth equation follows straight-forwardly
from the Bianchi identity for $\scdelB_{\idxtvec{m}}$, $\scdelB_{\idxtvec{n}}$
and $\scdelF_{\idxtspin{A}}$, the third one requires a bit more work.
Starting from the Bianchi identity for one bosonic
and two fermionic gauge covariant derivatives one finds
\begin{align}
\Gamma^{\idxtvec{n}}_{\idxtspin{A}\idxtspin{B}} \,  \sfstr_{\idxtvec{m} \idxtvec{n}}
=\sfrac{1}{2} \Gamma_{\idxtvec{m}, \idxtspin{B}\idxtspin{C}} \bigl\{\scdelF_\idxtspin{A}, \Psi^\idxtspin{C} \bigr\}
+\sfrac{1}{2} \Gamma_{\idxtvec{m}, \idxtspin{A}\idxtspin{C}} \bigl\{\scdelF_\idxtspin{B}, \Psi^\idxtspin{C} \bigr\} \, .
\label{eqn:Bianchi_Df_Df_Db}
\end{align}
By multiplying this equation with $\tilde{\Gamma}_{\idxtvec{o}}$
and taking the trace we obtain the following set of equations:
\begin{align}
\sfstr_{\idxtvec{m} \idxtvec{n}}
=-\sfrac{1}{16} \bigl(\tilde{\Gamma}_{\idxtvec{m} \idxtvec{n}}\bigr)^\idxtspin{A}{}_\idxtspin{C}
\left\{ \scdelF_\idxtspin{A} , \Psi^\idxtspin{C}\right\} \, ,\hspace{2cm} \left\{ \scdelF_\idxtspin{A} , \Psi^\idxtspin{A}\right\} =0 \, .
\end{align}
If we now expand the expression $\bigl\{ \scdelF_\idxtspin{A} , \Psi^\idxtspin{B}\bigr\}$
on a basis of gamma-matrices and make use of the two former equations we find
\begin{align}
\bigl\{ \scdelF_\idxtspin{A} , \Psi^\idxtspin{B}\bigr\}
=-\sfrac{1}{2} \bigl(\Gamma^{\hat{\mu} \idxtvec{n}}\bigr){}_\idxtspin{A}{}^\idxtspin{B} \sfstr_{\idxtvec{m} \idxtvec{n}}
+ \sfrac{1}{16} \sfrac{1}{4!} \bigsbrk{ \bigl( \Gamma_{\idxtvec{p} \idxtvec{o} \idxtvec{n} \idxtvec{m}} \bigr){}_\idxtspin{C}{}^\idxtspin{D}
\bigl\{ \scdelF_\idxtspin{D} , \Psi^\idxtspin{C} \bigr\} }
\bigl( \Gamma^{ \idxtvec{m} \idxtvec{n}  \idxtvec{o} \idxtvec{p}} \bigr){}_\idxtspin{A}{}^\idxtspin{B} \, .
\end{align}
Equation \eqref{eqn:third_relation} follows now by plugging this expression
back into equation \eqref{eqn:Bianchi_Df_Df_Db}
and noting that the coefficient multiplying $\Gamma^{ \idxtvec{m} \idxtvec{n}  \idxtvec{o} \idxtvec{p}}$ vanishes.

After having briefly recalled the derivation
of the relations \eqref{eqn:first_relation}--\eqref{eqn:forth_relation}
we now proceed by fixing the fermionic gauge invariance.
Following \cite{Harnad:1985bc}, we adopt the gauge fixing condition
\begin{align}
\theta^{\idxtspin{A}}\,\sgauge_{\idxtspin{A}}(x,\theta)=0 \, .
\label{eqn:theta-gauge}
\end{align}
Next, we define the operator
\begin{align}
\mathbf{D}:= \theta^{\idxtspin{A}} \, \scdelF_\idxtspin{A}=\theta^\idxtspin{A} \, \frac{\partial}{\partial \theta^{\idxtspin{A}}} \, ,
\end{align}
where the rightmost statement is a consequence
of the gauge fixing condition \eqref{eqn:theta-gauge}
and the fact that the Pauli matrices are symmetric.
By combining the gauge fixing condition \eqref{eqn:theta-gauge}
with the equations \eqref{eqn:first_relation}--\eqref{eqn:forth_relation}
one easily derives the following set of recursion relations
\begin{align}
\label{eqn:D-rec1}
(1+\mathbf{D}) \, \sgauge_{\idxtspin{A}} &= 2 (\theta\Gamma^{\idxtvec{m}})_{\idxtspin{A}}\, \sgauge_{\idxtvec{m}}  \, ,
\\
\mathbf{D} \, \sgauge_{\idxtvec{m}}&= - (\theta\, \Gamma_{\idxtvec{m}} \Psi )\, ,
\\
\mathbf{D} \,  \Psi^{\idxtspin{A}} &=
- \sfrac{1}{2} \, (\theta\Gamma^{\idxtvec{m} \idxtvec{n}})^{\idxtspin{A}} \sfstr_{\idxtvec{m} \idxtvec{n}} \, ,
\\
\mathbf{D} \, \sfstr_{\idxtvec{m} \idxtvec{n}}&=
(\theta\Gamma_{\idxtvec{m}}\scdelB_{\idxtvec{n}}\Psi) - (\theta\Gamma_{\idxtvec{n}}\scdelB_{\idxtvec{m}}\Psi) \,.
\label{eqn:D-rec4}
\end{align}
Using these relations we can now reconstruct the superfields
entirely from the lowest-order data
\begin{align}
\sgauge_{\idxtvec{m}}(x,\theta) = A_{\idxtvec{m}}(x)+\mathcal{O}(\theta)\, ,
\qquad
\sgauge_{\idxtspin{A}}(x,\theta) = \mathcal{O}(\theta) \, ,
\qquad
\Psi^\idxtspin{A}(x, \theta) = \psi^\idxtspin{A}(x)+\mathcal{O}(\theta)   \, .
\end{align}
The result reads
\begin{align}
\sgauge_{\idxtvec{m}}(x,\theta) &=
A_{\idxtvec{m}}(x) - \bigl(\theta \Gamma_{\idxtvec{m}} \psi(x) \bigr)
- \sfrac{1}{4} \bigl(\theta \Gamma_{\idxtvec{m}} \tilde{\Gamma}^{\idxtvec{o} \idxtvec{p}} \theta \bigr)
\Bigl( F_{\idxtvec{o} \idxtvec{p}} (x)
+ \sfrac{2}{3} \bigl(\theta \Gamma_{\idxtvec{o}} \cdel_{\idxtvec{p}} \psi(x) \bigr)
 \label{eqn:supcon_omega_mu}
\\ &\qquad
+\sfrac{1}{12} \bigl(\theta \Gamma_{\idxtvec{o}} \tilde{\Gamma}^{\idxtvec{n} \idxtvec{q}} \theta\bigr)
\cdel_{\idxtvec{p}} F_{\idxtvec{n} \idxtvec{q}}(x)
+\sfrac{1}{6} \bigl[ \bigl(\theta \Gamma_{\idxtvec{o}} \psi(x)\bigr),
\bigl(\theta \Gamma_{\idxtvec{p}} \psi(x) \bigr) \bigr] \Bigr)
+\mathcal{O}(\theta^{5}) \, ,
\nonumber \\
  \Psi^\idxtspin{A}(x, \theta) &=
 \psi^\idxtspin{A}(x) + \sfrac{1}{2} \bigl(\tilde{\Gamma}^{\idxtvec{m} \idxtvec{n}} \theta \bigr)^\idxtspin{A}
\Bigl(  F_{\idxtvec{m} \idxtvec{n}}(x) + \bigl(\theta \Gamma_{\idxtvec{m}} \cdel_{\idxtvec{n}} \psi(x) \bigr)
\\ & \qquad
  +\sfrac{1}{6}  \bigl(\theta \Gamma_{\idxtvec{m}} \tilde{\Gamma}^{\idxtvec{o} \idxtvec{p}} \theta\bigr)
\cdel_{\idxtvec{n}} F_{\idxtvec{o} \idxtvec{p}}(x)
    + \sfrac{1}{3}  \bigl[ \bigl(\theta \Gamma_{\idxtvec{m}} \psi(x)\bigr),
\bigl(\theta \Gamma_{\idxtvec{n}} \psi(x) \bigr) \bigr]  \Bigr) +\mathcal{O}(\theta^{4}) \, ,
\nonumber\\
\sgauge_{\idxtspin{A}}(x,\theta) &= \bigl(\theta\Gamma^{\idxtvec{m}}\bigr)_{\idxtspin{A}}
\Bigl( A_{\idxtvec{m}}(x) - \sfrac{2}{3} \bigl(\theta \Gamma_{\idxtvec{m}}\psi(x)\bigr)
- \sfrac 1 8 \bigl(\theta \Gamma_{\idxtvec{m}} \tilde{\Gamma}^{\idxtvec{o} \idxtvec{p}} \theta \bigr)
 F_{\idxtvec{o} \idxtvec{p}}(x) \Bigr)
+ \mathcal{O}(\theta^{4}) \, ,
\label{eqn:supcon_omega_A}
\end{align}
where $\cdel_{\idxtvec{n}}=\partial_{\idxtvec{n}}+A_{\idxtvec{n}}$ is the usual bosonic gauge covariant derivative.
In the linearized theory one can in fact easily find the complete theta expansion of $\sgauge_{\idxtvec{m}}$
and $\sgauge_{\idxtspin{A}}$. For this, note that at the linear level the recursion
relations \eqref{eqn:D-rec1}--\eqref{eqn:D-rec4} imply
\begin{align}
\mathbf{D}(\mathbf{D}-1) \sgauge^{\text{lin}}_{\idxtvec{m}}
=-\Sigma_{\idxtvec{m}}{}^{\idxtvec{n}} \sgauge^{\text{lin}}_{\idxtvec{n}},
\label{eqn:D-rec-2nd-order}
\end{align}
where we have introduced the operator
\begin{align}
\Sigma_{\idxtvec{m}}{}^{\idxtvec{n}}:=
\bigl(\theta\Gamma_{\idxtvec{m}}{}^{\idxtvec{o} \idxtvec{n}}\theta \bigr)\, \partial_{\idxtvec{o}}\, .
\end{align}
Given the second order relation \eqref{eqn:D-rec-2nd-order}
and the two lowest-order components of the bosonic superfield $\sgauge_{\idxtvec{m}}$
it is not too hard to see that in the linearized theory the superfield $\sgauge_{\idxtvec{m}}$
is to all orders in theta given by
\begin{align}
\sgauge^{\text{lin}}_{\idxtvec{m}}(x,\theta)= \biggl ( \sum_{n=0}^{8}\frac{(-1)^{n}}{(2n)!}\,
 \bigl(\Sigma^{n}\bigr)_{\idxtvec{m}}{}^{\idxtvec{n}}\,
\biggr) \, A_{\idxtvec{n}}(x) -\biggl( \sum_{n=0}^{7}\frac{(-1)^{n}}{(2n+1)!}\,
\bigl(\Sigma^{n}\bigr)_{\idxtvec{m}}{}^{\idxtvec{n}}\,
\biggr)\, \bigl(\theta\Gamma_{\idxtvec{n}}\psi(x)\bigr)\, ,
\label{eqn:omega-mu-all-order}
\end{align}
where
\begin{align}
\bigl(\Sigma^{n}\bigr)_{\idxtvec{m}}{}^{\idxtvec{n}}:=
\Sigma{}_{\idxtvec{m}}{}^{\idxtvec{p}} \Sigma{}_{\idxtvec{p}}{}^{\idxtvec{o}} \dots
\Sigma{}_{\idxtvec{r}}{}^{\idxtvec{q}}\Sigma{}_{\idxtvec{q}}{}^{\idxtvec{n}}
\hspace{1.5cm} \mbox{and} \hspace{1.5cm}
\bigl(\Sigma^0 \bigr){}_{\idxtvec{m}}{}^{\idxtvec{n}}:=
\delta_{\idxtvec{m}}^{\idxtvec{n}} \, .
\end{align}
With the help of equation \eqref{eqn:D-rec1}
we can now also write down an all-order expression for the linearized fermionic superfield $\sgauge^{\text{lin}}_{\idxtspin{A}}$:
\begin{align}
\sgauge^{\text{lin}}_{\idxtspin{A}}(x,\theta)=  2 \bigl( \theta \Gamma^{\idxtvec{m}} \bigr)_\idxtspin{A}
&\left\{ \biggl ( \sum_{n=0}^{7}\frac{(-1)^{n}}{(2n)!(2n+2)}\, \bigl(\Sigma^{n}\bigr)_{\idxtvec{m}}{}^{\idxtvec{n}}\,
\biggr)\, A_{\idxtvec{n}}(x) \right. \nonumber \\
& \hspace{0.4cm} \left. -\biggl( \sum_{n=0}^{7}\frac{(-1)^{n}}{(2n+1)!(2n+3)}\,
\bigl(\Sigma^{n}\bigr)_{\idxtvec{m}}{}^{\idxtvec{n}}\,
\biggr)\, \bigl(\theta\Gamma_{\idxtvec{n}}\psi(x)\bigr) \right\}\, .
\label{eqn:omega-A-all-order}
\end{align}
The generalization of these component field expansions to the full non-linear case
was recently  reported in \cite{Mafra:2015gia}.
Note however that all the fields in the above equations are on-shell since the
relations \eqref{eqn:first_relation}--\eqref{eqn:forth_relation}
do not only eliminate all the auxiliary fields,
but also impose the equations of motion \cite{Witten:1985nt}.

\subsection{Propagators and the One-Loop Vacuum Expectation Value}
\label{sec: one-loop-vev-theta-gauge}
After having established the component expansion of the super-connection
we will now compute the one-loop expectation value of the super Maldacena-Wilson loop operator \eqref{eq:superWL}
through quartic order in an expansion in the anti-commuting variables.
In what follows we write, for the sake of brevity, the super Maldacena-Wilson loop operator \eqref{eq:superWL} as
\begin{align}
\tilde{\swilson}= \frac{1}{N} \tr \pathord \,
\exp{ \left( \oint_{\tilde{Z}} \der \tau \left( p^{\idxtvec{m}} \sgauge_{\idxtvec{m}}
+ \dot\theta^\idxtspin{A} \sgauge_\idxtspin{A}  \right) \right) }
\, ,
\label{eqn:susymwloopop_short}
\end{align}
where we have reassembled the four and six vector that couple to $\sgauge_\mu$ and $\Phi_i$,
respectively, into one 10d vector being defined as follows:
\begin{align}
p^{\idxtvec{m}}= \begin{cases}
\dot{x}^{\mu} + \bigl(\theta \Gamma^{\mu} \dot\theta \bigr)
 \hspace{1,4cm} \text{for} \quad \idxtvec{m} = \mu = 0,\dots,3, \\
q^i= n^{i} \sqrt{p^{\nu} p_{\nu}} \hspace{1,265cm} \text{for} \quad \idxtvec{m} = i = 4,\dots,9.
 \end{cases}
\label{eqn:def-pi-hat-mu}
\end{align}
To get started let us compute the superfield propagators
to the relevant order in the anti-commuting variables.
Since we are eventually interested in $\mathcal{N}=4$ SYM
in 4d rather than in $\mathcal{N}=1$ SYM in 10d we will,
from now on, work with the reduced theory (i.e.\ $\partial_i A_{\idxtvec{m}}(x)=\partial_i \psi(x)=0$)
but keep the 10d notation for the spinors and vectors.
In order to have a self-contained presentation let us
once more list the component expansions
of the linearized superfields $\sgauge^{\text{lin}}_{\idxtvec{m}}$ and $\sgauge^{\text{lin}}_{\idxtspin{A}}$.
Spelling out the first few summands of \eqref{eqn:omega-mu-all-order} and \eqref{eqn:omega-A-all-order}
while discarding derivative terms with respect to the six extra coordinates yields
\begin{align}
\sgauge^{\text{lin}}_{\idxtvec{m}}(x,\theta) =& A_{\idxtvec{m}}(x) - \bigl(\theta\Gamma_{\idxtvec{m}}\psi(x)\bigr)
- \sfrac 1 2 \bigl(\theta\Gamma{}_{\idxtvec{m}}{}^{\rho \idxtvec{n}}\theta\bigr) \partial_{\rho} A_{\idxtvec{n}}(x)
+ \sfrac 1 6 \bigl(\theta\Gamma{}_{\idxtvec{m}}{}^{\rho \idxtvec{n}}\theta\bigr)
\bigl(\theta \Gamma_{\idxtvec{n}} \partial_{\rho} \psi(x)\bigr)  \\
&+\sfrac{1}{24} \bigl(\theta \Gamma{}_{\idxtvec{m}}{}^{\rho \idxtvec{n}}\theta\bigr)
\bigl(\theta\Gamma{}_{\idxtvec{n}}{}^{\sigma \idxtvec{q} } \theta\bigr) \partial_{\rho}
\partial_{\sigma} A_{\idxtvec{q}}(x) +\mathcal{O}(\theta^{5}) \, ,\nonumber\\
\sgauge^{\text{lin}}_{\idxtspin{A}}(x,\theta)=& \bigl(\theta\Gamma^{\idxtvec{m}}\bigr)_{\idxtspin{A}}
\Bigl( A_{\idxtvec{m}}(x) - \sfrac{2}{3} \bigl(\theta\Gamma_{\idxtvec{m}}\psi(x)\bigr)
- \sfrac 1 4 \bigl(\theta \Gamma{}_{\idxtvec{m}}{}^{\rho \idxtvec{n}}\theta\bigr) \partial_{\rho} A_{\idxtvec{n}}(x)
 \Bigr)+ \mathcal{O}(\theta^{4}) \, .
\end{align}
Here and in what follows, hatted vector indices run from $0$ to $9$
while unhatted ones run from $0$ to $3$.
The final ingredients needed to compute the superfield propagators
are the propagators of the component fields.
These can easily be derived from the dimensionally reduced version
of the $\mathcal{N}=1$ SYM action in 10d which in our conventions reads
\begin{align}
S_{\mathcal{N}=1}= -\frac{1}{g_{10}^2} \int \der^{10} x  \:
\tr \Bigl[ \sfrac{1}{2}  F_{\idxtvec{m} \idxtvec{n}} \, F^{\idxtvec{m} \idxtvec{n}}
+   \bigl(\psi  \Gamma_{\idxtvec{m}}  \cdel^{\idxtvec{m}}  \psi \bigr) \Bigr]\, .
\end{align}
In Feynman gauge one finds
\begin{align}
\left \langle A_{\idxtvec{m}}^\mathfrak{a}(x_1)  \, A_{\idxtvec{n}}^\mathfrak{b}(x_2) \right \rangle
= \frac{g^2 \delta^{\mathfrak{a}\mathfrak{b}}}{4 \pi^2}\frac{\eta_{\idxtvec{m} \idxtvec{n}}}{x_{12}^2} \, , \qquad
\left \langle \psi^{\idxtspin{A} \mathfrak{a}}(x_1) \,  \psi^{\idxtspin{B} \mathfrak{b}}(x_2) \right \rangle
= \frac{g^2 \delta^{\mathfrak{ab}}}{2 \pi^2}\frac{\tilde{\Gamma}_{\rho}^{\idxtspin{A}\idxtspin{B}} x_{12}^{\rho}}{x_{12}^4} \, ,
\end{align}
where we have abbreviated $x^{\rho}_{12}:=x^{\rho}_{1}-x^{\rho}_{2}$ and $\mathfrak{a},\mathfrak{b}$ are the
$\grp{SU}(N)$ color indices.
Using the expressions given above it is now a straight-forward exercise
to compute the leading order terms in the Gra{\ss}mann expansion of the superfield
propagators. Through quartic order in the anti-commuting variables we find the
following expression for the propagator of two bosonic superfields%
\footnote{%
Here we have dropped terms which are proportional to
$\square_2 \left( x_{12}^2\right)^{-1}$
and which do therefore vanish up to a delta-function $\delta^{(4)}(x(\tau_1)-x(\tau_2))$.
Explicitly, these terms are $\Bigl(  \sfrac{1}{12}  \bigl(\theta_{12}
 \Gamma_{\idxtvec{m} \idxtvec{n} \idxtvec{q}} \theta_{12}\bigr)
 - \sfrac{1}{12}
 \bigl(\theta_{1} \Gamma_{\idxtvec{m} \idxtvec{n} \idxtvec{q}} \theta_{2}\bigr)
 -\sfrac{1}{6} \eta_{\idxtvec{m} \idxtvec{n} }
 \bigl(\theta_{1} \Gamma_{\idxtvec{q}} \theta_{2}\bigr)
 + \sfrac{1}{2}
 \bigl(\theta_{1} \Gamma_{\idxtvec{m}} \theta_{2}\bigr) \eta_{\idxtvec{n} \idxtvec{q}}
\Bigr) \bigl(\theta_1 \Gamma^{\idxtvec{q}} \theta_2\bigr) \square_2 \bigl(x_{12}^2\bigr)^{-1}$.}
\begin{align}
\bigl\langle \sgauge^\mathfrak{a}_{\idxtvec{m}}(1) \, \sgauge^\mathfrak{b}_{\idxtvec{n}}(2) \bigr\rangle&=
\frac{g^2 \delta^{\mathfrak{ab}}}{4 \pi^2} \Bigl( \eta_{{\idxtvec{m}} {\idxtvec{n}}}+\eta_{{\idxtvec{m}} {\idxtvec{n}}}
\bigl(\theta_1 \Gamma_{\rho} \theta_2\bigr)\partial^{\rho}_2
- 2 \eta_{\rho ( {\idxtvec{n}}}\bigl(\theta_1 \Gamma_{{\idxtvec{m}})} \theta_2\bigr)\partial^{\rho}_2
 +  \sfrac{1}{2} \bigl(\theta_{12} \Gamma_{{\idxtvec{m}} \rho {\idxtvec{n}}} \theta_{12}\bigr) \partial^{\rho}_2
\nonumber\\&\qquad
 +\sfrac{1}{24} \bigl(\theta_{12} \Gamma{}_{{\idxtvec{m}} \sigma}{}^{\idxtvec{q}}
\theta_{12}\bigr)\bigl(\theta_{12} \Gamma_{{\idxtvec{q}} \rho {\idxtvec{n}}} \theta_{12}\bigr)
\partial^{\rho}_2 \partial^{\sigma}_2  -\sfrac{1}{6} \eta_{\sigma [ {\idxtvec{m}}}
\bigl(\theta_{1} \Gamma_{{\idxtvec{n}}] \rho {\idxtvec{q}}} \theta_{2}\bigr) \bigl(\theta_1 \Gamma^{\idxtvec{q}}
 \theta_2\bigr)\partial^{\rho}_2 \partial^{\sigma}_2
\nonumber\\&\qquad
  +\sfrac{1}{6} \eta_{\sigma [ {\idxtvec{m}}} \bigl(\theta_{12} \Gamma_{{\idxtvec{n}}] \rho {\idxtvec{q}}}
\theta_{12}\bigr) \bigl(\theta_1 \Gamma^{\idxtvec{q}} \theta_2\bigr)\partial^{\rho}_2
\partial^{\sigma}_2+\sfrac{1}{6} \eta_{\sigma  {\idxtvec{m}}}
\bigl(\theta_{2} \Gamma_{{\idxtvec{n}} \rho {\idxtvec{q}}} \theta_{2}\bigr) \bigl(\theta_1
\Gamma^{\idxtvec{q}} \theta_2\bigr)\partial^{\rho}_2 \partial^{\sigma}_2
\nonumber\\&\qquad
-\sfrac{1}{6} \eta_{\sigma  {\idxtvec{n}}} \bigl(\theta_{1}
\Gamma_{{\idxtvec{m}} \rho {\idxtvec{q}}} \theta_{1}\bigr) \bigl(\theta_1
\Gamma^{\idxtvec{q}} \theta_2\bigr)\partial^{\rho}_2 \partial^{\sigma}_2
+ \sfrac{1}{2}\bigl(\theta_{12} \Gamma_{{\idxtvec{m}} \sigma {\idxtvec{n}}} \theta_{12}\bigr)
\bigl(\theta_1 \Gamma_{\rho} \theta_2 \bigr) \partial^{\rho}_2 \partial^{\sigma}_2
\nonumber\\&\qquad
 +\sfrac{1}{6} \eta_{{\idxtvec{m}} \rho } \eta_{{\idxtvec{n}} \sigma }
 \bigl(\theta_{1} \Gamma_{{\idxtvec{q}}} \theta_{2}\bigr)
\bigl(\theta_1 \Gamma^{\idxtvec{q}} \theta_2\bigr)\partial^{\rho}_2
\partial^{\sigma}_2 + \sfrac{1}{2} \eta_{{\idxtvec{m}} {\idxtvec{n}}}
\bigl(\theta_{1} \Gamma_{\rho} \theta_{2}\bigr)
\bigl(\theta_1 \Gamma_{\sigma} \theta_2\bigr) \partial^{\rho}_2
\partial^{\sigma}_2
\nonumber\\&\qquad
  -\sfrac{1}{2} \eta_{{\idxtvec{m}} \rho } \bigl(\theta_{1} \Gamma_{{\idxtvec{n}}} \theta_{2}\bigr)
 \bigl(\theta_1 \Gamma_{\sigma} \theta_2\bigr)
\partial^{\rho}_2 \partial^{\sigma}_2 - \sfrac{1}{2} \eta_{\sigma {\idxtvec{n}}}
  \bigl(\theta_{1} \Gamma_{\rho} \theta_{2}\bigr)
\bigl(\theta_1 \Gamma_{{\idxtvec{m}}} \theta_2\bigr)\partial^{\rho}_2
\partial^{\sigma}_2
\nonumber\\&\qquad
+\mathcal{O}(\theta^6)\Bigr) \bigl(x_{12}^2 \bigr)^{-1} \, .
\label{eqn:corr_ommu_omnu}
\end{align}
In order to bring the propagator to this particular form we made repeated
use of the magic identity
$\eta_{\idxtvec{m} \idxtvec{n}}\Gamma^{\idxtvec{m}}_{(\idxtspin{A}\idxtspin{B}}
 \Gamma^{\idxtvec{n}}_{\idxtspin{C})\idxtspin{D}}=0$
as well as of the formula
\begin{align}
\Gamma^{\hat{\mu}_1 \ldots \hat{\mu}_p} \Gamma^{\hat{\nu}_1 \ldots \hat{\nu}_q}= \! \! \! \sum\limits_{k=0}^{\min\{ p,q\}} \! \!  k!
\, \binom{p}{k} \binom{q}{k} \eta^{{[\hat{\mu}_p|}^{\scriptstyle{[\hat{\nu}_1|}}}
\eta^{{|\hat{\mu}_{p-1}|}^{\scriptstyle{|\hat{\nu}_2|}}}\ldots\eta^{{|\hat{\mu}_{p+1-k}|}^{\scriptstyle{|\hat{\nu}_k|}}}
\Gamma^{ {|\hat{\mu}_1 \ldots \hat{\mu}_{p-k}]}^{\scriptstyle{|\hat{\nu}_{k+1} \ldots \hat{\nu}_q] }}}  .
\label{eqn:reduc_formula}
\end{align}
Note that in the last expression we have, for obvious reasons, given up the explicit distinction between the two types of Pauli matrices, cf. \eqref{eqn:10dPauli_antisym}.

The mixed superfield propagator can be computed to be
\begin{align}
\bigl\langle \sgauge^\mathfrak{a}_{\idxtvec{m}}(1) \, \sgauge^\mathfrak{b}_\idxtspin{A}(2) \bigr\rangle
&=
\frac{g^2 \delta^{\mathfrak{ab}}}{4 \pi^2} \bigl(\theta_2 \Gamma^{\idxtvec{n}} \bigr)_\idxtspin{A} \Bigl( \eta_{\idxtvec{m} \idxtvec{n}}
+   \sfrac{1}{4} \bigl(\theta_{2} \Gamma_{\idxtvec{m} \rho \idxtvec{n}} \theta_{2}\bigr)\partial^{\rho}_2
+\sfrac{1}{2} \bigl(\theta_{1} \Gamma_{\idxtvec{m} \rho \idxtvec{n}} \theta_{1} \bigr) \partial^{\rho}_2
\\&\quad \!
-\sfrac{4}{6} \bigl(\theta_{1} \Gamma_{\idxtvec{m} \rho \idxtvec{n}} \theta_{2}\bigr)\partial^{\rho}_2
- \sfrac{8}{6} \eta_{\rho (\idxtvec{n}} \bigl(\theta_{1} \Gamma_{\idxtvec{m})} \theta_{2}\bigr)\partial^{\rho}_2
 + \sfrac{4}{6} \eta_{\idxtvec{m} \idxtvec{n}} \bigl(\theta_{1} \Gamma_{\rho} \theta_{2}\bigr)  \partial^{\rho}_2
+\mathcal{O}(\theta^4) \Bigr) \bigl( x_{12}^2\bigr)^{-1}  , \nonumber
\end{align}
while the propagator between two fermionic superfields is, at leading order,
given by
\begin{align}
\bigl\langle \sgauge^\mathfrak{a}_\idxtspin{A}(1) \, \sgauge^\mathfrak{b}_\idxtspin{B}(2) \bigr\rangle=
\frac{g^2 \delta^{\mathfrak{ab}}}{4 \pi^2} \bigl(\theta_1 \Gamma^{\idxtvec{q}} \bigr)_\idxtspin{A}
 \bigl(\theta_2 \Gamma_{\idxtvec{q}} \bigr)_\idxtspin{B} \bigl( x_{12}^2 \bigr)^{-1} + \mathcal{O}(\theta^4) \, .
\end{align}
Using these results we can now compute the one-loop expectation value
of the loop operator \eqref{eqn:susymwloopop_short} up to quartic order in the anti-commuting variables.
Expanding the exponential in  \eqref{eqn:susymwloopop_short} leads to
\begin{align}
\bigvev{ \tilde{\swilson} } = 1+ \frac{\delta^{\mathfrak{ab}}}{4N}
\int \der \tau_1 \, \der \tau_2 \; \Bigl(&p_1^{\idxtvec{m}} p_2^{\idxtvec{n}}
\bigvev{ \sgauge^\mathfrak{a}_{\idxtvec{m}}(1) \, \sgauge^\mathfrak{b}_{\idxtvec{n}}(2) }
+ 2 p_1^{\idxtvec{m}} \dot\theta_2^\idxtspin{A} \bigvev{ \sgauge^\mathfrak{a}_{\idxtvec{m}}(1) \,
\sgauge^\mathfrak{b}_\idxtspin{A}(2)
}
\nonumber \\
&- \dot\theta_1^\idxtspin{A} \dot\theta_2^\idxtspin{B} \bigvev{ \sgauge^\mathfrak{a}_\idxtspin{A}(1) \,
\sgauge^\mathfrak{b}_\idxtspin{B}(2)
} \Bigr) \: + \, \: \ldots\, \, .
\end{align}
After plugging in the superfield propagators and performing a few manipulations
using the identity \eqref{eqn:reduc_formula}
as well as integration by parts, we obtain
\begin{align}
\langle \tilde{\swilson} \rangle = 1+ \frac{\lambda}{16 \pi^2} \int \der \tau_1 \, \der \tau_2 \;
\mathrm{F}(\tau_1, \tau_2) +\mathcal{O}(\lambda^2) \, ,
\label{eqn:one-loop-vev-theta4}
\end{align}
where the integrand is given by
\begin{align}
\label{eqn:oneloopintegr}
\mathrm{F}(\tau_1, \tau_2) = \Biggl\{& p_1^{\idxtvec{m}} p_2^{\idxtvec{n}} \biggl( \eta_{\idxtvec{m} \idxtvec{n}}
+ \sfrac{1}{2} \bigl(\theta_{12} \Gamma_{\idxtvec{m} \rho \idxtvec{n}} \theta_{12}\bigr)
\partial^{\rho}_2 +\sfrac{1}{24} \bigl(\theta_{12} \Gamma{}_{\idxtvec{m} \rho}{}^{\hat{n}}
\theta_{12} \bigr) \bigl(\theta_{12} \Gamma{}_{\hat{n} \sigma \idxtvec{n}} \theta_{12}\bigr)
\partial^{\rho}_2 \partial^{\sigma}_2 \biggr)  \nonumber\\
&+2 p_1^{\idxtvec{m}} \biggl(\eta_{\idxtvec{m} \idxtvec{q}} + \sfrac{1}{4} \bigl(\theta_{12}
\Gamma{}_{\idxtvec{m} \rho \idxtvec{q}} \theta_{12}\bigr)  \partial^{\rho}_2 \biggr)
 \bigl(\theta_{12} \Gamma^{\idxtvec{q}} \dot{\theta}_2 \bigr) \nonumber \\
&-\sfrac{2}{3}\bigl(\theta_{12} \Gamma^{\idxtvec{m}} \dot{\theta}_1 \bigr)
\bigl(\theta_{12} \Gamma_{\idxtvec{m}} \dot{\theta}_2 \bigr) \Biggr\}
\frac{1}{(x_{12}-\theta_1 \Gamma \theta_2)^2}
+ \mathcal{O}(\theta^6)  \, .
\end{align}
Note that for the sake of manifest super Poincar\'{e} invariance,
we have completed the distance $x_{12}$ in the denominator above.
Upon Taylor expansion all
terms which are of order 6 or higher in the anti-commuting variables of course need to be neglected.
Now, given the integrand \eqref{eqn:oneloopintegr}
it is tempting to speculate about the all-order form of the one-loop integrand.
In fact, by taking into account the all-order expressions \eqref{eqn:omega-mu-all-order} and \eqref{eqn:omega-A-all-order}
for the linearized superfields one can come up with an educated guess for the complete one-loop integrand
\begin{align}
\label{eqn:oneloopintegr-allorder}
\mathrm{F}(\tau_1, \tau_2) =
\Biggl\{& p_1^{\idxtvec{m}} \biggl ( \sum_{n=0}^{16}\frac{1}{(2n)!}\,
\bigl(\bar{\Sigma}^{n}_{12}\bigr)_{\idxtvec{m}}{}^{\idxtvec{n}}\, \biggr)  p_{2 \idxtvec{n}}
+4 p_1^{\idxtvec{m}} \biggl( \sum_{n=0}^{15}\frac{1}{(2n)!(2n+2)}\,
\bigl(\bar{\Sigma}^{n}_{12}\bigr)_{\idxtvec{m}}{}^{\idxtvec{n}}\,
\biggr) \bigl(\theta_{12} \Gamma_{\idxtvec{n}} \dot{\theta}_2 \bigr)
\nonumber \\
&-\frac{2}{3}\bigl(\theta_{12} \Gamma^{\idxtvec{m}} \dot{\theta}_1 \bigr)
\biggl ( \sum_{n=0}^{15} c_n \bigl(\bar{\Sigma}^{n}_{12}\bigr)_{\idxtvec{m}}{}^{\idxtvec{n}}\,
\biggr ) \bigl(\theta_{12} \Gamma_{\idxtvec{n}} \dot{\theta}_2 \bigr) \Biggr\}
\frac{1}{(x_{12}-\theta_1 \Gamma \theta_2)^2} \, ,
\end{align}
where we have used a similar notation as in \eqref{eqn:omega-mu-all-order}
with $\bar{\Sigma}_{12}$ being defined as
\begin{align}
\bigl(\bar{\Sigma}_{12} \bigr)_{\idxtvec{m}}{}^{\idxtvec{n}}
:=\bigl(\theta_{12}\Gamma_{\idxtvec{m}}{}^{\rho \idxtvec{n}}\theta_{12} \bigr) \,
\partial_{2 \rho} \hspace{1.5cm} \mbox{and}
\hspace{1.5cm}
\bigl(\bar{\Sigma}^0_{12} \bigr)_{\idxtvec{m}}{}^{\idxtvec{n}}
:=\delta_{\idxtvec{m}}^{\idxtvec{n}} .
\end{align}
Unfortunately there seems to be no easy way of determining the coefficients $c_n$
beyond explicit calculation so that at this point we only know that they are rational and that $c_0=1$.

\subsection{Finiteness}
\label{sec:finitetheta}

Although it has never been rigorously proven, the Maldacena-Wilson loop operator
as it was introduced in \cite{Maldacena:1998im}
is generally believed to have a finite expectation value as long
as the path on which it depends is smooth and non-intersecting.
In this section we will address the question whether a similar statement holds true for our super-loop operator \eqref{eq:superWL}.
While the question of finiteness is on the one hand very interesting in itself,
it is on the other hand also crucial for us with respect to the way
we will deal with the symmetries of the quantum object.
Since an all-loop analysis seems to be far beyond reach we will,
in what follows, restrict ourselves to a discussion at the one-loop level.
More precisely, we will only focus on the short-distance behavior of the one-loop integrand \eqref{eqn:oneloopintegr}
resp.~\eqref{eqn:oneloopintegr-allorder}
as the absence of short-distance singularities at the level of the integrand already guarantees,
at least for a wide class of contours%
\footnote{Due to the Minkowski structure of our superspace
it could in principal happen that the denominator of \eqref{eqn:oneloopintegr}
becomes zero although the two points $(x(\tau_1),\theta(\tau_1))$ and $(x(\tau_2),\theta(\tau_2))$ do not coincide.
However, for the time being, let us not consider this case
and instead restrict to curves for which the square of supertranslation invariant interval
does not vanish for any two distinct $\tau_1$ and $\tau_2$.}
\!, the finiteness of the one-loop expectation value.

To investigate the UV-behavior of our super-loop operator \eqref{eq:superWL}
we will study the short-distance limit of the one-loop integrand \eqref{eqn:oneloopintegr}.
For this we introduce a parameter
$\varepsilon$ being defined as $\varepsilon=\tau_2-\tau_1$
and expand the integrand \eqref{eqn:oneloopintegr}
\begin{align}
\mathrm{F}(\tau_1, \tau_2)=\mathrm{F}(\tau, \tau+\varepsilon)
\label{eqn:integrand_uv}
\end{align}
for small $\varepsilon$. If we can show that the resulting series
is free of poles, the integrand stays finite over the whole integration domain
thus leading to a finite vacuum expectation value.
The loop contour that we are going to consider is assumed to be a smooth,
non-intersecting,
closed path in superspace which is furthermore super light-like in 10d, i.e.,
\begin{align}
p^{\idxtvec{m}} p_{\idxtvec{m}}=p^{\mu} p_{\mu} -q^i q^{i}=  0
\qquad  \longrightarrow \qquad
p^{\idxtvec{m}} \dot{p}_{\idxtvec{m}}=0 \, ,
\label{eqn:super-lightlike-10d}
\end{align}
but nowhere light-like in 4d, i.e.\ $p^{\mu} p_{\mu}\neq 0$ for all $\tau$.
Here we have used the same definition for the 10d vector $p_{\idxtvec{m}}$
as in the previous section, see \eqref{eqn:def-pi-hat-mu}.
The expansions of the super-path variables $Z(\tau+\varepsilon)$
around $\varepsilon=0$ are given by
\begin{align}
\label{eqn:x-expansion}
x_1^\mu&=x^\mu, &
x_2^\mu&=x^\mu+ \varepsilon \dot{x}^\mu
+ \half\varepsilon^2 \ddot{x}^\mu + \mathcal{O}(\varepsilon^3),
\\
\label{eqn:theta-expansion}
\theta_1^\idxtspin{A}&=\theta^\idxtspin{A}, &
\theta_2^\idxtspin{A}&=\theta^\idxtspin{A}+\varepsilon \dot{\theta}^\idxtspin{A}
+ \half \varepsilon^2 \ddot{\theta}^\idxtspin{A} +\mathcal{O}(\varepsilon^3),
\\
q_1^i&=q^i, &
q^i_2&=
q^i +\varepsilon \dot {q}^i + \half \varepsilon^2 \ddot{q}^i + \mathcal{O}(\varepsilon^3) \, .
\label{eqn:q-expansion}
\end{align}
We start our investigation of the pole structure of \eqref{eqn:integrand_uv}
by focusing on the denominator term in \eqref{eqn:oneloopintegr} and its derivatives.
Using the expansions \eqref{eqn:x-expansion}--\eqref{eqn:q-expansion} we find
\begin{align}
 \frac{1}{(x_{12}-\theta_1 \Gamma \theta_2)^2}
=\frac{1}{\varepsilon^{2}}\left(\frac{1}{p^\mu p_{\mu}}
- \varepsilon\, \frac{p^\mu \dot{p}_{\mu}}{\left( p^\nu p_{\nu}\right)^2}
+\mathcal{O}(\varepsilon^2) \right) \,.
\label{eqn:expansions-denominator-0pd}
\end{align}
For the first derivative of the denominator term we obtain the following Taylor expansion
\begin{align}
 \partial_2^\rho\left( \frac{1}{(x_{12}-\theta_1 \Gamma \theta_2)^2}\right)
=\frac{1}{\varepsilon^{3}}\left(\frac{-2p^\rho}{\left(p^\mu p_{\mu}\right)^2}
+ \varepsilon\, \frac{4 p^\rho p^\mu \dot{p}_{\mu} - \dot{p}^\rho p^\mu p_{\mu} }{\left( p^\nu p_{\nu}\right)^3}
+\mathcal{O}(\varepsilon^2) \right) \, .
\label{eqn:expansions-denominator-1pd}
\end{align}
In general, each partial derivative increases the power of the leading divergence by one unit.
Therefore, we can write down the following formal Taylor series
for the $n$-th derivative of the inverse square of the supertranslation invariant interval
\begin{align}
 \left(\partial_2\right)^n\left( \frac{1}{(x_{12}-\theta_1 \Gamma \theta_2)^2}\right)
=\frac{1}{\varepsilon^{2+n}}\left(S_n(\tau)+\varepsilon T_n(\tau) +\mathcal{O}(\varepsilon^2) \right) \, ,
\label{eqn:expansions-denominator-npd}
\end{align}
where $S_n$ and $T_n$ denote two different rank-$n$ tensors.
Having discussed the denominator term and its derivatives,
we now need to compute the Taylor expansions of the numerator terms.
We start by focusing on the sum of all terms which do not involve a partial derivative.
For these terms we find the following result
\begin{align}
p_1^{\idxtvec{m}} \bigl( p_{2 \idxtvec{m}}
+ 2  \bigl(\theta_{12} \Gamma_{\idxtvec{m}} \dot{\theta}_{2} \bigr) \bigr)
-\sfrac{2}{3}\bigl(\theta_{12} \Gamma^{\idxtvec{m}} \dot{\theta}_1 \bigr)\bigl(\theta_{12} \Gamma_{\idxtvec{m}} \dot{\theta}_2 \bigr)
=\half \varepsilon^2  p^{\idxtvec{m}} \bigl( \ddot{p}_{\idxtvec{m}}
- 2 \bigl( \dot{\theta}\Gamma_{\idxtvec{m}} \ddot{\theta} \bigr) \bigr) +\mathcal{O}(\varepsilon^3) ,
\label{eqn:expansion-numerator-0pd}
\end{align}
where we have used the relations \eqref{eqn:super-lightlike-10d}
as well as the fact that the Pauli matrices are symmetric.
If we multiply the two expansions \eqref{eqn:expansions-denominator-0pd} and \eqref{eqn:expansion-numerator-0pd}
we immediately see that the resulting series starts at order $\varepsilon^0$
and does therefore have a finite limit as $\varepsilon \rightarrow 0$.
For the sum of all terms which multiply
the first derivative of the denominator term \eqref{eqn:expansions-denominator-1pd}
we obtain the following expansion
\begin{align}
\half p_1^{\idxtvec{m}} p_{2}^{\idxtvec{n}}
\bigl(\theta_{12} \Gamma_{\idxtvec{m} \rho \idxtvec{n}} \theta_{12}\bigr)
+ \half p_1^{\idxtvec{m}} \bigl(\theta_{12} \Gamma_{\idxtvec{m} \rho \idxtvec{q}} \theta_{12}\bigr)
\bigl(\theta_{12} \Gamma^{\idxtvec{q}} \dot{\theta}_{2} \bigr)
=\half \varepsilon^3 p^{\idxtvec{m}} \dot{p}^{\idxtvec{n}}
\bigl( \dot{\theta} \Gamma_{\idxtvec{m} \rho \idxtvec{n}} \dot{\theta} \bigr)
+ \mathcal{O}(\varepsilon^4) \, .
\label{eqn:expansion-numerator-1pd}
\end{align}
Here we have used the fact that the term at order $\varepsilon^2$ vanishes due
to the anti-symmetry of $\Gamma_{\idxtvec{m} \rho \idxtvec{n}}$.
Therefore, the product of the two expansions \eqref{eqn:expansions-denominator-1pd} and \eqref{eqn:expansion-numerator-1pd}
is also free of poles. The last term which
we will investigate explicitly is the one which multiplies the second derivative of the denominator term.
Its Taylor expansion reads
\begin{align}
\sfrac{1}{24} p_1^{\idxtvec{m}} p_{2}^{\idxtvec{n}} \bigl(\theta_{12} \Gamma{}_{\idxtvec{m} \rho}{}^{\hat{n}}
\theta_{12} \bigr) \bigl(\theta_{12} \Gamma{}_{\hat{n} \sigma \idxtvec{n}} \theta_{12}\bigr)
 =\sfrac{1}{24}\varepsilon^4 p^{\idxtvec{m}} p^{\idxtvec{n}} \bigl(\dot{\theta} \Gamma{}_{\idxtvec{m} \rho}{}^{\hat{n}}
\dot{\theta} \bigr) \bigl(\dot{\theta} \Gamma{}_{\hat{n} \sigma \idxtvec{n}} \dot{\theta}\bigr)
+ \mathcal{O}(\varepsilon^5) \, .
\end{align}
If we combine this expression with equation \eqref{eqn:expansions-denominator-npd} (for $n=2$)
we see that the resulting series is again UV-safe,
thus completing our proof that the first three terms in the Gra{\ss}mann expansion of the complete one-loop integrand are free of short-distance singularities. Given this result one can now of course raise the question whether this statement
also holds for the higher-order terms in the theta expansion.
In fact, provided that our conjecture about
the form of the complete one-loop integrand \eqref{eqn:oneloopintegr-allorder}
is structurally correct, one easily sees that all the higher-order terms
are also UV-finite since every operator $\bar{\Sigma}_{12}$
comes with two $\varepsilon$'s originating from the expansions of the two $\theta_{12}$
but contains only one partial derivative, cf.\ \eqref{eqn:expansions-denominator-npd}.

\subsection{Conformal Symmetry}

In \secref{sec:conformal-symmetry} we argued that our super-loop operator \eqref{eq:superWL}
is superconformal in the following sense: The action of a superconformal
transformation on the path of the loop is equal to minus the superconformal transformation
acting on the fields.
Assuming the superconformal invariance of the path integral measure this
implies a   Ward identity for the super Maldacena-Wilson loop: The path action of (say) the superconformal generators
$\gen{S}$ and $\bar{\gen{S}}$  on the vacuum expectation value of our super-loop operator
should vanish.
Using the results obtained in \secref{sec: one-loop-vev-theta-gauge}
we will now explicitly check this Ward identity at leading order in the perturbative expansion.
More precisely, we will compute the variation of the expectation value \eqref{eqn:one-loop-vev-theta4}
under a superconformal boost and then show that this expression vanishes modulo terms
which are at least cubic in the anti-commuting variables.

In 10d notation for the spinors the superconformal transformation laws
of the superspace coordinates
\eqref{eqn:s-sbar-x-transformation}, \eqref{eqn:s-sbar-theta-transformation}
and \eqref{eqn:s-sbar-q-transformation} take the following form
\begin{align}
  \delta x_{\mu} &= x^{\nu} \bigl( \theta \Gamma_{\mu}
 \tilde{\Gamma}_{\nu} \xi \bigr)  - \tfrac {1}{12} \bigl(\theta \Gamma^{\nu
    \rho} \xi \bigr) \bigl(\theta \Gamma_{\mu}
  \tilde{\Gamma}_{\nu \rho} \theta \bigr) + \tfrac {1}{12} \bigl(\theta
  \Gamma^{i j} \xi \bigr) \bigl(\theta \Gamma_{\mu} \tilde{\Gamma}_{i j}
  \theta \bigr) \, ,\\
  \delta \theta^\idxtspin{A} &= x^{\mu} \bigl( \xi \tilde{\Gamma}_{\mu} \bigr)^\idxtspin{A}  + \tfrac 1 2 \bigl(\theta
  \xi \bigr) \theta^\idxtspin{A} - \tfrac 1 4 \bigl(\theta \Gamma^{\mu \nu}
  \xi \bigr) \bigl(\theta \Gamma_{\mu \nu} \bigr)^\idxtspin{A} + \tfrac 1 4 \bigl(\theta
  \Gamma^{i j} \xi \bigr) \bigl(\theta \Gamma_{i j}\bigr)^\idxtspin{A} \, , \\
  \delta q^i&=2 q_j \bigl(\theta \Gamma^j \tilde{\Gamma}^i \xi \bigr) \, ,
\end{align}
where $\xi$ is the parameter which specifies the transformation. In order to be able to split the subsequent computation in smaller
but self-contained pieces we will work with the following form of the one-loop expectation value
\begin{align}
\bigl \langle \tilde{\swilson} \bigr \rangle_{(1)}=
\frac{\lambda}{16 \pi^2} \int \der \tau_1 \, \der \tau_2 \;
\biggl\{ I_{p}(\tau_1,\tau_2)+I_q(\tau_1,\tau_2) + \mathcal{O}(\theta^4)\biggr\} \, ,
\end{align}
where we have separated the integrand into a part which does not involve the coordinate $q$ at all
\begin{align}
I_{p}(\tau_1,\tau_2)&=\frac{ p_1^{\mu} p_{2}^\nu}{x^2_{12}}
\left( \eta_{\mu \nu}+2 \eta_{\mu \nu} \bigl( \theta_1 \Gamma_\rho \theta_2 \bigr)
\frac{x^\rho_{12}}{x^2_{12}}+ \bigl(\theta_{12} \Gamma_{\mu \rho \nu} \theta_{12}\bigr) \frac{x^\rho_{12}}{x^2_{12}}\right)
+\frac{2 p_1^\mu}{x_{12}^2} \bigl(\theta_{12} \Gamma_{\mu} \dot{\theta}_{2}\bigr) \, ,
\label{eqn:Inq}
\end{align}
and a piece which involves all the remaining terms
\begin{align}
I_q(\tau_1,\tau_2)=\mathord{}&\frac{q_1^{i} q_{2}^j}{x_{12}^2} \left( \eta_{ij}+2 \eta_{ij}
\bigl( \theta_1 \Gamma_\rho \theta_2 \bigr) \frac{x^\rho_{12}}{x^2_{12}}
+ \bigl(\theta_{12} \Gamma_{i \rho j} \theta_{12}\bigr) \frac{x^\rho_{12}}{x^2_{12}} \right)
\nonumber \\
&+\frac{2  q_1^i}{x_{12}^2}\left(p_2^{\mu} \bigl(\theta_{12} \Gamma_{i \rho \mu} \theta_{12}\bigr)\frac{x^\rho_{12}}{x^2_{12}}
+ \bigl(\theta_{12} \Gamma_{i}  \dot{\theta}_{2}\bigr)\right)\, .
\end{align}
This particular splitting turns out to be useful since the $p$-part
and the $q$-part do not mix under superconformal transformations.

Let us start by computing the variation of the integrand $I_{p}(\tau_1,\tau_2)$.
Under a superconformal boost the super-momentum $p^\mu$ transforms as
\begin{align}
\delta p_\mu=2 \dot{x}^\nu \bigl(\theta \Gamma_\mu \tilde{\Gamma}_\nu \xi \bigr)
+ \mathcal{O}(\theta^3)\, .
\end{align}
Using this expression we can now compute the variation of the integrand \eqref{eqn:Inq}.
If we only keep terms which are linear in theta after the variation we find
\begin{align}
\delta I_{p}(\tau_1,\tau_2)\biggl|_{\mathcal{O}(\theta)}\! \! \!\mathrel{\hat{=}}\mathord{}&
\frac{2 \dot{x}_1^\mu}{x_{12}^2}\biggl(2\dot{x}_2^\nu
 \bigl(\theta_2 \Gamma_\mu \tilde{\Gamma}_\nu \xi \bigr)
- x_{12}^\nu \bigl( \dot{\theta}_2 \Gamma_\mu \tilde{\Gamma}_\nu \xi \bigr)
+ \dot{x}_2^\nu \bigl( \theta_{12} \Gamma_\mu \tilde{\Gamma}_\nu \xi \bigr) \biggr)
\nonumber \\
&-\frac{2 \dot{x}_1^\mu \dot{x}_2^\nu}{x_{12}^4}\biggl( \eta_{\mu \nu} x_{12}^\rho x_{12}^\sigma
\bigl(\theta_1 \Gamma_\rho \tilde{\Gamma}_\sigma \xi \bigr) +  \eta_{\mu \nu} x_{12}^\rho x_{12}^\sigma
\bigl(\theta_2 \Gamma_\rho \tilde{\Gamma}_\sigma \xi \bigr) \nonumber \\
& \hspace{2cm}-x_{12}^\rho x_{12}^\sigma \bigl( \theta_{12} \Gamma_{\mu \rho \nu}
\tilde{\Gamma}_\sigma \xi \bigr)\biggr) \, ,
\end{align}
where by $\hat{=}$ we mean that the expression on the right-hand side gives the same result
when integrated over $\tau_1$ and $\tau_2$, i.e.,
we already made use of the freedom to relabel integration variables.
Using the Clifford algebra relation for the Pauli matrices as well as the identity \eqref{eqn:reduc_formula}
the former expression can be rewritten as follows:
\begin{align}
\delta I_{p}(\tau_1,\tau_2)\biggl|_{\mathcal{O}(\theta)}\! \! \!\mathrel{\hat{=}}\mathord{}&
\frac{2 \dot{x}_1^\mu}{x_{12}^2}\biggl(2\dot{x}_2^\nu \bigl(\theta_2 \Gamma_\mu \tilde{\Gamma}_\nu \xi \bigr)
- x_{12}^\nu \bigl( \dot{\theta}_2 \Gamma_\mu \tilde{\Gamma}_\nu \xi \bigr)
+ \dot{x}_2^\nu \bigl( \theta_{12} \Gamma_\mu \tilde{\Gamma}_\nu \xi \bigr) -\dot{x}_{2 \mu}
\bigl(\theta_1 \xi \bigr) \nonumber \\
&\hspace{1,1cm} - \dot{x}_{2 \mu} \bigl(\theta_2 \xi \bigr)
- \dot{x}_2^\nu \bigl(\theta_{12} \Gamma_{\mu \nu} \xi \bigr)
+\frac{2 \dot{x}_2^\rho x_{12 \rho}}{x_{12}^2}\, x_{12}^\nu \bigl(\theta_{12} \Gamma_\mu \tilde{\Gamma}_\nu \xi \bigr)\biggr)  \, .
\end{align}
By use of the Clifford algebra relation as well as integration by parts
the variation of $I_{p}(\tau_1,\tau_2)$ can be shown to be equivalent to
\begin{align}
\delta I_{p}(\tau_1,\tau_2)\biggl|_{\mathcal{O}(\theta)}\! \! \! \mathrel{\hat{=}}\mathord{}&
2 \dot{x}_1^\mu \biggl(-\frac{\dot{x}_2^\nu}{x_{12}^2} \bigl(\theta_{12} \Gamma_\mu  \tilde{\Gamma}_\nu \xi \bigr)
- \frac{x_{12}^\nu}{x_{12}^2} \bigl( \dot{\theta}_2 \Gamma_\mu \tilde{\Gamma}_\nu \xi \bigr)
+\!\frac{2 \dot{x}_2^\rho x_{12 \rho}}{x_{12}^4}\, x_{12}^\nu \bigl(\theta_{12} \Gamma_\mu \tilde{\Gamma}_\nu \xi \bigr)\biggr).
\end{align}
This expression can however easily be seen to be a total derivative with respect to $\tau_2$
\begin{align}
\delta I_{p}(\tau_1,\tau_2)\biggl|_{\mathcal{O}(\theta)}\! \! \!\mathrel{\hat{=}}
2\dot{x}_1^\mu \,\frac{\partial}{\partial \tau_2}\left( \frac{x_{12}^\nu}{x_{12}^2}
\bigl(\theta_{12} \Gamma_\mu \tilde{\Gamma}_\nu \xi   \bigr)  \right)
\end{align}
and does therefore vanish when integrated over. Let us note that the expression is also
not singular in the limit $1\to 2$.

We now turn to the variation of $I_q(\tau_1,\tau_2)$.
If we again discard terms which are of higher-order than linear in theta we find
\begin{align}
\delta I_q(\tau_1,\tau_2)\biggl|_{\mathcal{O}(\theta)}\! \! \!\mathrel{\hat{=}}\mathord{}&
\frac{2  q_1^i  q_{2}^j}{x_{12}^2} \biggl(  \bigl(\theta_{12} \Gamma_{ij} \xi \bigr)
+   \eta_{ij} \bigl(\theta_1 \xi \bigr)+ \eta_{ij} \bigl(\theta_2 \xi \bigr)
- \frac{1}{x_{12}^2} \biggl( \eta_{ij} x_{12}^\mu  x_{12}^\nu \bigl( \theta_1 \Gamma_\mu \tilde{\Gamma}_\nu \xi \bigr)
\nonumber \\
&\hspace{1,8cm}+ \eta_{ij} x_{12}^\mu x_{12}^\nu \bigl(\theta_2 \Gamma_\mu \tilde{\Gamma}_\nu \xi \bigr)
-   x_{12}^\mu x_{12}^\rho \bigl(\theta_{12} \Gamma_{i \rho j} \tilde{\Gamma}_\mu\xi \bigr) \biggr)\biggr)
\\
&+\frac{2q_1^i}{x^2_{12}} \biggl(  -x_{12}^\mu \bigl(  \dot{\theta_2}\Gamma_i \tilde{\Gamma}_\mu \xi\bigr)
+  \dot{x}_2^\mu \bigl(\theta_{12} \Gamma_i \tilde{\Gamma}_\mu \xi  \bigr)
+\frac{2 p_2^\nu}{x^2_{12}} \, x_{12}^\mu x_{12}^\rho
\bigl( \theta_{12} \Gamma_{i \rho \nu} \tilde{\Gamma}_\mu \xi  \bigr)  \biggr)
\nonumber \, .
\end{align}
Using the Clifford algebra relation as well as the reduction formula \eqref{eqn:reduc_formula}
one easily shows that the part being quadratic in $q$ vanishes without leaving behind a total derivative term.
Note that this is expected as the terms being quadratic in $q$
originate from the correlator of two scalar superfields $\Phi_i$
and the combination $q^i \Phi_i$ is superconformally invariant,
see section \eqref{sec:conformal-symmetry}.
The terms being linear in $q$ can be rewritten as
\begin{align}
\delta I_q(\tau_1,\tau_2)\biggl|_{\mathcal{O}(\theta)}\! \! \!\mathrel{\hat{=}}\mathord{}
2q_1^i \biggl( - \frac{x_{12}^\mu}{x_{12}^2}
\bigl(\dot{\theta_2} \Gamma_i \tilde{\Gamma}_\mu \xi   \bigr)
-  \frac{\dot{x}_2^\mu}{x_{12}^2}  \bigl(\theta_{12} \Gamma_i \tilde{\Gamma}_\mu \xi  \bigr)
+\frac{2 \dot{x}_2^\rho x_{12 \rho}}{x^4_{12}} \, x_{12}^\mu
\bigl( \theta_{12} \Gamma_i \tilde{\Gamma}_\mu \xi  \bigr)  \biggr)  \, ,
\end{align}
which is pretty much the same total derivative that we encountered
when varying the integrand $I_{p}(\tau_1,\tau_2)$
\begin{align}
\delta I_q(\tau_1,\tau_2)\biggl|_{\mathcal{O}(\theta)}\! \! \!\mathrel{\hat{=}}\mathord{}
2q_1^i\, \frac{\partial}{\partial \tau_2}\left( \frac{x_{12}^\nu}{x_{12}^2}
\bigl(\theta_{12} \Gamma_i \tilde{\Gamma}_\nu \xi   \bigr)  \right) \, .
\end{align}
Putting both partial results together yields
\begin{align}
\delta \bigl \langle \tilde{\swilson} \bigr \rangle_{(1)}=
\frac{\lambda}{8 \pi^2} \int \der \tau_1 \, \der \tau_2 \;
\biggl\{ \frac{\partial}{\partial \tau_2} \left( \frac{x_{12}^\nu}{x_{12}^2}
\bigl(\theta_{12} \bigl(\dot{x}_1^\mu \Gamma_\mu + q_1^i \Gamma_i \bigr) \tilde{\Gamma}_\nu \xi   \bigr) \right)
+ \mathcal{O}(\theta^3)\biggr\} \, ,
\end{align}
which shows that the vacuum expectation value of our super-loop operator \eqref{eq:superWL}
is annihilated by the generators of superconformal transformations modulo terms
which are at least cubic in the anti-commuting variables.

\section{The Super Wilson Loop in Non-Chiral Superspace using Light-Cone Gauge}
\label{sect:4}

In the previous section we showed that the Maldacena-Wilson loop
is finite and conformally invariant
at one loop and at the leading orders in the theta expansion.
Here we investigate the $\order{g^2}$ divergence structure
of a Wilson loop in full superspace.
This is not only relevant for the kappa-symmetric case,
which is expected to be finite, but also towards
properly renormalizing the remaining Wilson loops.

\subsection{Superspace Propagators}
\label{sec:superspace-prop}

At $\order{g^2}$ there is only one
term contributing to the expectation value of a Wilson loop:
The Wilson loop must be expanded to two fields
which are consequently connected by a propagator.
All effects of a non-abelian gauge group are
irrelevant at this level, and one can effectively
assume an abelian gauge group.

In order to discuss the singularities of a Wilson loop,
we need the propagators in superspace.
The propagators for the gauge connection clearly depend on the gauge chosen.
In \cite{Beisert2012} a full superspace
expression was derived for an abelian gauge group in a light-cone gauge.

\paragraph{Pre-potentials.}

First of all, the connection can be split up into chiral,
anti-chiral and pure gauge contributions
\[
\sgauge=\sgauge^+ + \sgauge^- + \der\Lambda.
\]
The chiral components $\sgauge^\pm$
can be expressed in terms of chiral pre-potentials $\spre^\pm$
\begin{align}
\sgauge^+ &=
\varepsilon^{\beta\gamma}  (\der x^+)^{\dot\alpha\alpha}
\frac{\partial}{\partial (x^+)^{\dot\alpha\gamma}}\, \spre^+_{\alpha\beta}
+\varepsilon^{\beta\gamma}  \der\theta^{a\alpha}
\frac{\partial}{\partial\theta^{a\gamma}}\, \spre^+_{\alpha\beta},
\nln
\sgauge^- &=
-\varepsilon^{\dot\beta\dot\gamma} (\der x^-)^{\dot\alpha\alpha}
\frac{\partial}{\partial (x^-)^{\dot\beta\alpha}}\, \spre^-_{\dot\alpha\dot\gamma}
-\varepsilon^{\dot\beta\dot\gamma} \der\bar\theta^{\dot\alpha}{}_a
\frac{\partial}{\partial\bar\theta^{\dot\beta}{}_a}\, \spre^-_{\dot\alpha\dot\gamma}.
\end{align}
It was then shown in \cite{Beisert2012} that the pre-potentials have the following propagators
\begin{align}
\label{eq:preprop}
\bigvev{
\spre^+_{\alpha\beta}(1)
\spre^+_{\gamma\delta}(2)
}
&\sim
\frac{\ell_{\alpha}\ell_{\beta}\ell_{\gamma}\ell_{\delta}\,
      \delta^{0|4}(\theta_{12}| x^+_{12} \cket{\bar \ell})}
     {\sbra{\ell} x^+_{12} \cket{\bar \ell}^4(x^+_{12})^2}\,,\nln
\bigvev{
\spre^+_{\alpha\beta}(1)
\spre^-_{\dot\gamma\dot\delta}(2)
}
&\sim
\frac{\ell_{\alpha}\ell_{\beta}\bar \ell_{\dot\gamma}\bar \ell_{\dot\delta}
     (x^{+-}_{12})^2\bigbrk{* + \log (x^{+-}_{12})^2}}{\sbra{\ell} x^{+-}_{12} \cket{\bar \ell}^2}
+ \frac{\ell_{\{\alpha}\varepsilon_{\beta\}\lambda}\varepsilon_{\dot\kappa\{\dot\gamma}\bar \ell_{\dot\delta\}}
(x^{+-}_{12})^{\lambda\dot\kappa}}{\sbra{\ell} x^{+-}_{12} \cket{\bar \ell}}\,,
\nln
\bigvev{
\spre^-_{\dot\alpha\dot\beta}(1)
\spre^-_{\dot\gamma\dot\delta}(2)
}
&\sim
\frac{\bar \ell_{\dot\alpha}\bar \ell_{\dot\beta}\bar \ell_{\dot\gamma}\bar \ell_{\dot\delta}\,
      \delta^{0|4}(\sbra{\ell} x^-_{12} |\bar\theta_{12})}
     {\sbra{\ell} x^-_{12} \cket{\bar \ell}^4(x^-_{12})^2}\,,
\end{align}
where $\ell$ and $\bar{\ell}$ are spinors corresponding
to a light-like gauge parameter and $*$ denotes an unspecified integration constant.
We also use the intervals \(x_{12}^\pm := x_1^\pm - x_2^\pm\) and
\(x_{12}^{+-} = -x_{21}^{-+} := x_1^+ - x_2^- + 4 \bar{\theta}_2 \theta_1\)
and the compact notations
\begin{gather}
  \langle \ell \vert x_{12}^+ \vert \bar{\ell}] = \ell^\alpha (\bar{x}_{12}^+)_{\alpha \dot{\alpha}} \bar{\ell}^{\dot{\alpha}}, \qquad
  (\theta_{12} \vert x_{12}^+ \vert \bar{\ell}])^a = \theta^{a \alpha}(\bar{x}_{12}^+)_{\alpha \dot{\alpha}} \bar{\ell}^{\dot{\alpha}},
\end{gather}
with \(\bar{x}_{\alpha \dot{\alpha}} = x_\mu \sigma_{\alpha \dot{\alpha}}^\mu\).

\paragraph{Potentials.}

Combining the above expressions and performing some spinor algebra,
we find the propagator for the mixed-chiral gauge connections
\begin{align}
\bigvev{\sgauge^+(1) \sgauge^-(2)}
&= \frac {g^2}{8 \pi^2} \Biggr\lbrace
\sbra{\ell}\der_1x_{12}^{+-} (x_{12}^{+-})^{-1} \der_2x_{12}^{+-}\cket{\bar\ell}
\frac{1}{\bra{\ell}x_{12}^{+-}\cket{\bar\ell}}
\nln&\qquad
-
\sbra{\ell}\der_1x^{+-}_{12}\cket{\bar\ell}
\sbra{\ell}\der_2x^{+-}_{12}\cket{\bar\ell}
\frac{\lrbrk{\#+\log (x_{12}^{+-})^2}}{\sbra{\ell}x_{12}^{+-}\cket{\bar\ell}^2}
\nln&\qquad
+
\sbra{\ell}\der_1\der_2x^{+-}_{12}\cket{\bar\ell}
\frac{\lrbrk{\#+\log (x_{12}^{+-})^2}}{\sbra{\ell}x_{12}^{+-}\cket{\bar\ell}}\,\Biggr\rbrace.
\end{align}
The factor $\#$ is related to the above integration constant $*$.
As such this propagator contains logarithmic terms, a common feature
of light-cone gauges.
However, one can show that they are merely due to the choice of gauge.
Namely, we have the freedom to add terms
which are a total derivative on at least one leg.
Adding the following total derivative terms is convenient to simplify the propagator
\[
\label{eq:proptotalder}
\half \der_2\lrsbrk{\log (x^{+-}_{12})^2\der_1\log\sbra{\ell}x^{+-}_{12}\cket{\bar\ell}}
    -\half \der_1\lrsbrk{\log (x^{+-}_{12})^2\der_2\log\sbra{\ell}x^{+-}_{12}\cket{\bar\ell}}
-\#\der_1 \der_2 \log \langle \ell \vert x_{12}^{+-} \vert \bar{\ell}].
\]
They remove the logarithmic terms as well as the integration constant,
and we are left with a purely rational propagator
\begin{align}
\label{eq:mixedchiralpropagator}
\bigvev{\sgauge^+(1) \sgauge^-(2)}
&= \frac {g^2}{8 \pi^2} \Biggl\lbrace
\frac{\sbra{\ell}\der_1x_{12}^{+-} (x_{12}^{+-})^{-1} \der_2x_{12}^{+-}\cket{\bar\ell}}{\sbra{\ell}x_{12}^{+-}\cket{\bar\ell}}
\nln&\qquad
-\frac{1}{2} \,\frac{\der_1\sbra{\ell}x^{+-}_{12}\cket{\bar\ell}}{\sbra{\ell}x^{+-}_{12}\cket{\bar\ell}}\,
\frac{\der_2 (x^{+-}_{12})^2}{(x^{+-}_{12})^2}
-\frac{1}{2} \,\frac{\der_1 (x^{+-}_{12})^2}{(x^{+-}_{12})^2}\,
\frac{\der_2\sbra{\ell}x^{+-}_{12}\cket{\bar\ell}}{\sbra{\ell}x^{+-}_{12}\cket{\bar\ell}} \Biggr\rbrace.
\end{align}

Several comments on this change of gauge are in order.
The above derivative terms do not follow from a specific gauge or a Lagrangian.
We merely added them to simplify the structure of the propagator for gauge potentials.
In particular, they have no influence on the one-loop expectation value
of Wilson loops because they can be integrated trivially.
There is, however, a subtlety in the previous statement:
Alike the propagator, these terms are singular in the limit
of coincident points.
This implies that the Wilson loop
must be regularized and renormalized in the UV.
Then the additional terms do not necessarily integrate to zero.
The change of gauge must thus be accompanied by
a compensating change of the renormalization.
Only then the expectation value does not change.

We have added the above terms to remove the logarithmic terms
in the propagator, which would lead to an inconvenient logarithmic
singularity structure. However, there are further terms
that could be added which preserve the rational structure of the propagator,
most notably
\[
\der_1\der_2\log (x^{+-}_{12})^2.
\]
Here we fix its coefficient such that
the singularity structure of the bosonic truncation
of the propagator agrees with
an ordinary bosonic propagator in Feynman gauge.
It turns out that no further contribution from this term
is needed, and \eqref{eq:mixedchiralpropagator} is the desired final result.

Here we only treated the mixed-chiral case.
It would be nice to find a convenient superspace expression
for the purely chiral propagator
$\vev{\sgauge^+(1) \sgauge^+(2)}$.
The fermionic delta-function in
$\vev{\spre^+(1)\spre^+(2)}$ \eqref{eq:preprop}
leads to more involved combinatorics,
yet the result will be rational by construction.
As we will not actually need the precise expression,
we shall not compute it here.

\paragraph{Field strength.}

Let us continue to evaluate correlators involving
the field strength $\sfstr_{\text{lin}} =\der\sgauge$.
It can be decomposed into chiral and anti-chiral components
\(\sfstr_{\text{lin}} = \sfstr^+ + \sfstr^-\),
where we consider only the linearized part of the field strength,
since the higher-order terms in the fields do not contribute at this order
(see \appref{sec:supersp-geom} for the formulas needed to prove this decomposition).
The chiral and anti-chiral field strengths then take the form
\begin{align}
  \sfstr^- &=
  \bigsbrk{ 2 (\svielF \epsilon \svielF^\trans)^{ab} +
  \sfrac 1 3 (\svielB \epsilon \svielF^\trans)^{\dot{\alpha} a} \bar{\sdel}^b{}_{\dot{\alpha}} +
  \sfrac 1 {96} (\svielB \epsilon \svielB^\trans)^{\dot{\alpha} \dot{\beta}}
\bar{\sdel}^a{}_{\dot{\alpha}} \bar{\sdel}^b{}_{\dot{\beta}}} \Phi_{ab},\\
  \sfstr^+ &=
  \bigsbrk{ 2 (\svielC^\trans \epsilon \svielC)_{ab} +
  \sfrac 1 {3} (\svielB^\trans \epsilon \svielC)^{\alpha}_{\hphantom{\alpha} a} \sdel_{b \alpha} +
  \sfrac 1 {96} (\svielB^\trans \epsilon \svielB)^{\alpha \beta} \sdel_{a \alpha} \sdel_{b \beta}} \bar{\Phi}^{ab}.
\end{align}
All the components of the field strength can be written as SUSY covariant derivatives
acting on the scalar superfield \(\Phi_{a b} = \tfrac 1 2 \epsilon_{a b c d} \bar{\Phi}^{c d}\).

The two-point functions of field strengths follow from
the above gauge propagator \eqref{eq:mixedchiralpropagator}
via $\sfstr^\pm=\der \sgauge^\pm$ which includes
the scalar correlator.
The latter takes the form%
\footnote{The numerical constant can not be determined by using symmetry considerations.
It can be obtained by comparing the two-point functions following
from it with the ones computed from canonically normalized gauge fields.}
\[
\label{eq:phi-phi-propagator}
\bigvev{\bar{\Phi}^{ab}(1) \Phi_{cd}(2)}
=
-\frac {g^2}{\pi^2}
\frac {(1-4 \theta_{12} x_{12}^{-+,-1} \bar{\theta}_{12})^a_{\hphantom{a} [c}
  (1-4 \theta_{12} x_{12}^{-+,-1} \bar{\theta}_{12})^b_{\hphantom{b} \vert d]}}{(x_{12}^{-+})^2}\,.
\]
This form can be easily derived
from elementary CFT considerations
by using translation symmetry to place one point
at the origin and then using the inversion transformations of the scalar fields,
as described in eq.~\eqref{eq:scalar-inv}.

For the other two-point functions we find
\begin{align}
  \bigvev{ \mathcal{F}^+(1) \Phi_{ab}(2)} &= -\frac {2 g^2}{\pi^2} \epsilon^{\alpha \beta}
  \der_1 \bigbrk{x_{12}^{-+,-1} \bar{\theta}_{12}}_{\alpha a}
  \der_1 \bigbrk{x_{12}^{-+,-1} \bar{\theta}_{12}}_{\beta b},
\label{eq:fp-phi-propagator}
\\
  \bigvev{ \mathcal{F}^+(1) \mathcal{F}^-(2)} &=
  \frac {g^2} {4 \pi^2} \Bigl[
  \tr\bigbrk{\der_1 (x_{12}^{-+,-1} \der_2 x_{12}^{-+}) \der_1 (x_{12}^{-+,-1} \der_2 x_{12}^{-+})}
\nln & \qquad
 - \tr\bigbrk{\der_1 (x_{12}^{-+,-1} \der_2 x_{12}^{-+})) \tr(\der_1 (x_{12}^{-+,-1} \der_2 x_{12}^{-+})}\Bigr]
  ,\label{eq:fp-fm-propagator}
\\
  \bigvev{ \mathcal{F}^+(1) \mathcal{F}^+(2)} &=
  -\frac{2 g^2}{\pi^2} \epsilon^{\alpha \beta} \epsilon_{\dot{\alpha} \dot{\beta}} \epsilon^{a b c d}
  \bigsbrk{x_{12}^{+-} \der_2 (x_{12}^{+-,-1} \bar{\theta}_{12})
  (1 - 4 \theta_{12} x_{12}^{-+,-1} \bar{\theta}_{12})}^{\dot{\alpha}}_{\hphantom{\alpha} a}
\nln &\qquad
  \cdot\bigsbrk{x_{12}^{+-} \der_2 (x_{12}^{+-,-1} \bar{\theta}_{12})
  (1 - 4 \theta_{12} x_{12}^{-+,-1} \bar{\theta}_{12})}^{\dot{\beta}}_{\hphantom{\beta} b}
\nln &\qquad
  \cdot\der_1 (x_{12}^{-+,-1} \bar{\theta}_{12})_{\alpha c}\,
  \der_1 (x_{12}^{-+,-1} \bar{\theta}_{12})_{\beta d}.
\label{eq:fp-fp-propagator}
\end{align}

\subsection{Wilson Loop Singularity Structure}

Now we are ready to investigate the UV-singularities of the Wilson loop
one-loop expectation value
in full superspace
\[
\bigvev{\tilde{\swilson}}_{(1)}
\sim
\int \int
\bigvev{(\sgauge^++\sgauge^-+\sscalar)(1) (\sgauge^++\sgauge^-+\sscalar)(2)}
.
\]
Here, $\sscalar := \der\tau q^i \Phi_i$ denotes the coupling of the scalars
as a one-form on the Wilson loop.
For simplicity, we regularize this expression by a short-distance cut-off
as in \secref{sec:finitetheta}.

We parametrize the points $\tau_1=\tau$ and $\tau_2=\tau+\varepsilon$
and expand the position variables in $\varepsilon$
\eqref{eqn:x-expansion}--\eqref{eqn:q-expansion}
\[
Z_1=Z(\tau),
\qquad
Z_2=Z(\tau+\varepsilon)=Z+\varepsilon \dot Z(\tau) + \half \varepsilon^2 \ddot Z(\tau)+\ldots\,.
\]
Typically, only the leading and sub-leading terms
in the expansion w.r.t.\ $\varepsilon$
will contribute to the divergences.
For the superspace intervals we obtain the expansions
\begin{align}
\theta_{12}&
= \theta_2-\theta_1
= \varepsilon \dot\theta+\half \varepsilon^2 \ddot \theta+\ldots\,,
\nln
x_{12}^{+-}&
= x_1^+ - x_2^- + 4 \bar\theta_2\theta_1
= -\varepsilon p - \half\varepsilon^2  \bigbrk{ \dot p + 4\dot{\bar\theta}\dot\theta}
+\ldots\,,
\nln
x_{12}^{-+}&
= x_1^- - x_2^+ - 4 \bar\theta_1\theta_2
= -\varepsilon p - \half\varepsilon^2 \bigbrk{\dot p - 4\dot{\bar\theta}\dot\theta}
+\ldots\,,
\end{align}
where \eqref{eq:supermomentum} $p=\dot x + 2\bar\theta\dot\theta - 2\dot{\bar\theta}\theta$.
The derivations act as
$\der_1 = \der\tau (\partial_\tau - \partial_\varepsilon)$ and
$\der_2 = \der\varepsilon\, \partial_\varepsilon$.

We are now ready to perform the expansion of the correlators.
It is straight-forward to show that the chiral part of the gauge propagator
$\vev{\sgauge^\pm(1)\sgauge^\pm(2)}$
is non-singular.
This follows from the propagator for the chiral pre-potentials \eqref{eq:preprop},
which is rational and of order $\varepsilon^2$.
Two derivatives are needed to turn it into
the propagator for the gauge fields.
At worst this reduces the short-distance behavior
to $\varepsilon^0$. In other words, there cannot be singularities.

Conversely, a singularity of $\varepsilon^{-2}$ is expected
for the mixed-chiral part \eqref{eq:mixedchiralpropagator} of the propagator.
Collecting all the terms, we find%
\footnote{The additional gauge terms \eqref{eq:proptotalder}
remove the dependence on the light-cone vector $\ell$.
While also other gauge terms are conceivable,
the absence of gauge artifacts
distinguishes the particular combination \eqref{eq:proptotalder}.}
\[
\bigvev{\sgauge^+(1) \sgauge^-(2)}
=
g^2 \frac{\der\tau\, \der\varepsilon}{8 \pi^2}
\lrsbrk{
\frac{1}{\varepsilon^2}
-\frac{1}{\varepsilon}\,
\frac{4(\dot{\bar\theta}\dot\theta)\cdot p}{p^2}
+\ldots
},
\]
where the notation $(\dot{\bar\theta}\dot\theta)\cdot p$ stands
for the dot product between the vector corresponding to the bi-spinor $\dot{\bar\theta}\dot\theta$ and $p$.
The opposite chirality assignment yields the
same term with the opposite sign for the sub-leading term
\[
\bigvev{\sgauge^-(1) \sgauge^+(2)}
=
g^2 \frac{\der\tau\, \der\varepsilon}{8 \pi^2}
\lrsbrk{
\frac{1}{\varepsilon^2}
+\frac{1}{\varepsilon}\,
\frac{4(\dot{\bar\theta}\dot\theta)\cdot p}{p^2}
+\ldots
}.
\]
All contributions of the gauge connections together yield a plain $\varepsilon^{-2}$ singularity
\[
\bigvev{\sgauge(1) \sgauge(2)}
=
g^2 \frac {\der\tau\, \der\varepsilon}{4 \pi^2}
\lrsbrk{
\frac{1}{\varepsilon^2}
+\ldots
}.
\]
This is precisely the same singularity as in a
non-supersymmetric Yang-Mills theory.

Next we consider the singularities involving the scalar fields.
The gauge-scalar propagator is non-singular.
This can be checked, for example,
by splitting up the gauge field into chiral and anti-chiral components
and then picking out the scalar contributions to
$\bigvev{\sgauge^\pm(1)\der \sgauge^\mp(2)}$
or equivalently
$\bigvev{\sgauge^\pm(1)\der \sgauge^\pm(2)}$.
All terms are of order at most $1/\varepsilon$,
and the sum of the most singular parts turns out to be zero.

The scalar-scalar propagator reads (see eq.~\eqref{eq:phi-phi-propagator})
\[
\bigvev{\sscalar(1) \sscalar(2)}
=
-\frac {g^2} {16 \pi^2} \frac{C^c{}_a C^d{}_b\; q_1^{ab} q_{2, cd}}{(x^{+-}_{12})^2}\, \der \tau_1 \der \tau_2,
\]
where \(\Phi(1) = \tfrac 1 4 \Phi_{a b}(1) q_1^{a b}\; \der \tau_1\) and
\[
\label{eq:scalarbridge}
C^b{}_a = \delta_a^b + 4 \bigbrk{\theta_{12} (x_{12}^{+-})^{-1}\bar\theta_{12}}^{b}{}{}_a.
\]
Again this term yields a singularity of order $\varepsilon^{-2}$
originating from the denominator.
To simplify the contributions from the matrices $C^b{}_a$,
we can use $\grp{SU}(4)$ representation theory.
Inspection of $C^b{}_a$ shows that
it differs from the identity matrix by a sub-leading term.
At this order we can approximate all other terms
by their leading order, in which case $q_1=q_2$
are two equal vectors and the other matrix $C^d{}_c$ is
the identity.
The symmetric product of two vector representations
splits into a singlet and 20-dimensional representation.
There is no 15-dimensional (adjoint) representation to accommodate
the non-trace elements of $C^b{}_a$.
Hence we can safely replace the matrices $C^b{}_a\to \delta_a^b C$ by their
trace components $C$
\[
C=
\sfrac{1}{4}C^a{}_a
= 1 - \tr\bigsbrk{\bar\theta_{12}\theta_{12} (x_{12}^{+-})^{-1}}
= 1 - \varepsilon\tr (\dot{\bar\theta}\dot\theta p^{-1})+\ldots
= 1 - 2 \varepsilon \frac{(\dot{\bar\theta}\dot\theta)\cdot p}{p^2}+\ldots\,.
\]
We thus obtain upon expansion in $\varepsilon$
\begin{align}
\bigvev{\sscalar(1) \sscalar(2)}
&=
-\frac {g^2} {32 \pi^2} \frac{C^2 \epsilon_{a b c d} q_1^{a b} q_2^{c d}}{(x^{+-}_{12})^2}\, \der \tau_1 \,\der\tau_2
=
-\frac {g^2} {4 \pi^2} \frac{(q_1\cdot q_2)C^2}{(x^{+-}_{12})^2} \der \tau_1\, \der \tau_2
\nln
&=-\frac {g^2} {4 \pi^2} \,\der\tau\,\der\varepsilon\,
\frac{q^2}{p^2}
\lrsbrk{
\frac{1}{\varepsilon^2}
+\frac{1}{\varepsilon}
\lrbrk{ \frac{\dot q\cdot q}{q^2}
- \frac{\dot p\cdot p}{p^2}
}
+\ldots
}
,
\end{align}
which again agrees with the result
for plain non-supersymmetric scalar fields
with the appropriate usage of the super-momentum $p$ instead of $\dot x$.
Here we used the definition \(q_1 \cdot q_2 = \frac 1 8 \epsilon_{a b c d} q_1^{a b} q_2^{c d}\).

Assembling all contributions we find the singularity structure%
\footnote{Notice that the $1/\varepsilon$ term is the derivative of the $1/\varepsilon^2$ term,
thus the $1/\varepsilon$ singularity implies no additional constraints.
In fact, both singular terms can be combined to
$\brk{\partial_\tau-2\partial_\varepsilon}\partial_\varepsilon
\sbrk{f(p,q)\log \varepsilon}$ with $f(p,q)=1-q^2/p^2$.
}
\begin{align}
\bigvev{(\sgauge(1)+\sscalar(1))(\sgauge(2)+\sscalar(2))}
&=
\frac {g^2} {4 \pi^2}\,\der\tau\,\der\varepsilon\,
\lrsbrk{
\frac{1}{\varepsilon^2}\,
\frac{p^2-q^2}{p^{2}}
-\frac{1}{\varepsilon}\,
\frac{q^2}{p^2}
\lrbrk{ \frac{\dot q\cdot q}{q^2}
- \frac{\dot p\cdot p}{p^2}
}
+\ldots
}
.
\end{align}
When \(p^2 = q^2\) all the divergences cancel and the Wilson loop is free of UV-divergences.
Note, however, that pure gauge terms do have an impact on the singularities,
see the discussion below \eqref{eq:mixedchiralpropagator}.

\subsection{Conformal Symmetry}

Finally, let us confirm the full conformal invariance of
our super-loop at $\order{g^2}$.
The propagator of gauge connections discussed in \secref{sec:superspace-prop}
is manifestly invariant under translations in superspace
and under internal symmetries.
The parameters $\ell$ and $\bar\ell$ of the light-cone gauge,
however, break Lorentz invariance. As this is merely a gauge artifact,
the propagator only changes by a gauge transformation, i.e.\
by a total derivative
\begin{align}
\genfield{L}\bigvev{\sgauge(1)\,\sgauge(2)}_{(1)} &=
\bigvev{\genfield{L}_1\sgauge(1)\,\sgauge(2)}_{(1)}
+\bigvev{\sgauge(1)\,\genfield{L}_2\sgauge(2)}_{(1)} \nonumber \\&=
-\delta_\ell \bigvev{\sgauge(1)\,\sgauge(2)}_{(1)} =
\der_1 G_{12}
+\der_2 G_{21}.
\label{eq:lorentzgaugeshift}
\end{align}
Here $G_{12}$ is a function of points 1 and 2 and a one-form at point 2.
The latter combination is due to the symmetry between points 1 and 2.
It remains to confirm invariance under special conformal transformations
(up to total derivatives).

The identification of the corresponding function $G$ is rather tedious,
let us therefore show the previous statement indirectly.
First, it suffices to consider a conformal inversion as in \secref{sec:conformal-symmetry}.
Secondly, we will consider the correlator of two field strengths.
At $\order{g^2}$, the theory is effectively abelian,
and we can consider $\sfstr_{\text{lin}}=\der \sgauge$.
The derivatives remove the gauge artifacts
from the r.h.s.\ of \eqref{eq:lorentzgaugeshift}
because $\sfstr_{\text{lin}}$ is invariant under linearized gauge transformations.
Thus we have to show
\[
I \bigvev{\sfstr_{\text{lin}}(1)\,\sfstr_{\text{lin}}(2)}_{(1)} =
\bigvev{\sfstr_{\text{lin}}(1)\,\sfstr_{\text{lin}}(2)}_{(1)}.
\]
Let us consider the mixed-chiral part of the correlator (see eq.~\eqref{eq:fp-fm-propagator}).
This correlator can be expressed in terms of the combination
$\der_1(x_{12}^{-+,-1}\der_2 x_{12}^{-+})$.
The inverse of the mixed-chiral superspace interval
reads $I[x_{12}^{-+}]=-\epsilon x_1^{+,\trans,-1} x_{12}^{+-,\trans} x_2^{-,\trans,-1}$.
The inversion of the relevant combination nicely
maps to a similar combination with chiralities interchanged
and conjugated by $x_2^{-,\trans}$,
\[
  I[\der_1 (x_{12}^{-+,-1} \der_2 x_{12}^{-+})]
= \epsilon x_2^{-,\trans} \der_1 (x_{12}^{+-,\trans,-1} \der_2 x_{12}^{+-,\trans}) x_2^{-,\trans,-1} \epsilon.
\]
Within the trace, the conjugation cancels out showing that
\[
I \bigvev{\sfstr^+(1)\,\sfstr^-(2)}_{(1)} =
\bigvev{\sfstr^-(1)\,\sfstr^+(2)}_{(1)}.
\]

Using the formulas in \appref{sec:inversion}
the analogous statement can be shown to hold for the chiral-chiral
two-point function \(\langle \mathcal{F}^+(1) \mathcal{F}^+(2)\rangle\).
Altogether this implies a similar statement for the correlator of gauge connections
but with additional total derivative terms on the r.h.s.\ as in \eqref{eq:lorentzgaugeshift}.

\section{Conclusions}
\label{sec:conclusions}

In this work we have
studied in detail a non-chiral, smooth Wilson loop operator in
$\mathcal{N}=4$ SYM theory. It represents the supersymmetric completion of the
Maldacena-Wilson loop operator from 1998 \cite{Maldacena:1998im,Rey:1998ik}
and enjoys remarkable properties.
Firstly, it is the natural non-local operator to consider in the AdS/CFT correspondence as it is dual to
a minimal surface of the IIB superstring ending on the superspace boundary.
Secondly, it enjoys superconformal and local kappa-symmetry leading to a large
equivalence class of loop operators with differing paths in non-chiral superspace
but identical vacuum expectation values. Moreover,
it is a finite observable of the 4d interacting QFT -- at least to the one-loop order.
This was proven both in a detailed component field expansion using a convenient Harnad-Shnider
gauge \cite{Harnad:1985bc}
as well as in the general non-chiral superspace setting employing a light-cone gauge.

The natural next question to address concerns the possibility of Yangian symmetry. If indeed
present
the super Maldacena-Wilson loop operator represents yet another integrable observable
within planar $\mathcal{N}=4$
super Yang-Mills theory. This is to be expected as (i) the classical dual superstring
 minimal super-surface problem is Yangian symmetric \cite{Munkler:2015gja} and (ii) the leading
$\theta$ terms in the component field expansion of the vacuum expectation value enjoy
this property \cite{Muller:2013rta}. It would be important to establish this
hidden symmetry in full superspace and at the operator level. This and related
questions will be answered in our upcoming work \cite{usyangian}.

The general super-loop depends on an infinite number of variables.
In order to obtain concrete expressions for the vacuum expectation value of such operators,
it is useful to restrict to special types of contours.
Apart from very symmetric shapes like the circle, the polygonal light-like contours are the simplest one can consider.

Unfortunately, the vacuum expectation values of the polygonal super-loops are UV-divergent.
In \secref{sect:geometrycosets} we have proposed some contours which
are still simple in the sense that they depend on a small number
of parameters and which yield UV-finite observables.
These contours are piecewise-quadratic, joined in a way which insures UV finiteness at one loop.
At this point we can not exclude the possibility that UV divergences will appear at higher loop orders.
However, if such divergences do appear, then they should result in interesting quantities,
just like for polygonal Wilson loops the divergences lead to the cusp anomalous dimension
(see ref.~\cite{Polyakov:1980ca}).

The piecewise-quadratic contours can be generalized to higher degrees,
with different smoothness conditions at the joining points.

In ref.~\cite{Alday:2009zm} a mass regularization was proposed for scattering amplitudes.
 In this regularization conformal symmetry is preserved if the symmetry generators are suitably modified.
 It would be interesting to investigate if scattering amplitudes
regularized in this way are dual to super-loops similar to the ones we considered in this paper.

\paragraph{Acknowledgments.}

DM and JP thank Hagen M\"unkler for discussions. DM gratefully acknowledges the support of the DFG-funded Graduate School GK 1504. The work of NB is partially supported
by grant no.\ 200021-137616 from the Swiss National Science Foundation,
through the SNSF NCCR SwissMAP
and by grant no.\ 615203 from the European Research Council under the FP7. JP thanks the Pauli Center for Theoretical Studies Z\"urich and the Institute for 
Theoretical Physics at the ETH Z\"urich for hospitality and support in the framework of a visiting
professorship.

\appendix

\section{Conventions}
\label{sec:conventions}

\subsection{Notations}
  \begin{itemize}
  \item 10d vector indices: \(\idxtvec{m}\), \(\idxtvec{n}, \idxtvec{o},\idxtvec{p}, \ldots=0,1,\ldots,9\)
  \item 10d spinor indices: \(\idxtspin{A}, \idxtspin{B}, \idxtspin{C}, \idxtspin{D}, \ldots=1,\ldots,16\)
  \item 4d vector indices: \(\mu, \nu, \rho, \ldots=0,\ldots,3\)
  \item 4d left spinor indices: \(\alpha, \beta, \gamma, \ldots=1,2\)
  \item 4d right spinor indices: \(\dot{\alpha}, \dot{\beta}, \dot\gamma, \ldots=\dot 1,\dot 2\)
  \item 6d vector indices: \(i, j, k, \ldots=1,\ldots,6\)
  \item 6d spinor indices: \(a, b, c, \ldots=1,\ldots,4\)
  \item Color indices: \(\mathfrak{a}\), \(\mathfrak{b}\)
  \item 10d $\mathcal{N}=1$ superspace: $\{x^{\idxtvec{m}}, \theta^{\idxtspin{A}}\}$
  \item 4d $\mathcal{N}=4$ non-chiral superspace: $\{x^{\mu},\theta^{a\alpha}, \bar\theta^{\dot\alpha}{}_{a}\}$
  \item 10d $\mathcal{N}=1$ super-connections: $\sgauge_{\idxtvec{m}}$, $\sgauge_{\idxtspin{A}}$
  \item 4d $\mathcal{N}=4$ super-connections and scalar superfield:
$\sgauge_{\mu}$, $\sgauge_{a\alpha}$,  $\sgauge^{a}{}_{\dot\alpha}$ and  $\Phi_{i}$ or $\Phi_{ab}$
  \item 10d super-momentum: $p^{\idxtvec{m}}=\dot x^{\idxtvec{m}}+ \theta\Gamma^{\idxtvec{m}}\dot\theta$
  \item 4d super-momentum: $p^{\mu}=\dot x^{\mu}+ \theta \sigma^{\mu}\dot{\bar\theta} - \dot\theta\sigma^{\mu}\bar\theta$
  \item Scalar coupling: $q^{i}=n^{i}\sqrt{p^{\mu}p_{\mu}}$ with $q^{i}q^{i}=p^{\mu}p_{\mu}$ and $n^{i}n^{i}=1$
\end{itemize}

\subsection{Gamma-Matrices}
\label{sec:gamma-matrices}

We use the mostly minus signature.  The four-dimensional Pauli
matrices are given by \(\sigma_{\mu} = (\mathbf{1}, \vec{\sigma})\)
and \(\bar{\sigma}_{\mu} = (\mathbf{1}, -\vec{\sigma})\), where
\(\sigma_{1} = \left(\begin{smallmatrix}
    0&1\\1&0\end{smallmatrix}\right)\), \(\sigma_{2} =
\left(\begin{smallmatrix}0&-i\\i&0\end{smallmatrix}\right)\) and
\(\sigma_{3} =
\left(\begin{smallmatrix}1&0\\0&-1\end{smallmatrix}\right)\).  They
satisfy the usual algebra
\begin{equation}
  \sigma_{\mu} \bar{\sigma}_{\nu} + \sigma_{\nu} \bar{\sigma}_{\mu} =
  2 \eta_{\mu \nu}, \qquad
  \bar{\sigma}_{\mu} \sigma_{\nu} + \bar{\sigma}_{\nu} \sigma_{\mu} =
  2 \eta_{\mu \nu}.
\end{equation}

The Pauli matrices transform between the four-dimensional chirality
left and right spinors.  We index the left chirality spinors by Greek
letters from the beginning of the alphabet while the right chirality
spinors by dotted Greek letters from the beginning of the the
alphabet.  We denote the four-dimensional space-time directions by
Greek variables \(\mu, \nu, \rho, \tau\) while the same hatted
variables denote space-time directions in ten dimensions.  Then the
index structure of the four-dimensional Pauli matrices is
\(\sigma_{\mu, \alpha \dot{\alpha}}\),
\(\bar{\sigma}_{\mu}^{\dot{\alpha} \alpha}\).

We define
\begin{gather}
  \sigma_{\mu \nu} = \half (\sigma_{\mu}\bar{\sigma}_{\nu} -
     \sigma_{\nu} \bar{\sigma}_{\mu}), \qquad
  \bar{\sigma}_{\mu \nu} = \half (\bar{\sigma}_{\mu} \sigma_{\nu} -
     \bar{\sigma}_{\nu} \sigma_{\mu}),\\
  \sigma_{\mu \nu \rho} = \sigma_{[\mu}\bar{\sigma}_{\nu} \sigma_{\rho]},
  \qquad
  \bar{\sigma}_{\mu \nu \rho} = \bar{\sigma}_{[\mu} \sigma_{\nu} \bar{\sigma}_{\rho]},
\end{gather} where the anti-symmetrization is done with unit strength.
If \(\epsilon_{\mu \nu \rho \tau}\) is the anti-symmetric tensor
normalized as \(\epsilon_{0123} = 1\), \(\epsilon^{0123} = -1\), then
we have
\begin{gather}
  \sigma_{\mu \nu} = \frac i 2 \epsilon_{\mu \nu \rho \tau} \sigma^{\rho \tau},
  \qquad
  \bar{\sigma}_{\mu \nu} = -\frac i 2 \epsilon_{\mu \nu \rho \tau} \bar{\sigma}^{\rho \tau},\\
  \sigma_{\mu} = \frac i {3!} \epsilon_{\mu \nu \rho \tau} \sigma^{\nu \rho \tau},
  \qquad
  \sigma^{\mu \nu \rho} = i \epsilon^{\mu \nu \rho \tau} \sigma_{\tau}, \\
  \bar{\sigma}_{\mu} = -\frac i {3!} \epsilon_{\mu \nu \rho \tau} \bar{\sigma}^{\nu \rho \tau}, \qquad
  \bar{\sigma}^{\mu \nu \rho} = -i \epsilon^{\mu \nu \rho \tau} \bar{\sigma}_{\tau}.
\end{gather}

The anti-symmetric \(2 \times 2\) matrix\footnote{There are two such
  anti-symmetric matrices, one with undotted indices and one with
  dotted indices.  They act on different spaces, but we do not
  distinguish them by notation since they are numerically equal.}
\(\epsilon =
\left(\begin{smallmatrix}0&1\\-1&0\end{smallmatrix}\right)\) is useful
when performing contractions.  It also appears when expressing symmetry properties
\begin{gather}
  \epsilon^{\trans} = -\epsilon, \qquad
  (\sigma_{\mu} \epsilon)^{\trans} = -\bar{\sigma}_{\mu} \epsilon, \qquad
  (\sigma_{\mu \nu} \epsilon)^{\trans} = \sigma_{\mu \nu} \epsilon, \qquad
  (\bar{\sigma}_{\mu \nu} \epsilon)^{\trans} = \bar{\sigma}_{\mu \nu} \epsilon.
\end{gather}

Other useful identities involving the four-dimensional Pauli
matrices are
\begin{gather}
  \sigma_{\mu, \alpha \dot{\alpha}} \bar{\sigma}^{\mu, \dot{\beta} \beta} =
  2 \delta_{\alpha}^{\beta} \delta_{\dot{\alpha}}^{\dot{\beta}}, \qquad
  (\sigma^{\mu \nu})^{\hphantom{\alpha} \beta}_{\alpha}
  (\bar{\sigma}_{\mu \nu})^{\dot{\beta}}_{\hphantom{\dot{\beta}} \dot{\alpha}} = 0, \qquad
  (\sigma^{\mu \nu})_{\alpha}^{\hphantom{\alpha} \beta}
  (\sigma_{\mu \nu})_{\gamma}^{\hphantom{\gamma} \delta} =
  -8 \delta_{\alpha}^{\delta} \delta_{\gamma}^{\beta} + 4
  \delta_{\alpha}^{\beta} \delta_{\gamma}^{\delta}.
\end{gather} An easy way to see why the contraction \(\sigma^{\mu \nu}
\bar{\sigma}_{\mu \nu} = 0\) is to notice that it is a contraction
between a selfdual and an anti-selfdual rank two tensor.

If we do not insist on the reality conditions,\footnote{If we do
  insist on preserving the reality of the ten-dimensional Pauli
  matrices, then only an \(\grp{SO}(4) \subset \grp{SU}(4) \simeq
  \grp{Spin}(6)\) symmetry is manifest.} we can write the
ten-dimensional Pauli matrices as
\begin{alignat}{2}
  \Gamma_{\mu, (a \alpha), \genfrac(){0pt}{}{b}{\dot{\alpha}}}
  &= \delta_{a}^{b} \sigma_{\mu, \alpha \dot{\alpha}}, &\qquad
  \tilde{\Gamma}_{\mu}^{(a \alpha),
    \genfrac(){0pt}{}{b}{\dot{\alpha}}} &= \delta_{b}^{a}
  \bar{\sigma}_{\mu}^{\dot{\alpha} \alpha},\\
  \Gamma_{[ab], (c \alpha), (d \beta)} &= \epsilon_{\alpha \beta}
  \epsilon_{a b c d}, &\qquad
  \tilde{\Gamma}_{[a b]}^{(c \alpha), (d \beta)} &= \epsilon^{\alpha
    \beta} (\delta_{a}^{c} \delta_{b}^{d} - \delta_{b}^{c}
  \delta_{a}^{d}),\\
  \Gamma_{[a b], \genfrac(){0pt}{}{c}{\dot{\alpha}},
    \genfrac(){0pt}{}{d}{\dot{\beta}}} &= \epsilon_{\dot{\alpha}
    \dot{\beta}} (\delta_{a}^{c} \delta_{b}^{d} - \delta_{b}^{c}
  \delta_{a}^{d}), &\qquad
  \tilde{\Gamma}_{[a b]}^{\genfrac(){0pt}{}{c}{\dot{\alpha}},
    \genfrac(){0pt}{}{d}{\dot{\beta}}} &= \epsilon^{\dot{\alpha}
    \dot{\beta}} \epsilon_{a b c d}.
\end{alignat} Here we have identified the six-dimensional vectors
\(\Gamma_{i}\) for \(i = 4, \dotsc, 9\) with \(\Gamma_{[a b]}\) where
\([a b] = -[b a]\), for \(a, b = 1, \dotsc, 4\) is an anti-symmetric
index.  We have not written all the matrix elements, but the missing
ones can either be obtained by symmetry or they vanish.

The metric in the extra six dimensions can be obtained from the Pauli
matrices.  We find
\begin{equation}
  \Gamma_{[a b]} \tilde{\Gamma}_{[c d]} + \Gamma_{[c d]}
  \tilde{\Gamma}_{[a b]} = 2 \eta_{[a b] [c d]} = - \epsilon_{a b c
    d},\qquad
  \tilde{\Gamma}_{[a b]} \Gamma_{[c d]} + \tilde{\Gamma}_{[c d]}
  \Gamma_{[a b]} = 2 \eta_{[a b] [c d]}.
\end{equation}  Whenever we contract a pair of upper anti-symmetric
indices with a pair of lower anti-symmetric indices we introduce a
factor of \(\tfrac 1 2\) by hand, to avoid double counting.  For
example, from above we have \(\eta_{[a b] [c d]} = -\tfrac 1 2
\epsilon_{a b c d}\), \(\eta^{[c d] [e f]} = -2 \epsilon^{c d e f}\).
Their contraction is \(\delta_{[a b]}^{[e f]} = \tfrac 1 2 \eta_{[a b]
[c d]} \eta^{[c d] [e f]} = \delta_{a}^{e} \delta_{b}^{f} -
\delta_{b}^{e} \delta_{a}^{f}\).

We sometimes use anti-symmetrized products of ten-dimensional \(\Gamma\) matrices such as
\begin{gather}
  \Gamma^{\hat{\mu} \hat{\nu}}
= \frac 1 2 (\Gamma^{\hat{\mu}} \tilde{\Gamma}^{\hat{\nu}} - \Gamma^{\hat{\mu}} \tilde{\Gamma}^{\hat{\nu}}),
\qquad
  \tilde{\Gamma}^{\hat{\mu} \hat{\nu}}
= \frac 1 2 (\tilde{\Gamma}^{\hat{\mu}} \Gamma^{\hat{\nu}} - \tilde{\Gamma}^{\hat{\mu}} \Gamma^{\hat{\nu}}),
\label{eqn:10dPauli_antisym}
\end{gather}
and analogous formulas for higher rank products.
As a rule, such a matrix has no tilde if the row index
is lower and it has a tilde if the row index is upper.

\subsection{Superspace Geometry}
\label{sec:supersp-geom}

The four-dimensional superspace has coordinates \((x, \theta,
\bar{\theta})\).  We will use a matrix notation similar to the one in
ref.~\cite{Beisert:2012gb}.  The bosonic coordinate \(x\) is
represented by a \(2 \times 2\) matrix with index structure
\(x^{\dot{\alpha} \alpha}\), while \(\theta^{a \alpha}\) is \(4 \times
2\) and \(\bar{\theta}^{\dot{\alpha}}{}_{a}\) is a \(2 \times 4\)
matrix.  Here lowercase Latin letters from the beginning of the
alphabet represent \(\grp{SU}(4)\) R-symmetry indices taking
values \(1\) through \(4\).

Under supersymmetry transformations the superspace coordinates transform as
\begin{gather}
  \label{eq:susy-transf}
  \delta \theta = \varrho, \qquad
  \delta \bar{\theta} = \bar{\varrho}, \qquad
  \delta x = 2 \bar{\theta} \varrho - 2 \bar{\varrho} \theta,
\end{gather} where \(\varrho\), \(\bar{\varrho}\) are the parameters
of the supersymmetry transformations.  It is useful to introduce
\(x^\pm = x \mp 2 \bar{\theta} \theta\), which transform in a simpler
way
\begin{gather}
  \delta x^+ = -4 \bar{\varrho} \theta, \qquad
  \delta x^- = 4 \bar{\theta} \varrho.
\end{gather}

Note that \(x^{\pm}\) satisfy some constraints
\begin{gather}
  x = \frac 1 2 (x^+ + x^-), \qquad
  \bar{\theta} \theta = \frac 1 4 (x^- - x^+).
\end{gather}
These constraints are preserved by the supersymmetry transformations.

So far we have taken the superspace coordinates \((x, \theta,
\bar{\theta})\) to be complex.  We can impose the following reality
conditions
\begin{gather}
  x^\ddagger = x, \qquad
  \theta^\ddagger = \bar{\theta}, \qquad
  \bar{\theta}^\ddagger = \theta, \qquad
  (x^{\pm})^\ddagger = x^{\mp},
\end{gather}
where, for Gra{\ss}mann variables \(\chi, \psi\) we use the convention
that \((\chi \psi)^\ddagger = -\psi^\ddagger \chi^\ddagger\).  This
convention, which is not so frequent in the physics literature, allows
us to avoid introducing many factors of \(i\) in the equations.  For
transposition we use the same convention as for \(\ddagger\).

Let us now consider two points with coordinates \((x_1, \theta_1,
\bar{\theta}_1)\) and \((x_2, \theta_2, \bar{\theta}_2)\).  Using the
supersymmetry transformations in eq.~\eqref{eq:susy-transf}, we can
translate the second point to zero while the first point acquires
coordinates \((x_1 - x_2 + 2 \bar{\theta}_2 \theta_1 - 2
\bar{\theta}_1 \theta_2, \theta_{12}, \bar{\theta}_{12})\).  Here we
have used the short-hand notation \(\theta_{12} = \theta_1-\theta_2\)
and similarly for \(\bar{\theta}_{12}\).  It is then natural to define
the bosonic supersymmetric interval \(x_{12} = x_1 - x_2 + 2
\bar{\theta}_2 \theta_1 - 2 \bar{\theta}_1 \theta_2\).  The analogs of
the chiral and anti-chiral coordinates for the translated point are
\begin{align}
  x_{12}^{+-} = x_1^+ - x_2^- + 4 \bar{\theta}_2 \theta_1 =
     x_{12} - 2 \bar{\theta}_{12} \theta_{12},\\
  x_{12}^{-+} = x_1^- - x_2^+ - 4 \bar{\theta}_1 \theta_2 =
     x_{12} + 2 \bar{\theta}_{12} \theta_{12}.
\end{align}

The supersymmetry covariant derivatives are defined as
\begin{equation}
  \sdel_{a \alpha} = \partial_{a \alpha} + \sigma^{\mu}_{\alpha \dot{\alpha}}
  \bar{\theta}^{\dot{\alpha}}{}_{a} \partial_{\mu}, \qquad
  \bar{\sdel}^{a}{}_{\dot{\alpha}} = \bar{\partial}^{a}{}_{\dot{\alpha}} +
  \theta^{a \alpha}\sigma_{\alpha \dot{\alpha}}^{\mu} \partial_{\mu}.
\end{equation}
The chiral covariant derivative \(\sdel\) annihilates
functions depending on \((x^-, \bar{\theta})\) while the anti-chiral
covariant derivative \(\bar{\sdel}\) annihilates functions depending on
\((x^+, \theta)\).  The quantities \(x_{12}^{+-}\) and \(x_{12}^{-+}\)
are mixed; for instance, \(x_{12}^{+-}\) is chiral in the coordinates
of point \(1\) and anti-chiral in the coordinates of point \(2\).

The transformations under superconformal boosts (\(\gen{S}\) and
\(\bar{\gen{S}}\)) are as follows
\begin{gather}
\label{eqn:s-sbar-x-transformation}
  \delta x = 2 \bar{\theta} \bar{\rho} x^- - 2 x^+ \rho \theta,
\\
  \label{eqn:s-sbar-theta-transformation}
  \delta \theta = \bar{\rho} x^+ - 4 \theta \rho \theta, \qquad
  \delta \bar{\theta} = x^- \rho + 4 \bar{\theta} \bar{\rho} \bar{\theta},
\\
  \delta x^+ = -4 x^+ \rho \theta, \qquad
  \delta x^- = 4 \bar{\theta} \bar{\rho} x^-.
  \label{eqn:s-sbar-x-theta-transformation}
\end{gather}

In working with differentials in superspace we adopt the bi-graded
point of view: each expression has a form grading and a Gra{\ss}mann
grading.  For example, \(x\), \(\theta\) and \(\bar{\theta}\) have
zero form grading while \(\der x\), \(\der \theta\) and \(\der \bar{\theta}\)
have form grading one.  When permuting two such expressions, we pick
up a sign from the differential grading and another from the
Gra{\ss}mann grading.  This means that \(\der \theta\) commutes with \(\der
\theta\) and \(\der \bar{\theta}\) but anti-commutes with \(\der x\).

The superspace comes equipped with vielbeine
\begin{equation}
  \svielB^{\dot\alpha\alpha}
= \der x^{\dot\alpha\alpha}
- 2 \der \bar{\theta}^{\dot\alpha}{}_a \theta^{a\alpha}
+ 2 \bar{\theta}^{\dot\alpha}{}_a \der \theta^{a\alpha},
\qquad
  \sviel^{a\alpha} = \der \theta^{a\alpha},
\qquad
  \sviel^{\dot\alpha}{}_a = \der \bar{\theta}^{\dot\alpha}{}_a
\end{equation} whose differentials are%
\footnote{In the following we
will mostly omit the wedge symbol in the wedge product of differential
forms.}
\begin{equation}
  \der \svielB = 4 \svielC \wedge \svielF, \qquad
  \der \svielF = 0, \qquad \der \svielC = 0.
\end{equation}

The total differential in superspace can be written using normal or
supersymmetry covariant derivatives
\begin{equation}
  \der = \half \der x^{\dot{\alpha} \alpha} \partial_{\alpha \dot{\alpha}} +
   \der \theta^{a \alpha} \partial_{a \alpha} +
   \der \bar{\theta}^{\dot{\alpha}}{}_a \bar{\partial}^a{}_{\dot{\alpha}} =
   \half \svielB^{\dot{\alpha} \alpha} \partial_{\alpha \dot{\alpha}} +
   \svielF^{a \alpha} \sdel_{a \alpha} +
   \svielC^{\dot{\alpha}}{}_{a} \bar{\sdel}^a{}_{\dot{\alpha}}.
\end{equation}

The gauge connection is a one-form in superspace which can be
decomposed on the vielbein basis as
\begin{equation}
  \sgauge = \half \svielB^{\dot{\alpha} \alpha} \sgauge_{\alpha \dot{\alpha}} +
   \svielF^{a \alpha} \sgauge_{a \alpha} +
   \svielC^{\dot{\alpha}}{}_{a} \sgauge^a{}_{\dot{\alpha}}.
\end{equation}

The supersymmetry and gauge covariant derivatives are defined as
\begin{equation}
  \scdelB_{\alpha \dot{\alpha}} = \partial_{\alpha \dot{\alpha}} + \sgauge_{\alpha \dot{\alpha}},
\qquad
  \scdelF_{a \alpha} = \sdel_{a \alpha} + \sgauge_{a \alpha},
\qquad
  \scdelC^a{}_{\dot{\alpha}} = \bar{\sdel}^a{}_{\dot{\alpha}} + \sgauge^a{}_{\dot{\alpha}}.
\end{equation}
They satisfy some constraints
\begin{equation}
  \label{eq:constraints-4d-language}
  \{\scdelF_{a \alpha}, \scdel_{b \beta}\} = -4 \epsilon_{\alpha \beta} \Phi_{a b}, \qquad
  \{\scdelF_{a \alpha}, \scdelC^b{}_{\dot{\beta}}\}
= 2 \delta_a^b \scdelB_{\alpha \dot{\beta}}, \qquad
  \{\scdelC^a{}_{\dot{\alpha}}, \scdelC^b{}_{\dot{\beta}}\}
= -2 \epsilon_{\dot{\alpha} \dot{\beta}} \epsilon^{a b c d} \Phi_{c d}.
\end{equation}

The components of the field strength can be computed by taking
commutators or anti-commutators of gauge covariant derivatives.  They
can all be expressed in terms of covariant derivatives acting on the
scalar fields, as follows
\begin{align}
  [\scdelF_{a \alpha}, \scdelB_{\beta \dot{\beta}}]
&=
  \frac 2 3 \epsilon_{\alpha \beta} [\scdelC^b{}_{\dot{\beta}}, \Phi_{a b}],
\\
  [\scdelC^a{}_{\dot{\alpha}}, \scdelB_{\beta \dot{\beta}}]
&=
  \frac 1 3 \epsilon_{\dot{\alpha} \dot{\beta}} \epsilon^{abcd} [\scdelF_{b \beta}, \Phi_{c d}],
\\
  [\scdelB_{\alpha \dot{\alpha}}, \scdelB_{\beta \dot{\beta}}]
&=
  \frac 1 {24} \epsilon_{\dot{\alpha} \dot{\beta}}
\epsilon^{abcd} \{\scdelF_{a \alpha}, [\scdelF_{b \beta}, \Phi_{cd}]\}
  +\frac 1 {12} \epsilon_{\alpha \beta}
\{\scdelC^a{}_{\dot{\alpha}}, [\scdelC^b{}_{\dot{\beta}}, \Phi_{ab}]\}.
\end{align}

\section{Cosets}
\label{sec:cosets}

The \(\mathcal{N}=1\) ten-dimensional superspace can be described as coset of the super Poincar\'e group by the right action of the Lorentz
group.  A representative of this coset is \(g(x,\theta) = \exp (x \gen{P} +
\theta \gen{Q})\).  The left action yields the supersymmetry transformations
\begin{equation}
  e^{\epsilon \gen{Q}} g(x, \theta) = g(x - \epsilon \Gamma \theta, \theta + \epsilon).
\end{equation}

We can define the vielbeine as
\begin{equation}
  g^{-1}(x, \theta) \der g(x, \theta) = (\der x + \theta \Gamma \der \theta) \gen{P} + \der \theta \gen{Q}.
\end{equation}
Since the left-hand side is invariant under the left action it follows
that the vielbeine are invariant under supersymmetry.  The bosonic
vielbein can be used to define a supersymmetric momentum \(p^{\idxtvec{m}} =
\dot{x}^{\idxtvec{m}} + \theta \Gamma^{\idxtvec{m}} \dot{\theta}\).

The kappa-symmetry can be thought of as a \emph{right} action, in
the following sense (see also ref.~\cite{McArthur:1999dy} for a
similar analysis for \(p\)-branes).  We pick a path in superspace,
defined parametrically by \(Z(\tau) = (x(\tau), \theta(\tau))\).  Then, we
introduce an equivalence relation
\begin{equation}
  g(Z(\tau)) \sim g(Z_1(\tau)) = g(Z(\tau)) \exp \bigsbrk{\kappa(\tau) p^{\idxtvec{m}}(Z(\tau)) \tilde{\Gamma}_{\idxtvec{m}} \gen{Q}}.
\end{equation}

Actually, in order for this to be an equivalence relation, we need to
show that the right action transformations compose correctly, in a
group-like fashion.  More concretely,
\begin{align}
  g(Z(\tau)) \sim g(Z_1(\tau))
= g(Z(\tau)) \exp \bigsbrk{p^{\idxtvec{m}}(Z(\tau)) \kappa_1(\tau) \tilde{\Gamma}_{\idxtvec{m}} \gen{Q}} \\
 \sim g(Z_2(\tau)) = g(Z_1(\tau)) \exp \bigsbrk{p^{\idxtvec{m}}(Z_1(\tau)) \kappa_2(\tau) \tilde{\Gamma}_{\idxtvec{m}} \gen{Q}},
\end{align}
where the outer transformation acts at the transformed point \(Z_1\).

The reparametrization of the path can also be described by a right
action.  Defining
\begin{equation}
  g(Z(\tau)) \sim g(Z_1(\tau))
= g(Z(\tau)) \exp \bigsbrk{\sigma_1(\tau) p^{\idxtvec{m}}(Z(\tau)) \gen{P}_{\idxtvec{m}}
+ \sigma_1(\tau) \dot{\theta}(\tau) \gen{Q}},
\end{equation}
we obtain \(x_1(\tau) = x(\tau) + \sigma(\tau) \dot{x}(\tau)\), \(\theta_1(\tau) =
\theta(\tau) + \sigma(\tau) \dot{\theta}(\tau)\), which are just the
re\-pa\-ra\-me\-tri\-zation transformations \(\tau \to \tau_1 = \tau - \sigma(\tau)\).

In order to show that the right action is consistent, we proceed as
follows.  We compute the composition of two infinitesimal right
actions to second order in the small parameters.  Then, we take the
logarithm and subtract the corresponding quantity obtained by doing
the composition in the opposite order.  This corresponds to taking the
commutator.  If this commutator has the same structure as another
right action, then the right action is consistent.

Let us show this explicitly for the path reparametrization.  We find
\begin{multline}
  \exp \bigsbrk{\sigma_1 p^{\idxtvec{m}}(Z_1) \gen{P}_{\idxtvec{m}} + \sigma_1 \dot{\theta}_1 \gen{Q}}
  \exp \bigsbrk{\sigma_2 p^{\idxtvec{m}}(Z_2) \gen{P}_{\idxtvec{m}} + \sigma_2 \dot{\theta}_2 \gen{Q}} \\
= \exp \lrsbrk{\bigbrk{(\sigma_1 + \sigma_2 + \sigma_2 \dot{\sigma}_1) p^{\idxtvec{m}}
+ \sigma_1 \sigma_2 \dot{p}^{\idxtvec{m}}} \gen{P}_{\idxtvec{m}}
+ \bigbrk{(\sigma_1 + \sigma_2 + \sigma_2 \dot{\sigma}_1) \dot{\theta} + \sigma_1 \sigma_2 \ddot{\theta}} \gen{Q}}.
\end{multline}
Computing the commutator we find
\begin{equation}
  [\delta_{\sigma_1}, \delta_{\sigma_2}] = \delta_{\sigma_2 \dot{\sigma}_1 - \sigma_1 \dot{\sigma}_2},
\end{equation}
which is the expected rule for the algebra of diffeomorphisms.
Following the same steps we find that the commutator of a
reparametrization and a kappa-transformation is another
kappa-transformation
\begin{equation}
  [\delta_{\sigma_1}, \delta_{\kappa_2}] = \delta_{\kappa(\sigma_1, \kappa_2)},
\end{equation}
with \(\kappa(\sigma_1, \kappa_2) = \dot{\sigma}_1 \kappa_2 - \sigma_1 \dot{\kappa}_2\).

Finally, the composition of two kappa-transformations is more
involved.  We only give the final answer, without going through the
derivation.
\begin{equation}
  [\delta_{\kappa_1}, \delta_{\kappa_2}] = \delta_{\kappa(\kappa_1, \kappa_2)} + \delta_{\sigma(\kappa_1, \kappa_2)},
\end{equation} where
\begin{equation}
\kappa(\kappa_1, \kappa_2) = 4 (\dot{\theta} \kappa_2) \kappa_1 -
  4 (\dot{\theta} \kappa_1) \kappa_2 +
  2 (\kappa_1 \tilde{\Gamma}_{\idxtvec{m}} \kappa_2) (\Gamma^{\idxtvec{m}} \dot{\theta}),\qquad
  \sigma(\kappa_1, \kappa_2) = - 4 p^{\idxtvec{m}} (\kappa_1 \tilde{\Gamma}_{\idxtvec{m}} \kappa_2).
\end{equation}

The spinors \(\kappa\) which appear in the kappa-transformations
are defined up to an equivalence \(\kappa \sim \kappa + p \cdot \Gamma \xi\).
We now show that this equivalence relation is compatible with the commutation relations.
 In other words, we need to show that
\begin{align}
  [\delta_{\sigma_1}, \delta_{\kappa_2 + p \cdot \Gamma \xi_2}] &=
  [\delta_{\sigma_1}, \delta_{\kappa_2}] + \delta_{p \cdot \Gamma \xi(\sigma_1, \xi_2)},
\\
  [\delta_{\kappa_1 + p \cdot \Gamma \xi_1}, \delta_{\kappa_2}] &=
  [\delta_{\kappa_1}, \delta_{\kappa_2}] + \delta_{p \cdot \Gamma \xi(\kappa_2, \xi_1)}.
\end{align}

When computing a commutator like \([\delta_{\sigma_1}, \delta_{\kappa_2 + p \cdot \Gamma \xi_2}]\)
there are two contributions%
\footnote{A similar subtlety appears when checking the Jacobi identities
of kappa-transformations and re\-pa\-ra\-me\-tri\-za\-tions.}%
; one computed as if \(\sigma_1\)
did not act on \(p\) and the second taking into account the action of reparametrizations on \(p\).
A similar statement holds for kappa transformations.  Hence, we obtain
\begin{gather}
  [\delta_{\sigma_1}, \delta_{\kappa_2 + p \cdot \Gamma \xi_2}] =
  \delta_{\kappa(\sigma_1, \kappa_2 + p \cdot \Gamma \xi_2)} +
  \delta_{(\delta_{\sigma_1} p) \cdot \Gamma \xi_2},\\
  [\delta_{\kappa_1 + p \cdot \Gamma \xi_1}, \delta_{\kappa_2}] =
  \delta_{\kappa(\kappa_1 + p \cdot \Gamma \xi_1, \kappa_2)} -
  \delta_{(\delta_{\kappa_2} p) \cdot \Gamma \xi_1},
\end{gather}
where
\begin{align}
  \kappa(\sigma_1, \kappa_2 + p \cdot \Gamma \xi_2) &= \kappa(\sigma_1, \kappa_2) +
  p \cdot \Gamma (\dot{\sigma}_1 \xi_2 - \sigma_1 \dot{\xi}_2) -
  \sigma_1 \dot{p} \cdot \Gamma \xi_2,
\\
  \kappa(\kappa_1 + p \cdot \Gamma \xi_1, \kappa_2) &= \kappa(\kappa_1, \kappa_2) +
  p \cdot \Gamma (4 (\xi_1 \kappa_2) \dot{\theta} -
    2 (\dot{\theta} \Gamma_\mu \xi_1) \tilde{\Gamma}^\mu \kappa_2)
  -2 (\kappa_2 \tilde{\Gamma}_\mu p \cdot \Gamma \dot{\theta}) (\Gamma^\mu \xi_1).
\end{align}
Now, using
\(\delta_\sigma p = \dot{\sigma} p + \sigma \dot{p}\)
and \(\delta_{\kappa} p = -2 \dot{\theta} \Gamma^\mu p \cdot \tilde{\Gamma} \kappa\),
we find that indeed the commutators are compatible with the equivalence \(\kappa \sim \kappa + p \cdot \Gamma \xi\) and
\begin{equation}
  \xi(\sigma_1, \xi_2) = 2 \dot{\sigma}_1 \xi_2 - \sigma_1 \dot{\xi}_2, \qquad
  \xi(\kappa_2, \xi_1) = 4 (\xi_1 \kappa_2) \dot{\theta}
+ 4 (\dot{\theta} \kappa_2) \xi_1 - 2 (\dot{\theta} \Gamma_\mu \xi_1) \tilde{\Gamma}^\mu \kappa_2.
\end{equation}


\section{Inversion}
\label{sec:inversion}

Under inversion the superspace coordinates transform as follows
\begin{gather}
  I(x^{\pm}) = \epsilon (x^{\mp})^{\trans,-1} \epsilon, \\
  I(\theta) = -M \bar{\theta}^{\trans} (x^{-})^{\trans,-1} \epsilon, \\
  I(\bar{\theta}) = \epsilon (x^{+})^{\trans,-1}
  \theta^{\trans} M^{-1}.
\end{gather}  It can be checked that the constraint \(x^{-}-x^{+}=4
\bar{\theta} \theta\) is preserved by inversion.  The reality
conditions \(\theta^{\ddagger} = \bar{\theta}\) and \((x^{+})^{\ddagger} =
x^{-}\) are also preserved.  The inversion for \(x\) can be obtained
from \(x = \tfrac 1 2 (x^{+} + x^{-})\)
\begin{equation}
  I(x) = \epsilon (x^{-})^{\trans, -1} x^{\trans} (x^{+})^{\trans, -1}
  \epsilon.
\end{equation}

For the purpose of checking the conformal invariance of various two-point functions,
the following transformations under inversion are useful
\begin{align}
  x_{12}^{+-} &\to -\epsilon x_2^{+,\trans,-1} x_{12}^{-+,\trans} x_1^{-,\trans,-1} \epsilon,
\\
  \bar{\theta}_{12} &\to \epsilon x_2^{+,\trans,-1}
 (\theta_{12}^\trans (1 + 4 \theta_1 x_1^{+,-1} \bar{\theta}_1)^{\trans}
- x_{12}^{-+,\trans} x_1^{+,\trans,-1} \theta_1^{\trans}) M^{-1},
\\
  \bar{\theta}_{12} &\to \epsilon x_1^{+,\trans,-1}
(\theta_{12}^\trans (1 + 4 \theta_2 x_2^{+,-1} \bar{\theta}_2)^{\trans}
- x_{12}^{+-,\trans} x_2^{+,\trans,-1} \theta_2^{\trans}) M^{-1},
\\
  \der_1 (x_{12}^{-+,-1} \bar{\theta}_{12}) &\to \epsilon x_2^{-,\trans} \der_1 (\theta_{12} x_{12}^{+-,-1})^{\trans}
(1 + 4 \theta_2 x_2^{+,-1} \bar{\theta}_2)^{\trans} M^{-1},
\\
  (1 + 4 \theta_{12} x_{12}^{+-,-1} \bar{\theta}_{12}) &\to
  M (1 - 4 \theta_2 x_2^{-,-1} \bar{\theta}_2)^{\trans}
  (1 - 4 \theta_{12} x_{12}^{-+,-1} \bar{\theta}_{12})^{\trans}
  (1 + 4 \theta_1 x_1^{+,-1} \bar{\theta}_1)^{\trans} M^{-1},\\
  \der_1 (x_{12}^{-+,-1} \der_2 x_{12}^{-+})
&\to \epsilon x_2^{-,\trans} \der_1 (x_{12}^{+-,\trans,-1} \der_2 x_{12}^{+-,\trans}) x_2^{-,\trans,-1} \epsilon.
\end{align}
Some other useful transformations are obtained trivially by relabeling
\(1 \leftrightarrow 2\) or by \(\ddagger\) conjugation.  It is noteworthy
that the quantities \(\der_1 (x_{12}^{-+,-1} \bar{\theta}_{12})\) and \(\der_1 (x_{12}^{-+,-1} \der_2 x_{12}^{-+})\) which
appear in the correlation functions transform homogeneously under
inversion.

\bibliographystyle{nb}
\bibliography{Superwilson10d}

\end{document}